\documentclass[a4paper, 11pt]{article}
\usepackage[margin=1in]{geometry}

\usepackage{algorithm}
\usepackage{algpseudocode}
\usepackage{amsmath}
\usepackage{amsfonts}
\usepackage{amssymb}
\usepackage{amsthm} 
\usepackage{multicol}
\usepackage{graphicx}
\usepackage{enumerate}
\usepackage[inline,shortlabels]{enumitem}
\usepackage{bbm}
\usepackage[usenames,dvipsnames]{color}
\usepackage[colorlinks=true,linkcolor=blue, citecolor=blue, urlcolor=blue]{hyperref}
\usepackage{fancyhdr}
\usepackage{epsfig}
\usepackage{epstopdf}
\usepackage{caption}
\usepackage[textfont=footnotesize,justification=centering]{subcaption}
\usepackage{multirow}
\usepackage{appendix}
\usepackage{tabularx}
\usepackage{threeparttable}
\usepackage{float}
\usepackage{amsfonts}
\usepackage{url}
\usepackage{comment}

\usepackage{titling}
\usepackage{sectsty}
\subsubsectionfont{\normalfont\itshape}

\newtheorem*{theorem-non}{Theorem}
\newtheorem{theorem}{Theorem}

\newtheorem{proposition}{Proposition}

\newtheorem{corollary}{Corollary}
\newtheorem{remark}{Remark}

\newtheorem{example}{Example}

\usepackage[sort]{natbib}
\usepackage[UTF8,scheme=plain]{ctex}

\providecommand{\keywords}[1]
{
  	
  \textbf{\textit{Keywords ---}} #1
}

\def \E{\mathbb{E}}
\def \P{\mathbb{P}}

\DeclareMathOperator*{\argmax}{argmax}

\usepackage[utf8]{inputenc}

\title{
Optimal Investment to Reach a Financial Goal:\\ A Stochastic Control Framework}
\author{Gechun Liang\thanks{Department of Statistics, University of Warwick. Email: \href{mailto:g.liang@warwick.ac.uk}{g.liang@warwick.ac.uk}} \and Moris S. Strub\thanks{Warwick Business School, University of Warwick. Email: \href{mailto:moris.strub@wbs.ac.uk}{moris.strub@wbs.ac.uk}} \and Yuwei Wang\thanks{Corresponding author. School of Mathematics, Shanghai University of Finance and Economics. Email: \href{mailto:wangyuwei@shufe.edu.cn}{wangyuwei@shufe.edu.cn}}
\and Zhaojun Yang\thanks{Department of Finance, Southern University of Science and Technology. Email: \href{mailto:yangzj@sustech.edu.cn}{yangzj@sustech.edu.cn}}}

\date{}
\begin{document}   
\maketitle

\begin{abstract}
We develop a framework for an investor who trades until she either reaches a financial goal or an exogenous deadline arrives. 
Analogous to utility functions over wealth, we measure satisfaction with the timing of reaching a goal by a discount function.
For a continuous-time market where a stochastic factor drives the dynamics of stock prices and the financial goals, the investor maximizes the expected discount at the goal reaching time and the expected utility of the funding ratio if the goal remains unreached by the deadline.  
This setup leads to a new class of stochastic control problems. 
We establish Bellman's principle of optimality and characterize the value function as a viscosity solution to the associated Hamilton-Jacobi-Bellman equation. 
When the deadline is infinite and the goal is constant, the HJB equation reduces to a form related to the backward heat equation, for which we provide a complete characterization of smooth solutions.
When analytical solutions are unavailable, we develop an approach based on Howard's algorithm to obtain numerical solutions. 
Our analysis shows that optimal investment policies for goal reaching problems can be decreasing in the drift of risky assets and need not converge to full risk-free investment even as volatility diverges to infinity.
\end{abstract}

\noindent \keywords{discount function; time risk preferences; active portfolio management; benchmarking; time-inconsistent discounting}

\newpage
\section{Introduction}
The classical paradigm for portfolio selection in continuous time is the Merton problem pioneered in \cite{merton1975optimum}. 
Its structure is defined by three features. 
First, the investment horizon $T$ is fixed and exogenous. 
Second, wealth at the investment horizon, $X_T^\pi$, is random and controlled by the investor through her trading strategy. 
Third, a utility function $u$ measures the satisfaction with the wealth achieved at the horizon, and the investor maximizes expected utility $\mathbb{E}\left[u(X_T^\pi)\right]$ over admissible strategies. 
In this paradigm, the investor faces risk about which wealth level will be realized at an exogenous time horizon, and preferences are defined in terms of realized wealth.

Many investment situations, however, do not fit the structure of the Merton problem. 
Investors often pursue a specific financial goal rather than a fixed date: a family may invest to purchase a house; an entrepreneur builds up capital until it suffices to launch her business; or a saver invests until they accumulate sufficient wealth to retire early. 
In such problems, the roles of time and state are opposite to the ones in the Merton problem. 
The goal reaching time $\tau^\pi = \inf\{t > 0 : X_t^\pi = G_t\}$ is random and controlled by the investor, whereas wealth at that time, $X_{\tau^\pi}^\pi = G_{\tau^\pi}$, is stochastic but exogenous, described by a stochastic process $G$ representing the value of the goal. 
In analogy to a utility function over wealth, we measure satisfaction with the goal reaching time by a discount function $\rho$, and consider the goal reaching problem
\begin{align*}
\sup_{\pi \in \mathcal{A}} \mathbb{E}\left[\rho(\tau^\pi)\right].
\end{align*}
We, herein, introduce and study this problem, along with a variant in which an exogenous deadline $T$ is imposed, and the investor derives utility from the achieved fraction of the goal if the goal remains unreached by the deadline.

Beyond its relevance for investment practice, our framework is also motivated by the paper \textit{Decision Making When Things Are Only a Matter of Time} of \cite{ebert2020decision} laying out a theory of time-risk preferences. 
In this theory, expected discounted utility (EDU) preferences $\succeq$ over lotteries $(\widetilde{\tau}, \widetilde{x})$ are represented by
$\mathbb{E}\left[\rho(\widetilde{\tau}) u(\widetilde{x})\right]$, 
where $u$ is a utility function over outcomes and $\rho$
is a discount function over time. 
Time-outcome risk duality then states that delay-certain outcome risks are ranked by expected utility, while outcome-certain delay risks are ranked by expected discount. 
The discount function, thus, plays an analogous role for risk over \textit{when} an outcome is realized, as the utility function does for \textit{what} outcome is realized at a specific future time. 
The curvature of the discount function $\rho$ encodes attitudes towards time risk, see \cite{ebert2020decision} for a detailed exposition and discussion. 
Empirical evidence on such attitudes is provided in, for example,
\cite{chesson2003commonalities}, \cite{onay2007intertemporal}, \cite{chen2013effect},  \cite{dejarnette2020time}, or \cite{Ebert21:IER}.

This paper studies optimal investment in continuous time when things are, in the sense of \cite{ebert2020decision}, a matter of time.
This gives rise to a new family of stochastic control problems in which the objective contains a controlled hitting time, which goes beyond classical optimal control or optimal stopping.
We introduce this problem in an incomplete stochastic factor model for the financial market, where the stochastic factor influences both dynamics of risky assets and of the financial goal of the investor. 

Our main contributions are as follows. First, we establish regularity properties of the value function, prove Bellman's principle of optimality, derive the associated Hamilton-Jacobi-Bellman equation together with a verification theorem, and characterize the value function as a viscosity solution in case no sufficiently differentiable value function exists. 
Second, for the special case without exogenous deadline, constant goal, and no stochastic factor, we provide a complete characterization of the class of discount functions leading to an analytical value function building on the analysis of \cite{musiela2010portfolio}. These are parametrized, via Widder's theorem, by positive Borel measures, and we provide a construction scheme for goal reaching problems with an analytical solution. 
Third, we develop a numerical scheme based on Howard's algorithm for the problem with finite deadline, and we show that both value functions and optimal strategies of finite-horizon problems converge to their infinite-horizon counterparts as the deadline tends to infinity, which allows us to extend the algorithm to the infinite-horizon case. 

We investigate properties of optimal strategies, both analytically and numerically. 
Two findings stand out because they are qualitatively distinct from the predictions of classical expected utility theory: 
First, the optimal allocation to risky assets can be decreasing in the drift of risky assets, and, second, the optimal strategy need not converge towards investing everything into the risk-free asset as volatility diverges to infinity. 
The intuition is inherent to a goal reaching problem: when market conditions deteriorate, investing fully into the risk-free asset preserves wealth but guaranties that the goal is never attained. Thus, with worsening market conditions, the investor must keep and, in some scenarios, even increase her position in the risky assets to maintain a chance of reaching her financial goal within a reasonable amount of time. 
We also find that greater impatience leads to more aggressive risk taking and a more polarised outcome distribution, with a higher probability of reaching the goal early but also of ending with a low funding ratio. 

Our paper connects to several strands of literature, most directly to goal-based wealth management. 
\cite{Karatzas97:AJM}, \cite{HeZhou11:MF}, \cite{HeZhou16:MF}, \cite{DasOstrovRadhakrishanSrivastav18:JIM}, \cite{DasOstrovRadhakrishanSrivastav20:CMS}, and \cite{DaiKouQianWan22:MS} maximize the probability of reaching a single goal by a deadline, whereas \cite{CvitanicKouWanWilliams23:SSRN} optimize a weighted average of probabilities of achieving goal wealth levels and specified loss levels at a common date. \cite{DasOstrovRadhakrishanSrivastav22:JBF} and \cite{capponi2024continuous} study dynamic investment problems for multiple goals at different times with expected utility from partial fulfillment.  
A meta reinforcement learning approach to goal-based wealth management with multiple goals is developed in \cite{DasKhadilkarMittalOstrovSrivastavWang26:JFDS}.
The framework of \cite{capponi2024continuous} was extended in \cite{BayraktarHanXX:arxiv} to a setting where there is a separate mental account for each goal and the investor faces a mental cost when transferring funds allocated to different goals, and in \cite{BayraktarHanZhangXX:arxiv} to a setting with transaction costs.  
\cite{CampbellCapponiParasharXX:arxiv} study optimal portfolio choice for an investor simultaneously seeking to reach a fixed-deadline goal and a random-deadline goal transpiring at an exogenous random time.
In all of the above literature, the goal reaching time is exogenous and determines when preferences over wealth, or the probability of goal achievement, apply. 
Our goal reaching problem is fundamentally different in that the time when the goal is reached is controlled and itself the object of preferences. 

In the literature on stochastic control, 
\cite{pestien1985continuous}, \cite{orey1987reaching}, and \cite{heath1987minimizing} study how to control the coefficients of a diffusion process to maximize the probability of reaching a goal or to optimize the expected time to reach it. 
In the context of portfolio selection, \cite{browne1997survival,browne1999beating,browne1999reaching,browne2000risk} studies investment in geometric Brownian motion markets, where the objective is to exceed the performance of a given stochastic benchmark, considering the probability of beating the benchmark, the expected time until it is beaten, and the expected discounted rewards under exponential discounting.
Also related is \cite{yang2004optimal}, studying the minimization of the discounted penalty of bankruptcy in a related model. 

Finally, while our problem formulation is time-consistent, it relates to the literature on time-inconsistent stochastic control under non-exponential discounting going back to \cite{strotz1955myopia}. 
We refer to \cite{bjork2014theory}, \cite{BjorkKhapkoMurgoci17:FS}, \cite{BjorkKhapkoMurgoci21:book}, or \cite{HeZhou22:InBook} for a general overview of time-inconsistent stochastic control;
\cite{Karnam2017} for the dynamic utility approach to time-inconsistent control problems helpful in understanding why our formulation is time-consistent; and \cite{HamaguchiTseXX} or \cite{CaoWuXu25:AMO} for recent work under non-exponential discounting.

\section{Problem Setup}\label{sec:ProblemSetup}

Consider a financial market consisting of $N \in \mathbb{N}$ risky assets and one risk-free bond.
Let $M \in \mathbb{N}$ with $M \geq N$ and let $W$ be an $M$-dimensional Brownian motion with natural filtration $\mathbb{F}$. 
Asset prices are influenced by a non-tradeable, $d$-dimensional stochastic factor  $Y = (Y_1, \dots, Y_d)'$ which evolves according to
\begin{align*}
    dY_{t} = m_Y (Y_t) dt +  \Sigma_{Y} (Y_t) d W_{t} , \qquad t \geq 0,
\end{align*} 
$Y_{0} = y_{0} \in \mathbb{R}^d$, for Lipschitz continuous drift and volatility coefficients $m_Y: \mathbb{R}^d \rightarrow \mathbb{R}^d$ and $\Sigma_Y: \mathbb{R}^d \rightarrow \mathbb{R}^{d\times M}$.
The stochastic factor process $Y$ captures the evolution of macroeconomic factors, for example GDP growth, that influence asset prices, as well as factors relevant for the evolution of the price of the goal of the investor, for example housing prices if the goal is a house. 
We summarize both macroeconomic factors and goal-specific factors in a single process $Y$ of dimension $d$ for the sake of parsimoniousness and to allow for factors that influence both asset prices and the price of a goal, for example inflation if the goal is a required wealth level allowing the investor to retire and maintain a desired level of consumption. 

The prices of risky assets are denoted by $S = ( S_1, \dots, S_N)'$ and follow the dynamics
\begin{align*}
dS_{k,t}= S_{k,t} \mu_{k} (Y_t) dt +  S_{k,t} \sigma_{k} (Y_t) dW_{t}, \qquad t \geq 0,
\end{align*}
$S_{k,0} = s_{k,0}$, $k = 1, \dots, N$,
where the drift $\mu_{k}: \mathbb{R}^d \rightarrow \mathbb{R}$ and volatility $\sigma_{k} : \mathbb{R}^d \rightarrow \mathbb{R}^{1\times M}$ are bounded and Lipschitz continuous, $k = 1, \dots, N$. 
The risk-free bond generates interest at rate $r \geq 0$.


Throughout the paper, $\mathbb{R}_{+}=[0,\infty)$ denotes the set of
non-negative real numbers and $\mathbb{R}_{-}=(-\infty,0]$ denotes the
set of non-positive real numbers. Let
$\mu(y)=(\mu_1(y),\ldots,\mu_N(y))^\top$ be the vector of drift
coefficients and let
$\Sigma(y)=(\sigma_1(y),\ldots,\sigma_N(y))^\top\in\mathbb{R}^{N\times M}$
be the volatility matrix. Assume that there exists a constant $\underline a>0$ such that
$\Sigma(y)\Sigma(y)^\top\succeq\underline a I_N$ for every
$y\in\mathbb{R}^d$. Then $\Sigma(y)$ has full row rank $N$, and hence
the market price of risk equation
$\mu(y)-r\mathbf{1}_N=\Sigma(y)\kappa(y)$ admits solutions. We select
its minimum-norm solution,
\[
    \kappa(y)
    =
    \Sigma(y)^\top
    \bigl(\Sigma(y)\Sigma(y)^\top\bigr)^{-1}
    \bigl(\mu(y)-r\mathbf{1}_N\bigr),
    \qquad y\in\mathbb{R}^d,
\]
where $\mathbf{1}_N=(1,\ldots,1)^\top\in\mathbb{R}^N$. We further
assume that $\kappa$ is bounded. Therefore, on every finite horizon
$[0,T]$, Novikov's condition holds and
$\mathcal{E}(-\int_0^\cdot\kappa(Y_u)^\top dW_u)$ is a true martingale.
Consequently, it defines an equivalent (local) martingale measure, and
the market is free of arbitrage on $[0,T]$.

The investor starts with wealth $x_0 > 0$ and seeks to achieve a financial goal $G$. 
The value of the goal at time $t \geq 0$ is given by $G_t$ and follows the dynamics  
\begin{align*}
    dG_t &=  G_t \left( \alpha ( Y_t) dt + \beta(Y_t) dW_t \right), \qquad t \geq 0,
\end{align*}
$G_0 = g_0 > x_0$,
where $\alpha:\mathbb{R}^d \rightarrow \mathbb{R}$ and $\beta: \mathbb{R}^d \rightarrow \mathbb{R}^{1\times M}$ are Lipschitz continuous. 
We interpret $G_t$ as the price at time $t$ of an object which the investor desires, for example, a house or a car. 
When $M=N$ and $\Sigma$ is invertible, the goal can be replicated by admissible trading. However, in general, this is not the case. We will sometimes consider the case where the value of the goal is constant, $G \equiv g_0 > x_0$. A constant goal can reflect, for example, the goal of becoming a millionaire.  

The investor trades in the financial market to achieve her goal as early as possible. 
Trading strategies are described by progressively measurable $\mathbb{R}^N$-valued processes
$\pi = \left(\pi_t \right)_{t\geq 0}$, where $\pi_t=(\pi_{1,t},\dots,\pi_{N,t})^{'}$, and $\pi_{i,t}$ denotes the fraction of wealth invested in the $i$-th risky asset at time $t$, $i=1,\dots,N$. 
The remainder $1-\sum_{i=1}^N \pi_{i,t}$ is invested in the risk-free bond. The wealth of the investor adopting strategy $\pi$ is governed by the controlled stochastic differential equation 
\begin{align*}
dX^\pi_t
&=  \left(  X^\pi_t \pi_t^{'} \mu (Y_t) +  X^\pi_t \left(1 - \sum_{i=1}^N \pi_{i,t} \right)r \right) dt + X^\pi_t \pi_t^{'}\Sigma (Y_t) dW_t, \qquad t \geq 0,
\end{align*}
with $X^\pi_0 = x_0$. We next introduce the \textit{funding ratio process}, $Z^\pi=(Z_t^\pi)_{t\geq 0}$, where $Z_t^\pi:=X^\pi_t/G_t$ represents the ratio of goal $G$ funded at time $t \geq 0$ when trading according to strategy $\pi$.

For $T\in(0,\infty]$, $0\leq s<T$, and given $(Y_s,Z_s)=(y,z)$, we denote by
\[
\mathcal{A}_{s,T}(y,z)
=
\left\{
\pi \text{ is } \mathbb{F}\text{-progressively measurable}:
\E\left[\left.\int_s^t |\pi_u|^2\,du\right|Y_s=y,Z_s=z\right]<\infty,
s<t<T
\right\}
\]
the set of admissible trading strategies over $[s,T)$. We suppress the dependence on $(y,z)$ when no confusion arises and write $\mathcal{A}_{s,T}$, and abbreviate $\mathcal{A}=\mathcal{A}_{0,\infty}$.
For any admissible strategy $\pi\in\mathcal{A}$, one can obtain the following expression 
\begin{align*}
    X_t^{\pi}=x_0e^{\int_s^t(\pi_u' \mu(Y_u) +\left(1 - \sum_{i=1}^N \pi_{i,u} \right)r
       -\frac{1}{2}\left|\pi_u'\Sigma(Y_u)\right|^2du+ \int_s^t  \pi_u' \Sigma (Y_u) dW_u},
\end{align*}
and deduce that the wealth process $X^\pi$ remains strictly positive. 

Applying It\^{o}'s lemma, we find that $Z^\pi$ satisfies
\begin{align}\label{RatioProcess}
dZ_t^\pi  = Z_t^\pi \left( (r +  \pi_t' \left( \mu(Y_t) - r \mathbf{1}_N\right) - \alpha (Y_t) )dt+(\pi_t' \Sigma (Y_t)- \beta(Y_t))(dW_t-\beta(Y_t)'dt)\right),\quad t \geq 0, 
\end{align}
$Z_0^\pi = z_0 = x_0/g_0 \in (0,1)$. 

The \textit{goal reaching time} when trading according to $\pi \in \mathcal{A}$ is 
\begin{align*}
    \tau^{\pi} : = \inf \left\{ t > 0: Z_t^{\pi} = 1 \right\}  = \inf \left\{ t > 0: X_t^{\pi} = G_t \right\} .
\end{align*} 
Consider a fixed, exogenous deadline $T \in (0, + \infty]$, possibly infinity, and let $\widehat{\tau}^{\pi} = \tau^{\pi} \wedge T$ be the stopping time that either equals the goal reaching time or the deadline, whichever comes first. 
The objective of the investor is to reach her goal as soon as possible and to have the largest possible share funded in case she cannot achieve her goal by the exogenous deadline. 
Formally, the investor chooses a trading strategy $\pi \in \mathcal{A}$ to optimize the \textit{goal reaching problem}
\begin{align}\label{prob:GeneralObjective}
    \sup_{\pi \in \mathcal{A}} \E \left[ \rho \left( \widehat{\tau}^{\pi} \right) + \rho (T) u \left( Z^{\pi}_{\widehat{\tau}^{\pi}} \right) \right] - \rho (T) 
    = \sup_{\pi \in \mathcal{A}} \E \left[ \rho \left( \tau^{\pi} \right) \mathbf{1}_{\left\{ \tau^\pi \leq T \right\} }  
    + \rho (T) u \left( Z^{\pi}_{T} \right) \mathbf{1}_{\left\{ \tau^\pi > T \right\} } \right] ,  \tag{GRP}
\end{align}
where $\rho: \mathbb{R}_+ \rightarrow (0,1]$ is a \textit{discount function} and $u: (0,1] \rightarrow (0,1]$ is a \textit{utility function}. 
We assume that the discount function $\rho:\mathbb{R}_{+}\rightarrow (0,1]$ is
continuously differentiable, strictly decreasing, and satisfies $\rho(0)=1$ and $\lim\limits_{t\rightarrow +\infty}\rho(t)=0$, and that the utility function $u: (0,1]\rightarrow (0, 1]$ is strictly increasing, concave, and satisfies $\lim\limits_{z\rightarrow0}u(z)=0$ and $u(1)=1$.
The normalization $u(1) = 1$ ensures that the objective is continuous. 
If the goal could not be reached by the exogenous time horizon, the investor receives $\rho (T) u(Z^\pi_T)$, the utility from partially funding the goal at the ratio $Z^\pi_T$. 
The discount function $\rho$ represents the investor's attitudes towards \textit{time risk}. 
In particular,  the curvature of $\rho$ determines whether the investor is
risk-seeking ($\rho^{''}>0$) or risk-averse ($\rho^{''}<0$), prudent ($\rho^{'''}<0$) or imprudent ($\rho^{'''}>0$), and temperate ($\rho^{''''}<0$) or intemperate ($\rho^{''''}>0$) over time risk, see \cite{ebert2020decision} for a derivation and discussion of the above results and a comprehensive treatment of time-risk preferences.

When $T = + \infty$, Problem \eqref{prob:GeneralObjective} reduces to 
\begin{align}\label{prob:InfiniteHorizon}
    \sup_{\pi \in \mathcal{A}} \E \left[ \rho \left( \widehat{\tau}^{\pi} \right) \right],
\end{align}
with the convention $\rho(+\infty)=0$.
In this case, the investor trades until she reaches her goal. 
Problem \eqref{prob:InfiniteHorizon} is thus the continuous-time portfolio selection problem  \textit{when things are only a matter of time} \citep{ebert2020decision}. 
The outcome is certain, but the duration the investor has to wait until she achieves the outcome is not.  

When $T < + \infty$, the investor trades until she either achieves her goal or the deadline to do so arrives. 
On the one hand, if the investor achieves her goal before the deadline, she receives the reward $\rho (\tau^{\pi})$ exactly as in the case without a deadline. Indeed, $\rho \left( \widehat{\tau}^{\pi} \right) + \rho (T) u \left( Z^{\pi}_{\widehat{\tau}^{\pi}} \right)  - \rho (T) = \rho (\tau^{\pi})$ on $\{ \widehat{\tau}^{\pi} = \tau^{\pi} \}$. 
On the other hand, if the deadline hits before the goal is reached, she receives the reward $\rho(T) u(Z^{\pi}_T)$ where $Z^{\pi}_T$ is the fraction of the goal funded at time $T$. Indeed, $\rho \left( \widehat{\tau} \right) + \rho (T) u \left( Z^{\pi}_{\widehat{\tau}} \right)  - \rho (T) = \rho(T) u(Z^{\pi}_T)$ on $\{ \widehat{\tau}^{\pi} = T\}$.
In this case, the investor receives utility from partial achievement of the goal at the deadline similar to \cite{capponi2024continuous}.

\section{Characterization of the Value Function and Optimal Strategy}\label{sec:HJBEqn}
In this section, we characterize the value function of the goal reaching problem \eqref{prob:GeneralObjective}, along with the optimal strategy.
For $(s,y,z) \in [0,T) \times \mathbb{R}^d \times (0,1)$ and $\pi  \in \mathcal{A}_{s,T}$,  we consider the ratio process $(Z_t^{\pi})_{t\geq s}$
starting from $Z_s = z$ with the goal reaching time 
\begin{align*}
    \tau_{s,z}^{\pi} = \inf \left\{ t > s : Z^{\pi}_t = 1 \right\}.
\end{align*}
Set $\widehat{\tau}_{s,z}^{\pi} = \tau_{s,z}^{\pi} \wedge T$. We often drop the explicit dependence of the goal reaching time on the initial data and write $\widehat{\tau}^{\pi}$.
The objective function is defined as
\begin{align}\label{eq:Objective}
    J (s,y,z;\pi) = \E \left[ \rho \left( \widehat{\tau}^{\pi} \right) + \rho (T) u \left( Z^{\pi}_{\widehat{\tau}^{\pi}} \right) \big\vert Y_s = y, Z_s =z \right] - \rho (T) ,
\end{align}
and define the value function of the goal reaching problem  \eqref{prob:GeneralObjective} as 
\begin{align}\label{eq:ValueFunction}
    V (s,y,z) = \sup_{\pi \in \mathcal{A}_{s,T}} J (s,y,z;\pi) .
\end{align}
It is clear that $V(0,Y_0,Z_0)$ is equal to the value of \eqref{prob:GeneralObjective}. 

\begin{remark}\label{rem:TimeConsistency}
Non-exponential discounting typically leads to time-inconsistency \citep{strotz1955myopia}. 
It is thus remarkable that the family of problems in \eqref{eq:ValueFunction} is \textit{time-consistent}, a fact that we will formally show in Proposition \ref{Prop:DPP}.
The reason is that we do not reset the discount function to the present when formulating the dynamic problem \eqref{eq:Objective}. 
To illustrate this point, consider the problem of maximizing expected utility from terminal wealth with a general discount function 
\begin{align*}
    \sup_{\pi\in \mathcal{A}_{0,T}} \E \left[ \rho(T) U(X^{\pi}_T ) \right],
\end{align*}
for a utility function $U$.
The dynamic version of the above problem is typically formulated as 
\begin{align*}
    \widetilde{V} (s,y,x) = \sup_{\pi\in \mathcal{A}_{s,T}} \E \left[ \rho(T - s) U(X^{\pi}_T ) \big\vert Y_s=y, X_s = x \right] .
\end{align*}
Expected utility maximization under non-exponential discounting as formulated above is indeed time-inconsistent and has been studied, for example, in \cite{BjorkKhapkoMurgoci17:FS} or \cite{HamaguchiTseXX}. 
Similarly, the dynamic objective 
\begin{align*}
    \widetilde{J} (s,y,z;\pi) = \E \left[ \rho \left( \widehat{\tau}^{\pi} - s \right) + \rho (T-s ) u \left( Z^{\pi}_{\widehat{\tau}^{\pi}} \right) \big\vert Y_s = y, Z_s =z \right] - \rho (T-s) 
\end{align*}
would lead to a time-inconsistent problem formulation for the goal reaching problem \eqref{prob:GeneralObjective}. 
We can thus interpret \eqref{eq:Objective} as a \textit{dynamic utility} in the sense of \cite{Karnam2017} that allows us to obtain the pre-committed solution to a  time-inconsistent formulation of the goal reaching problem under non-exponential discounting. 
We refer to \cite{CaoWuXu25:AMO} for a time-consistent dynamic formulation of an expected utility maximization problem with non-exponential discounting. 
\end{remark}

Let $\mathbb{D}=[0,T)\times \mathbb{R}^{d}\times(0,1)$ and $\overline{\mathbb{D}}$ denote the closure of $\mathbb{D}$, i.e., $\overline{\mathbb{D}}=[0,T]\times \mathbb{R}^{d}\times[0,1]$ for finite $T$ and $\overline{\mathbb{D}}=[0,+\infty)\times \mathbb{R}^{d}\times[0,1]$ for $T=+\infty$. 
We have the following first result. 

\begin{proposition}\label{prop:ContinuityofV}
    The value function $V$ defined in \eqref{eq:ValueFunction} is continuous and bounded on  $\overline{\mathbb{D}}$.
\end{proposition}


Next, we show that the value function satisfies Bellman's principle of optimality. This in particular implies that the goal reaching problem is time-consistent, see Remark \ref{rem:TimeConsistency} for a discussion. 
\begin{proposition}\label{Prop:DPP}
(Bellman's principle of optimality) For any $(s,y,z)\in \mathbb{D}$ and for all stopping times $s\leq \alpha\leq \widehat{\tau}^{\pi}$, the value function $V$ defined in \eqref{eq:ValueFunction} satisfies
\begin{align*}
    V(s,y,z) = \sup_{\pi \in \mathcal{A}_{s,T}} \E\left[V(\alpha,Y_{\alpha},Z_{\alpha}^{\pi})\big\vert Y_s=y,Z_s=z \right] .
\end{align*}
\end{proposition}

Before turning to the Hamilton-Jacobi-Bellman (HJB) equation, we establish two fundamental monotonicity properties of the value function.
\begin{proposition}\label{Prop:MonotonicityofV}
    Suppose that an optimal strategy exists for every initial state $(s,y,z)$. Then the value function $V$ is strictly increasing in the funding ratio $z$ for any $0 \leq s < T$ and $y\in \mathbb{R}^{d}$. Furthermore, suppose that $Y_t=y_0$ for all $t\in\mathbb{R}_{+}$, $V$ is decreasing in time $s$ for any $z \in (0,1)$. 
\end{proposition}


Let $C^{1,2,2}(\mathbb{D})$ denote the set of continuous functions $w(s,y,z):\mathbb{D}\rightarrow \mathbb{R},$ where $y=(y_1,\dots,y_d)^{'}$, such that the partial derivatives $w_s$, $w_{y_i}$, $w_{y_iy_j}$, $w_{zy_i}$, $w_z$, and $w_{zz}$ are all continuous on $\mathbb{D}$ for $i,j=1,\dots,d$. 
Introduce for simpler notation $\widehat{\mu}(y)=\mu(y) - r \mathbf{1}_N - \Sigma (y) \beta (y)'$, $\widehat{r}(y)=r-\alpha(y)+\beta(y)\beta(y)'$, $A(y)=\Sigma(y)\Sigma(y)^{'}$, $B(y)=\Sigma(y)\Sigma_Y(y)^{'}$, $C(y)=\Sigma_Y (y)\Sigma_Y(y)^{'}$, and $D(y)=\beta(y)\Sigma(y)'$.
For $\pi \in \mathbb{R}^N$ and $\varphi \in C^{1,2,2}(\mathbb{D})$ define the infinitesimal generator by
\begin{align}\label{infGenerator}
\begin{split}
\mathcal{L}^{\pi}\varphi(s,y,z)=&\varphi_s(s,y,z)+\left( \widehat{r}(y) +  \pi' \widehat{\mu}(y) \right)z\varphi_z(s,y,z)\\&+\frac{1}{2}\left( \pi' A(y)\pi-2D(y)\pi+\beta(y)\beta(y)'  \right)z^2\varphi_{zz}(s,y,z)
    +m_Y(y)^{'}\varphi_y(s,y,z)\\&+\frac{1}{2}C(y)\varphi_{yy}(s,y,z)+\left( \pi' B (y) -  \beta(y)\Sigma_Y (y)^{'} \right)z\varphi_{yz}(s,y,z),
\end{split}
\end{align}
where $\varphi_y(s,y,z)=(\varphi_{y_1}(s,y,z),\dots, \varphi_{y_d}(s,y,z))^{'}$, $\varphi_{yz}(s,y,z)=(\varphi_{y_1z}(s,y,z),\dots, \varphi_{y_dz}(s,y,z))^{'}$, and $\varphi_{yy}(s,y,z)$ is the $d\times d$ Hessian matrix where the $(i,j)$-th entry is the mixed partial derivative $\varphi_{y_i y_j}$.
The following result establishes the concavity of the value function with respect to the funding ratio directly from the dynamic programming principle.
\begin{proposition}\label{prop:VConcave}
Suppose that the assumptions of Proposition~\ref{Prop:MonotonicityofV}
hold and that
\(V\in C^{1,2,2}(\mathbb{D})\cap C(\overline{\mathbb{D}}).\)
Then
\(V_{zz}(s,y,z)\leq0,\) for any \((s,y,z)\in\mathbb{D}.\)
If, in addition,
\(B(y)=0\) and \(\widehat{\mu}(y)\neq0\) for all \(y\in\mathbb{R}^d,\)
then
\(V_{zz}(s,y,z)<0, \) for any \((s,y,z)\in\mathbb{D}.\)
\end{proposition}

We are now ready to characterize the value function through the associated HJB equation
\begin{align}\label{DPE0-0}
    \sup_{\pi \in \mathbb{R}^N}\mathcal{L}^{\pi} V(s,y,z) = 0,
\end{align}
and establish a corresponding verification theorem.

\begin{theorem}\label{thm:Verification}
    Suppose that $V \in C^{1,2,2}(\mathbb{D})\cap C(\overline{\mathbb{D}})$, $V_{zz}<0$ and
    \begin{align*}
\E\left[\int_{s}^{\alpha}|\mathcal{L}^{\pi}V(t,Y_{t},Z^{\pi}_{t})|dt\right]<\infty,
    \end{align*}
    for any $(s,y,z)\in \mathbb{D}$, $\pi \in \mathcal{A}_{s,T}$ and  bounded stopping times $s\leq \alpha \leq \widehat{\tau}^{\pi}$.
    Then, writing $v=V$, the value function satisfies
\begin{equation}\label{DPE}
\begin{aligned}
&v_s(s,y,z)+m_Y(y)'v_y(s,y,z)
+\frac12\operatorname{Tr}\!\left(C(y)v_{yy}(s,y,z)\right)
+\widehat r(y)zv_z(s,y,z)\\
&\quad+\frac12\beta(y)\beta(y)'z^2v_{zz}(s,y,z)
-\beta(y)\Sigma_Y(y)'zv_{yz}(s,y,z)\\
&\quad-\frac{1}{2v_{zz}(s,y,z)}
\Big(
\widehat{\mu}(y)v_z(s,y,z)
+B(y)v_{yz}(s,y,z)
-D(y)'zv_{zz}(s,y,z)
\Big)'A(y)^{-1}\\
&\qquad\qquad\times
\Big(
\widehat{\mu}(y)v_z(s,y,z)
+B(y)v_{yz}(s,y,z)
-D(y)'zv_{zz}(s,y,z)
\Big)
=0.
\end{aligned}
\end{equation}
    with boundary condition $v(s,y,1)=\rho(s)$, and, if $T<+\infty$, $v(T,y,z)=\rho(T)u(z)$. 
    
    Vice versa, suppose that 
    $v(s,y,z) \in C^{1,2,2}(\mathbb{D})\cap C(\overline{\mathbb{D}}), v_{zz}<0$ and $v: \overline{\mathbb{D}}\rightarrow (0,1]$ is a solution to \eqref{DPE}
    with boundary condition $v(s,y,1)=\rho(s)$, and, if $T<+\infty$, $v(T,y,z)=\rho(T)u(z)$.
Then $v(s,y,z)$ is the value function, i.e., $v(s,y,z)=V(s,y,z)$ for any $(s,y,z) \in \mathbb{D}$, and the optimal trading strategy is given by 
    \begin{align}\label{Optimalf}
    \begin{split}
        \pi^*(s,y,z)=&A(y)^{-1}\left(-\frac{\widehat{\mu}(y)v_z(s,y,z)+B(y)v_{yz}(s,y,z)}{zv_{zz}(s,y,z)}+D(y)^{'}\right)\\
        =&-A(y)^{-1} \left( \mu(y) - r \mathbf{1}_N \right) \frac{v_z(s,y,z)}{zv_{zz}(s,y,z)}\\
        & \quad +A(y)^{-1} D(y)^{'}  \left(1+ \frac{v_z(s,y,z)}{zv_{zz}(s,y,z)} \right)-A(y)^{-1}B(y)\frac{v_{yz}(s,y,z)}{zv_{zz}(s,y,z)} .
    \end{split}        
    \end{align}   
\end{theorem}

The optimal portfolio decomposes into the sum of three components. The first term $$-A(y)^{-1} \left( \mu(y) - r \mathbf{1}_N \right) \frac{v_z(s,y,z)}{zv_{zz}(s,y,z)}$$ resembles the classical Merton portfolio. 
The second term corresponds to the portfolio that minimizes the local variance of the funding ratio process, $A(y)^{-1} D(y)^{'}$, scaled by a factor $1+\frac{v_z(s,y,z)}{zv_{zz}(s,y,z)}$,
which reflects the curvature of the value function with respect to the funding ratio $z$. The third term captures how the investor adjusts the portfolio because the same shock can affect both current wealth and future investment opportunities. Accordingly, a deterministic goal ($\beta=0$) eliminates the second component, while a deterministic factor process ($\Sigma_Y=0$) eliminates the third, leaving only the Merton-type component when both are deterministic.

\begin{remark}
Because the funding ratio process  $Z$ is a Dol\'{e}ans-Dade exponential, it is positive for any admissible investment policy. 
Thus, $\{z=0\}$ acts as a natural boundary. That is, a boundary condition for \eqref{DPE} at $\{z=0\}$ is redundant, as it is inherently implied by equation \eqref{DPE0-0} within the interior domain.
Indeed,
Fichera theory provides guidelines for imposing boundary conditions for second-order PDEs that degenerate at the boundary, see for example Section 5 of \cite{Jiang2007Fichera} for details.
Examining Fichera theory applied to our setting, it turns out that we only need to provide boundary conditions at the boundary $\{z=1\}$, and $\{s=T\}$ if $T<+\infty$. 
\end{remark}

For tractability and simplicity, we will often consider the special case where coefficients are constant, $Y_t \equiv y_0$. 
To further simplify notation, we let 
$\widehat{\mu} = \widehat{\mu}(y_0)$, $\widehat{r} = \widehat{r}(y_0)$, $A = A (y_0)$, $B = B(y_0)$, $C=C(y_0)$, and $D=D(y_0)$. 
The infinitesimal generator reduces to 
\begin{align}\label{infGenerator_CstCoefficients}
\begin{split}
\mathcal{L}^{\pi}\varphi(s,z)=&\varphi_s(s,z)+\left( \widehat{r} +  \pi' \widehat{\mu} \right)z\varphi_z(s,z) + \frac{1}{2}\left( \pi' A\pi-2D \pi+\beta \beta'  \right) z^2 \varphi_{zz}(s,z)
\end{split}
\end{align}
for $\varphi \in C^{1,2}\left([0,T)\times (0,1)\right)$ and Theorem \ref{thm:Verification} simplifies as in the following corollary. 
\begin{corollary}\label{corollary-HJB}
    Suppose that 
    $Y_t \equiv y_0$,
    $V \in C^{1,2}([0,T)\times (0,1))\cap C([0,T]\times [0,1]),$ $V_{zz}<0$ and
    \begin{align*}
\E\left[\int_{s}^{\alpha}|\mathcal{L}^{\pi}V(t,Z^{\pi}_{t})|dt\right]<\infty,
    \end{align*}
    for any $(s,z)\in [0,T)\times (0,1)$, $\pi \in \mathcal{A}_{s,T}$ and  bounded stopping times $s\leq \alpha \leq \widehat{\tau}^{\pi}$.
    Then the value function $V$ satisfies
    \begin{align}\label{DPE-Corr}
    \begin{split}
    v_s(s,z)  -\frac{1}{2}\widehat{\mu}'A^{-1}\widehat{\mu} \frac{v_z^2(s,z)}{v_{zz}(s,z)}  
     + \left( \widehat{r} +D A^{-1} \widehat{\mu} \right) z v_z(s,z) +\frac{1}{2}\left(-D A^{-1} D'+\beta \beta' \right)z^2v_{zz}(s,z)=0
    \end{split}
    \end{align}
    with boundary condition $v(s,1)=\rho(s)$, and, if $T<+\infty$, $v(T,z)=\rho(T)u(z)$. 
    
    
    Vice versa, suppose that 
    \(v\in C^{1,2}([0,T)\times(0,1))
    \cap C([0,T]\times[0,1]),
    v_{zz}<0,\)
    and \(v:[0,T]\times[0,1]\rightarrow(0,1]\)
    is a solution to \eqref{DPE-Corr}
    with boundary condition $v(s,1)=\rho(s)$, and, if $T<+\infty$, $v(T,z)=\rho(T)u(z)$.
    Then $v(s,z)$ is the value function, i.e., $v(s,z)=V(s,z)$ for any $(s,z) \in [0,T)\times (0,1)$, and the optimal trading strategy is given by
    \begin{align}\label{Optimalf-Corr}
        \pi^*(s,z)=A^{-1}\left(- \widehat{\mu} \frac{v_z(s,z)}{zv_{zz}(s,z)}+D^{'}\right).
    \end{align}
Furthermore, when $r=0$ and $G_t \equiv g_0$, \eqref{DPE-Corr} and \eqref{Optimalf-Corr} reduce to 
\begin{align}\label{PDE:r=0-ConstantG}
    v_s(s,z)-\frac{1}{2}\mu'A^{-1}\mu\frac{v_z^2(s,z)}{v_{zz}(s,z)}=0,
\end{align}
respectively,
\begin{align}
        \pi^*(s,z)=-A^{-1} \mu \frac{v_z(s,z)}{zv_{zz}(s,z)}.
\end{align}
\end{corollary}

In the following, we provide three examples showing how the above theory applies in the case of $T<+\infty$, where the investment problem concludes when the goal is achieved or the terminal horizon is reached, whichever occurs first. 
To obtain smooth solutions for the value functions, preferences in terms of discount and utility function must be coordinated and, at least one of them, connected with market parameters. 
For the first two examples, we set $r = 0$, $Y_t \equiv 0$, and $G_t \equiv g_0$ for simplicity. In the third, we relax only the assumptions on the interest rate and $G_t$, allowing $r\neq 0$ and a stochastic benchmark, while keeping $Y_t\equiv 0$.
All examples are obtained by straightforward computation of derivatives of suggested value functions, showing that they satisfy \eqref{DPE-Corr} or \eqref{PDE:r=0-ConstantG}, and then
applying Corollary \ref{corollary-HJB} to argue that these functions are indeed the value function as claimed. 

\begin{example}
Let the discount function and utility function be given by
\begin{align*}
    \rho(s) & = 1 -\frac{s}{T + \xi}, \quad s \in [0,T], \\
    u(z) & = 1 + \frac{2}{\xi \kappa' \kappa} 
    \log \left(\left( 1-e^{-\frac{1}{2} \kappa' \kappa \xi} \right) z + e^{-\frac{1}{2} \kappa' \kappa \xi} \right) , \quad z \in (0,1],
\end{align*} 
where $\xi > 0$. 
Then the optimal value function is given by 
\begin{align*}
V(s,z)=1 - \frac{s}{T+\xi} + \frac{2}{(T + \xi) \kappa' \kappa}  \log \left( \left( 1-e^{-\frac{1}{2} \kappa' \kappa \xi} \right) z + e^{-\frac{1}{2} \kappa' \kappa\xi} \right) ,
\end{align*}
with the corresponding optimal strategy given by
$\pi^*(s,z)=A^{-1}\mu \left(1+\frac{1}{\left( e^{\frac{1}{2} \kappa' \kappa \xi} -1\right) z} \right) $ for $(s,z)\in [0,T)\times(0,1)$.
\end{example}

\begin{example}\label{ex:FiniteTExprhoU}
Let the discount function and utility function be given by
\begin{align*}
\rho(s) & = \frac{1-e^{-\gamma-\frac{1}{2} \kappa' \kappa(T-s)}}{1-e^{-\gamma-\frac{1}{2} \kappa' \kappa T}}, \quad s \in [0,T], \\
u(z) & =\frac{1-e^{-\gamma z}}{1-e^{-\gamma}}, \quad z\in(0,1],
\end{align*}
where $\gamma>0$. Then the optimal value function is given by
\begin{align*}
V(s,z)= \frac{1-e^{-\gamma z-\frac{1}{2} \kappa' \kappa(T-s)}}{1-e^{-\gamma-\frac{1}{2} \kappa' \kappa T}},
\end{align*}
with the corresponding optimizer given by
$\pi^*(s,z)=A^{-1}\mu\frac{1}{\gamma z}$ for $(s,z)\in [0,T)\times(0,1)$. 
\end{example}
In Example \ref{ex:FiniteTExprhoU}, all orders of derivatives of discount function $\rho$ are negative. This indicates that the investor can be classified as risk-averse, prudent, and temperate towards time risks, as described in \cite{ebert2020decision}.
We now turn to an example considering a nonzero interest rate and a stochastic benchmark.

\begin{example}\label{ex:FiniteTExprhoUPower}
Let 
$\rho(s)=e^{-cs}$, and $u(z)=z^{\gamma}$ for $z\in(0,1]$, where $c>0$ and $0<\gamma<1$ are chosen such that the following relationship holds $c=\frac{\gamma}{2(1-\gamma)}\widehat{\mu}'A^{-1}\widehat{\mu}+ \gamma(\widehat{r}+DA^{-1}\widehat\mu)+\frac{1}{2}\gamma(\gamma-1)(-DA^{-1}D^{'}+\beta\beta^{'})$. 
Then the optimal value function is given by
$$V(s,z)=z^{\gamma}e^{-cs},$$
with the corresponding optimizer given by
$\pi^*(s,z)=A^{-1}D'
+\frac{1}{1-\gamma}A^{-1}\widehat{\mu},$ for $(s,z)\in [0,T)\times(0,1)$. 
\end{example}

The above theory assumes that the value function is sufficiently smooth.
As is common in stochastic control, such regularity is difficult to verify a priori.
Nevertheless, the value function can still be characterized in the viscosity sense.

\begin{proposition}\label{Prop:ViscositySol}
The value function $V$ is a viscosity solution of equation \eqref{DPE0-0} on $\mathbb{D}$.
\end{proposition}

A detailed proof of Proposition \ref{Prop:ViscositySol} is provided in
\cite[Propositions~4.3--4.4]{warwick189223Yuwei} and is omitted here.
Under conditions ensuring a comparison principle, $V$ is the unique bounded
continuous viscosity solution satisfying the corresponding boundary conditions.
For example, \cite[Theorem~7.6]{touzi2012optimal} applies when the stochastic factor
$Y$ is bounded, or when $\mu$, $\Sigma$, $m_Y$, $\Sigma_Y$, $\alpha$, and $\beta$
are all bounded.

\section{Infinite Time Horizon}\label{sec:InfiniteT}

In this section, we consider the goal reaching
problem \eqref{prob:GeneralObjective} without exogenous horizon, i.e., when $T = + \infty$. 
In this case, $\widehat{\tau}_{s,z}^{\pi}=\tau_{s,z}^{\pi}$ and
\eqref{prob:GeneralObjective}
reduces to 
\begin{align*}
    \sup_{\pi \in \mathcal{A}} \E \left[ \rho \left( \widehat{\tau}^{\pi} \right) \right] ,
\end{align*}
the continuous-time portfolio selection problem  when things are only a matter of time \citep{ebert2020decision}. 
The agent is determined to succeed, and will persist until her goal is reached.

To obtain analytical results, we focus on the case where the benchmark is a constant, $G_t\equiv g_0$, the market offers zero interest rate $r= 0$, and the drift and volatility coefficient of the stock price are deterministic and continuous functions of time. Equivalently, we may take a one-dimensional factor $Y_t\equiv t$.
Under these assumptions, the funding ratio process, $Z_t^{\pi}=X^{\pi}_t/g_0$ simplifies to 
\begin{align}\label{RatioProcessConstantB}
dZ_t^{\pi}  = Z_t^{\pi} \pi_t'  \mu(t)    dt +  Z_t^{\pi}  \pi_t' \Sigma (t)  dW_t=Z_t^{\pi} \pi_t'  \Sigma (t)(  \kappa(t)  dt + dW_t),
\end{align}
where $\kappa(t)$ is the market price of risk.
The associated HJB equation \eqref{DPE} reduces to
\begin{align}\label{DPETime}
    v_s(s,z)-\frac{1}{2} \kappa(s) ' \kappa(s)\frac{v_z^2(s,z)}{v_{zz}(s,z)}=0,
\end{align}
for $0<z<1$ with boundary condition $v(s,1)=\rho(s)$. 
To the best of our knowledge, this equation has not yet been studied in the literature. 

As outlined in \cite{musiela2010portfolio}, if $\kappa(s)' \kappa(s)= 1$, there is a one-to-one correspondence (modulo normalization constants) between strictly increasing functions $h(s,z)$ satisfying the backward heat equation
\begin{align}\label{PDEforFunctionh}
    h_s(s,z)+\frac{1}{2}h_{zz}(s,z)=0,
\end{align}
and strictly increasing concave solutions to the HJB equation \eqref{DPETime}.
We herein build on these results and extend the correspondence to the PDE \eqref{DPETime} with general Lipschitz continuous functions $\kappa$.
Let $\Theta: [0,\infty ) \rightarrow \mathbb{R}$ be defined as $\Theta(s) := \int_0^s \frac{1}{2} \kappa(t)' \kappa (t) dt$. 

\begin{theorem}\label{thm:InfHorizonCharacterization}
Let $v \in C^{1,2}\left( \mathbb{R}_{+}\times(0,1] \right)$ be the value function to the infinite horizon goal reaching problem \eqref{prob:InfiniteHorizon} with discount function $\rho$. 
Then, there exists a
strictly increasing $h:\mathbb{R}_+ \times \mathbb{R} \rightarrow \mathbb{R}$ solving \eqref{PDEforFunctionh} such that
\begin{align}\label{Lowerbound}
    0<L=\lim\limits_{s\rightarrow +\infty} \frac{1}{2}\int_0^{2\Theta(s)}e^{-h^{(-1)}(t,1)+\frac{t}{2}}h_z\left(t,h^{(-1)}(t,1)\right)dt 
    < \infty ,
\end{align}
and
$v$ and $\rho$ jointly satisfy 
\begin{align}\label{valuefunctionAfterNormalization}
    v(s,z)= 1 - \frac{1}{2L}\int_0^{2\Theta(s)}e^{-h^{(-1)}(t,z)+\frac{t}{2}}h_z\left(t,h^{(-1)}(t,z)\right)dt -\frac{1}{L}\int_z^1 e^{-h^{(-1)}(0,x)}dx, 
\end{align}
\begin{align}\label{rhoAfterNormalization}
\rho(s)=1 - \frac{1}{2L}\int_0^{2\Theta(s)}e^{-h^{(-1)}(t,1)+\frac{t}{2}}h_z\left(t,h^{(-1)}(t,1)\right)dt. 
\end{align}

Vice versa, if $h:\mathbb{R}_+ \times \mathbb{R} \rightarrow \mathbb{R}$ is a strictly increasing solution to \eqref{PDEforFunctionh} satisfying \eqref{Lowerbound}, let functions $v$ and $\rho$ be defined by \eqref{valuefunctionAfterNormalization} and \eqref{rhoAfterNormalization} respectively, with $v(s,z)>0$ for all $(s,z)\in\mathbb{R}_+\times(0,1]$. Then $v$ is the value function to the infinite horizon goal reaching problem \eqref{prob:InfiniteHorizon} with discount function $\rho$ and $v \in C^{1,2}\left( \mathbb{R}_{+}\times(0,1] \right)$. 

Furthermore, in both cases, the optimal strategy is given by \begin{align*}
\pi^*(s,z) = A(s)^{-1} \widehat{\mu} (s) \frac{h_z\left(2\Theta(s), h^{(-1)}(2\Theta(s),z)\right)}{z}.
\end{align*}
\end{theorem}

Theorem \ref{thm:InfHorizonCharacterization} provides a complete characterization of infinite horizon goal reaching problems with sufficiently smooth value functions.  
These are parametrized by increasing solutions $h:\mathbb{R}_+ \times \mathbb{R} \rightarrow \mathbb{R}$ to the (backward) heat equation \eqref{PDEforFunctionh} satisfying \eqref{Lowerbound}. 
Increasing solutions to the (backward) heat equation \eqref{PDEforFunctionh} were extensively studied in \cite{musiela2010portfolio} and can be parametrized by measures in 
\begin{align}\label{eq:SetOfMeasuresNu}
    \mathcal{B}^{+}(\mathbb{R})= \left\{\nu\ \text{Borel}\  {\text{measure}} \bigg\vert \text{for all }\ B\in \mathcal{B}, \nu(B)\geq 0 \ \text{and}\ \int_{\mathbb{R}}e^{yx}\nu(dy)<\infty, x\in \mathbb{R} \right\},
\end{align}
where $\mathcal{B}$ denotes the Borel sigma-algebra on $\mathbb{R}$.
We recall the following version of \cite[Propositon 9]{musiela2010portfolio}, originating from Widder's theorem, adapted to our notation. 
\begin{proposition}\label{Lemma:RepresentforFunctionh}
    (i) Let $\nu \in \mathcal{B}^{+}(\mathbb{R})$. Then, the function $h$ defined, for $(s,z)\in\mathbb{R}_{+}\times \mathbb{R}$, by 
    \begin{align}\label{Representationforh}
        h(s,z)=\int_{\mathbb{R}}\frac{e^{yz-\frac{1}{2}y^2s}-1}{y}\nu(dy)+C,
    \end{align}
    for any constant $C$, is a strictly increasing solution to (\ref{PDEforFunctionh}).\\(ii) Conversely, let $h:\mathbb{R}_{+}\times \mathbb{R}\rightarrow \mathbb{R}$ be a strictly increasing solution to (\ref{PDEforFunctionh}). Then, there exists $\nu\in\mathcal{B}^{+}(\mathbb{R})$ such that $h$ is given by (\ref{Representationforh}).
\end{proposition}

Theorem \ref{thm:InfHorizonCharacterization} and Proposition \ref{Lemma:RepresentforFunctionh} provide ways to construct discount functions $\rho$ that lead to infinite horizon goal reaching problems with analytical solutions. 
Whether one starts with a strictly increasing solution to \eqref{PDEforFunctionh} or first constructs one via a measure $\nu \in \mathcal{B}^{+}(\mathbb{R})$, the crucial condition to verify is \eqref{Lowerbound}. 
If the market price of risk $\kappa$ is constant, the following condition on $h$ is sufficient for \eqref{Lowerbound} to hold: There exist constants $M > 0$,  $t_0 > 0$, and  $\alpha > 1$ such that for all $t \geq t_0$,
\begin{align*}
    \left| e^{-h^{(-1)}(t,1) + \frac{t}{2}} h_z\left(t, h^{(-1)}(t,1)\right) \right| \le M t^{-\alpha}.
\end{align*}

We next provide two examples of families of infinite horizon goal reaching problems with analytical solutions building on the theory above. 

\begin{example}\label{ex2}
Consider the exponential discount function $\rho(s)=e^{-\gamma \Theta(s)}$, $\gamma> 0$,
corresponding, via \eqref{rhoAfterNormalization}, to the solution to the (backward) heat equation \eqref{PDEforFunctionh} given by \linebreak $h(s,z)=\frac{1}{1+\gamma}e^{(1+\gamma) z-\frac{1}{2}(1+\gamma)^2s}$. 
By Theorem \ref{thm:InfHorizonCharacterization}, the value function and optimal strategy to the infinite horizon goal reaching problem \eqref{prob:InfiniteHorizon} are given by 
\begin{align*}
V(s,z)=z^{\frac{\gamma}{1+\gamma}}e^{- \gamma \Theta(s)} \quad \mathrm{and} \quad \pi^*(s,z)=A(s)^{-1}\mu(s) (1+\gamma), \qquad (s,z)\in [0,\infty)\times(0,1).
\end{align*}
When the market price of risk $\kappa$ is constant and $\rho (s) = e^{-\gamma s }$, the solution is given by
\begin{align*}
V(s,z)=z^{\frac{\gamma}{\frac{1}{2} \kappa' \kappa+\gamma}}e^{- \gamma s} \quad \mathrm{and} \quad \pi^*(s,z)=A^{-1}\mu \left( 1 + 2\frac{\gamma}{\kappa' \kappa} \right), \qquad (s,z)\in [0,\infty)\times(0,1),
\end{align*}
see, also, \cite{browne1999beating}.
In one dimension, where the investor trades between a risk-free bond and a single stock, and assuming $\mu>0$, the optimal proportion of wealth invested in the stock is $\pi^* (s,z) = \frac{\mu}{\Sigma^2}+\frac{2 \gamma }{\mu}$. 
In particular, the optimal proportion of wealth invested in the stock is \textit{decreasing} in the drift of the stock when $\mu < \sqrt{2 \gamma} \Sigma$, and increasing in the drift of the stock only when $\mu > \sqrt{2 \gamma}\Sigma$.
This stands in sharp contrast to the Merton problem, where the optimal fraction of wealth invested in the risky asset increases linearly with the drift, given fixed risk aversion and volatility.
The underlying mechanism is intuitive in the context of goal reaching problems: when the drift of the market becomes small, the investor must allocate a greater share of wealth to the risky asset to preserve a sufficiently high probability of attaining the goal within a reasonable time horizon.

Another interesting feature of the optimal strategy is that $\lim_{\Sigma \rightarrow \infty} \pi^* (s,z) = \frac{2 \gamma }{\mu} > 0$.
The investor retains a strictly positive exposure to the stock even as the volatility of the stock goes to infinity. 
This behavior contrasts with the optimal strategy for the Merton problem, where  $\lim_{\Sigma \rightarrow \infty} \pi^* (s,z) = 0$. 
When maximizing expected utility of wealth at a fixed terminal horizon, the investor can allocate all wealth to the risk-free asset, thereby preserving current wealth. 
Under risk-aversion, this strategy is optimal in scenarios of extreme market volatility. 
However, this is not a good strategy for reaching a financial goal, as it ensures the goal will never be attained. The investor must put some of her money at risk, even under extremely adverse market conditions. Indeed, the lower bound $\lim_{\Sigma \rightarrow \infty} \pi^* (s,z) = \frac{2 \gamma }{\mu} > 0$ that the investor maintains in the stock when volatility goes to infinity is itself decreasing in the drift of the stock. 

\end{example}

We propose the following sufficient condition on the function $h$ ensuring that the inequality \eqref{Lowerbound} holds. In other words, we examine when a given defining element $h$ gives rise to a well-defined discount function and an associated value function.


\begin{proposition}\label{SufficientCondforLfinite}
Let $h$ be of the form $h(s,z)=\sum_{i=1}^N h^{(i)}(s,z)$ for some $N \in \mathbb{N}$,
where each $h^{(i)}$ is an increasing solution to \eqref{PDEforFunctionh}.
Assume that
\begin{itemize}
    \item[i)] There exist $\gamma>1$, $C>0$ such that $h(s,z)\geq \frac{C}{\gamma}\,e^{\gamma z-\frac12\gamma^2 s}$.
    \item[ii)] For every $i = 1, \dots , N$, there exist $\gamma_i>\gamma-1$, $K_i\ge0$ and $T_i \geq 0$ such that for all $t\ge T_{i}$ and $z\in\mathbb{R}$,
\begin{align*}
\partial_z h^{(i)}\bigl(t,z\bigr)\;\le\;K_i\,e^{\gamma_i z-\frac12\gamma_i^2 t},
\end{align*}
\end{itemize}
Then, \eqref{Lowerbound} holds. 
\end{proposition}

Next, we give an example that satisfies the required conditions of Proposition \ref{SufficientCondforLfinite}. This permits a direct application of the proposition and demonstrates its applicability.

\begin{example}\label{ex:ConstructingFromh}
Consider a class of discount functions $\rho(s)$ of which the defining element functions $h$ are given by $h(s,z)=\sum\limits_{i=1}^N\frac{C_i}{\gamma_i}e^{\gamma_i z - \frac{1}{2}\gamma_i^2 s},
$ where $C_i>0, \gamma_i>1$ and $N\in\mathbb{N}_{+}$.
Since each term is positive and the derivative of function $h$ in $z$ is computed by
$h_z(s,z)=\sum\limits_{i=1}^NC_ie^{\gamma_i z - \frac{1}{2}\gamma_i^2 s},$
the conditions of Proposition \ref{SufficientCondforLfinite} are readily satisfied and
\begin{align*}
    L=\lim\limits_{s\rightarrow +\infty} \frac{1}{2}\int_0^{2\Theta(s)}\sum\limits_{i=1}^NC_ie^{(\gamma_i-1)h^{(-1)}(t,1)  - \frac{1}{2}(\gamma_i^2-1) t}dt<\infty,
\end{align*}
since for large $t$, the term with smallest $\gamma_i$ dominates. 
Then the value function is given by 
\begin{align}
    v(s,z)= 1 - \frac{1}{2L}\sum\limits_{i=1}^NC_i\int_0^{2\Theta(s)}e^{(\gamma_i-1)h^{(-1)}(t,z)  - \frac{1}{2}(\gamma_i^2-1) t}dt -\frac{1}{L}\int_z^1 e^{-h^{(-1)}(0,x)}dx,
\end{align}
and the optimal strategy is given by
\begin{align*}
\pi^*(s,z)
=
\frac{1}{z}A(s)^{-1}\widehat{\mu}(s)
\sum_{i=1}^N
C_i
e^{\gamma_i h^{(-1)}(2\Theta(s),z)-\gamma_i^2\Theta(s)}.
\end{align*}

\end{example}

\section{Numerical Algorithm and Comparative Statics}


Analytical solutions to goal reaching problems are, in general, difficult to derive. 
Even for the infinite horizon case, where we have a complete characterization of all goal reaching problems with sufficiently smooth value functions, it is typically difficult to check whether a given discount function $\rho$ admits a representation \eqref{rhoAfterNormalization} for a suitable function $h$ satisfying all requirements. 

In this section, we turn towards the numerical analysis of  investment behavior for goal reaching problems and provide comparative statistics. 
For the problem with finite deadline $T$ we employ an approximation algorithm known as Howard's algorithm, which decouples the optimization from the boundary value problem.
We then extend the algorithm to the inifinte horizon case through a convergence argument. 

\begin{algorithm}[ht]
\caption{Howard's Algorithm}\label{alg:Howard}
\begin{algorithmic}
\State \textbf{Input:} Terminal horizon $T$, discount function $\rho$, market coefficients $m_Y, \Sigma_Y, \mu, \Sigma,\alpha$ and $\beta$, and error bound $\epsilon^H>0$.

\State \textbf{Initialization:} Set $k=0$ and choose an arbitrary initial control $f^0 \in \mathcal{A}$.\\
\textbf{Repeat}
    \State \hspace{1.8em} \parbox[t]{\dimexpr\linewidth-3em\relax}{\textbf{(i)} Solve the boundary value problem for the fixed control $f^k$ using the implicit finite difference method:
    \begin{align*}
    \mathcal{L}^{f^k}J^k(s,y,z)=0,
    \end{align*}
    with boundary conditions $J^k(s,y,1)=\rho(s)$, $J^k(T,t,z)=\rho(T)u(z)$, and $J^k(s,y,0)=0$.
    
    \State \textbf{(ii)} Compute the succeeding control $f^{k+1}$ by solving:
    \begin{align*}
    f^{k+1} := \argmax_{f \in \mathcal{A}} \left\{\mathcal{L}^{f^k}J^k(s,y,z)\right\}.
    \end{align*}
    
    \State \textbf{(iii)} Set $k := k+1$ and repeat from step (i).}\\~\\
\textbf{Until}{ the convergence criterion: $$\sup\limits_{(s,y,z)\in[0,T]\times\mathbb{R}^d\times(0,1]}\vert f^{k+1}(s,y,z)-f^{k}(s,y,z)\vert<\epsilon^H$$ is met.}

\State \textbf{Return} the optimal control $\pi^*=f^{k+1}$.
\end{algorithmic}
\end{algorithm}


The HJB equation is solved using an implicit finite-difference method combined with Howard's policy iteration outlined in Algorithm \ref{alg:Howard}. The state and time domains are discretized on uniform grids with $\Delta z=\Delta t=5\times10^{-4}$. At each time step, the control is initialized by the optimal policy obtained at the next time level (with the terminal step initialized by the constant control $f^{(0)}(z)\equiv1$). Given the current policy, the policy evaluation step consists of solving the resulting linear system obtained from the fully implicit finite-difference discretization, with boundary conditions imposed directly as $w(t,0)=0$ and $w(t,1)=\rho(t)$, and terminal condition $w(T,z)=\rho(T)u(z)$. The policy is then updated pointwise according to the analytical maximizer $\eqref{Optimalf}$. The policy evaluation and policy improvement steps are repeated until the policy converges, after which the algorithm proceeds to the next time step.
When the number of iterations increases, we will have $J^k\leq J^{k+1}$, and that $f^k$ and $J^k$ jointly
converge to the optimal feedback control $\pi^*$ and the value function $V(s,y,z)$ of our optimal control problem; see, for example, \cite{bokanowski2009some}.

To extend the numerical algorithm to the infinite horizon case, we first link the finite-horizon and infinite-horizon goal reaching problems and show that the former converges towards the latter as the time horizon goes to infinity. 

\begin{proposition}\label{Prop:Convergence}
Let  $\rho:\mathbb{R}_{+}\rightarrow (0,1]$ be a discount function and $u(z):(0,1]\rightarrow (0,1]$ be a terminal utility function. For $T > 0$, let $V^T(s,y,z) : [0,T]\times\mathbb{R}^{d}\times(0,1] \rightarrow \mathbb{R}$ be the value function of the finite horizon problem with horizon $T$  given in \eqref{eq:ValueFunction}. 
Analogously,  let $V^\infty(s,y,z) : [0,\infty) \times\mathbb{R}^{d}\times(0,1] \rightarrow \mathbb{R}$ be the value function of the infinite horizon problem \eqref{prob:InfiniteHorizon}. 
Then, for any $(s,y,z)\in [0,\infty) \times\mathbb{R}^{d}\times(0,1]$, the value function $V^T(s,y,z)$ converges to $V^{\infty}(s,y,z)$ as $T\rightarrow +\infty$.
\end{proposition}

When optimal strategies for all goal reaching problems with finite horizon $T$ are unique, and the optimal strategy for the infinite horizon problem satisfies a strong version of uniqueness, then optimal strategies also converge. 
\begin{proposition}\label{prop:ConvergenceStrat}
Assume that, for each $(s,y,z)\in [0,\infty)\times\mathbb{R}^{d}\times(0,1]$ and all $T\ge s$,
there exist unique maximizers
$\pi^{*,T}\in\arg\max\limits_{\pi\in\mathcal A_{s,T}} J^{T}(s,y,z;\pi)$ and $\pi^{*,\infty}\in\arg\max\limits_{\pi\in\mathcal A_{s,\infty}}J^{\infty}(s,y,z;\pi)$.
Moreover, let $d$ be a metric on $\mathcal{A}$ such that for every $\varepsilon>0$ there exists $\eta:=\eta(s,y,z;\varepsilon)>0$ such that
\begin{align}\label{eq:StrongUniqueness}
    J^{\infty}(s,y,z;\pi^{*,\infty})\ \ge\ \sup_{d(\pi,\pi^{*,\infty})\ge \varepsilon} J^{\infty}(s,y,z;\pi)\ +\ \,\eta.
\end{align} 
Then $\pi^{*,T}$ converges to $\pi^{*,\infty}$ in $d$ as $T\to\infty$.
\end{proposition}


We herein approximate the solution to the infinite-horizon problem using a sequence of finite-horizon solutions. 
Alternatively, one can employ the Trotter-Kato approximation scheme to solve the HJB equation \eqref{DPE}. This scheme is based on a splitting of the HJB equation into two sub-equations, see \cite{nadtochiy2013approximation} for an application to an optimal investment problem. 
We refer interested readers to \cite{warwick189223Yuwei} for details on the Trotter-Kato approximation method used to construct approximations of the value function and optimal policies for the goal reaching problem.


\subsection{Benchmark Model and Parameters}
In the reminder of this section, we provide a numerical analysis using Howard's algorithm to illustrate optimal strategies and investigate their sensitivities with regards to preference and model parameters. 
We consider two commonly used families of discount functions, a hyperbolic discount function, $\rho_h(s)=\frac{1}{1+ks}$, and exponential discount function, $\rho_e(s)=e^{-ks}$. 
For both families, the parameter $k$  quantifies an investor's aversion towards time risk, and determines the rate of discounting.
In the legends of the numerical results below, $k$ denotes the hyperbolic discount parameter $k_h$. For the exponential specification, we set $k_e=\log(1+k)$, so that both discount functions discount time $s=1$ by the same amount. This normalization allows for a direct comparison between the two discounting schemes.
Unless otherwise stated, we use the utility function \(u(z)=(1-e^{-\gamma z})/(1-e^{-\gamma})\), with \(\gamma>0\).

\begin{figure}[H]
   \includegraphics[width=7.5cm, height=6cm]{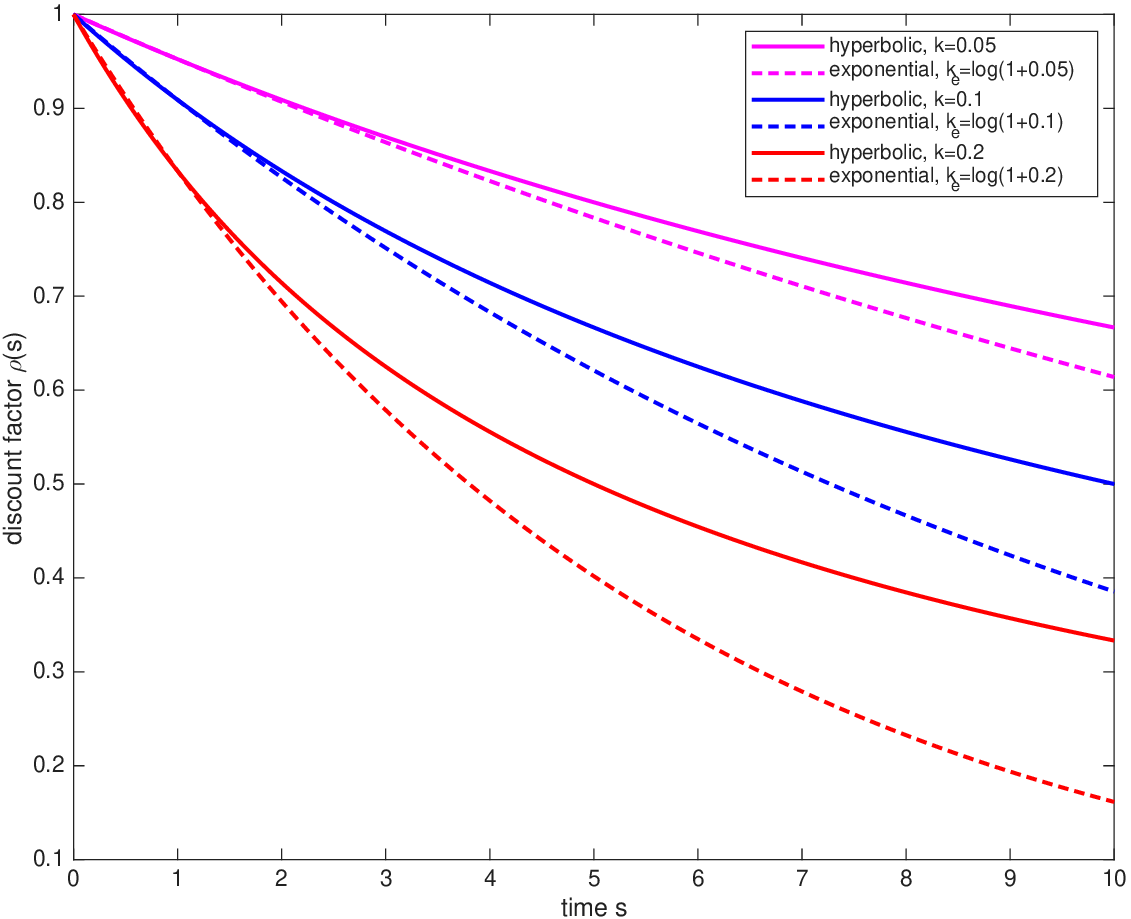} 
   \includegraphics[width=7.5cm, height=6cm]{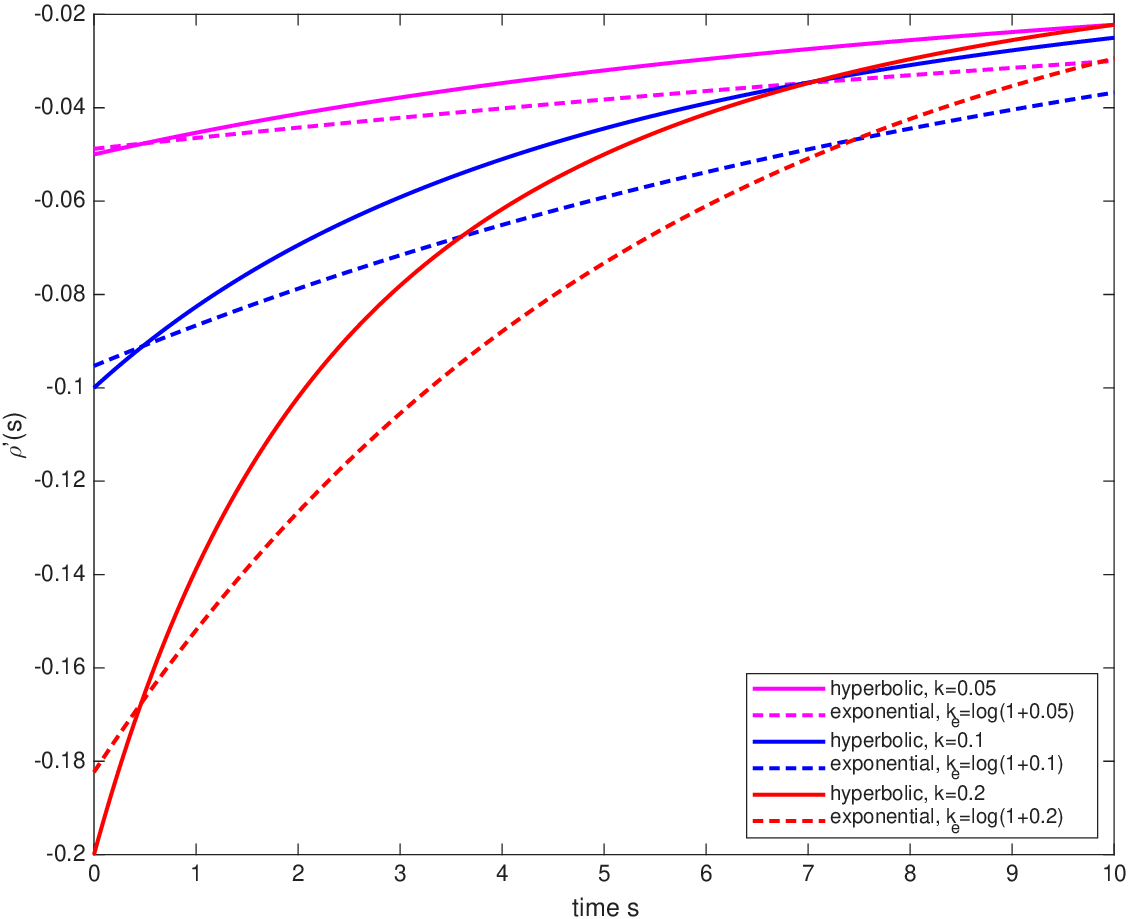}
   \caption{Hyperbolic and exponential discount functions and their first derivative
   }
   \label{figDiscountFunctions}
 \end{figure}
 
Figure \ref{figDiscountFunctions} compares the hyperbolic and exponential discount functions and their first derivatives for $k=0.05, k=0.1$ and $k=0.2$. 
The left panel shows that a larger $k$ leads to stronger discounting and reflects greater impatience toward delayed goal achievement.
Although the exponential parameters are chosen to match the corresponding hyperbolic discount factors at $s=1$, the two specifications differ at longer horizons. Hyperbolic discounting declines more slowly over time and therefore assigns a larger relative weight to distant goal reaching times.

The right panel plots $\rho'(s)$, which measures the marginal loss in value from additional delay. A larger $k$ leads to a more negative initial derivative, indicating a stronger marginal penalty for short delays. Under hyperbolic discounting, the first derivative is very steep near the origin but increases rapidly toward zero as $s$ grows. By contrast, the exponential specification exhibits a more persistent decline, reflecting its constant proportional discount rate, $\frac{\rho'(s)}{\rho(s)}$. 
Therefore, hyperbolic discounting is more sensitive to near-term delays but becomes more tolerant of long delays, whereas exponential discounting imposes a marginal delay penalty, measured by $-\rho'(s)$, that declines more gradually over time.

The benchmark model for stock dynamics is the Black and Scholes model 
\begin{align*}
    dS_t = S_t (\mu d_t + \sigma dW_t ) ,\quad t \geq 0,
\end{align*}
with one risky asset and benchmark parameters for drift and volatility given by $\mu=0.09$, and $\sigma=0.35$. Benchmark preference parameters are $k = 0.05,$ $k = 0.1$ or $k = 0.2$, and $\gamma = 2$. 
We model the stochastic goal by
$dG_t=G_t(\alpha dt+\beta_1 dW_t+\beta_2 d\hat{W}_t), \ t\geq 0,$ where $W$ and $\hat{W}$ are independent standard Brownian motions.
and set the initial funding ratio $z_0=0.6$ and $\alpha=\beta_2=0.01,\beta_1=0$. 
The exogenous time horizon is given by $T = 10$. 
For notational simplicity, in all subsequent figures we report only the value of \(k\), with the exponential parameter understood to satisfy \(k_e=\log(1+k)\).

\begin{figure}[H]
    \includegraphics[width=7.5cm, height=6cm]{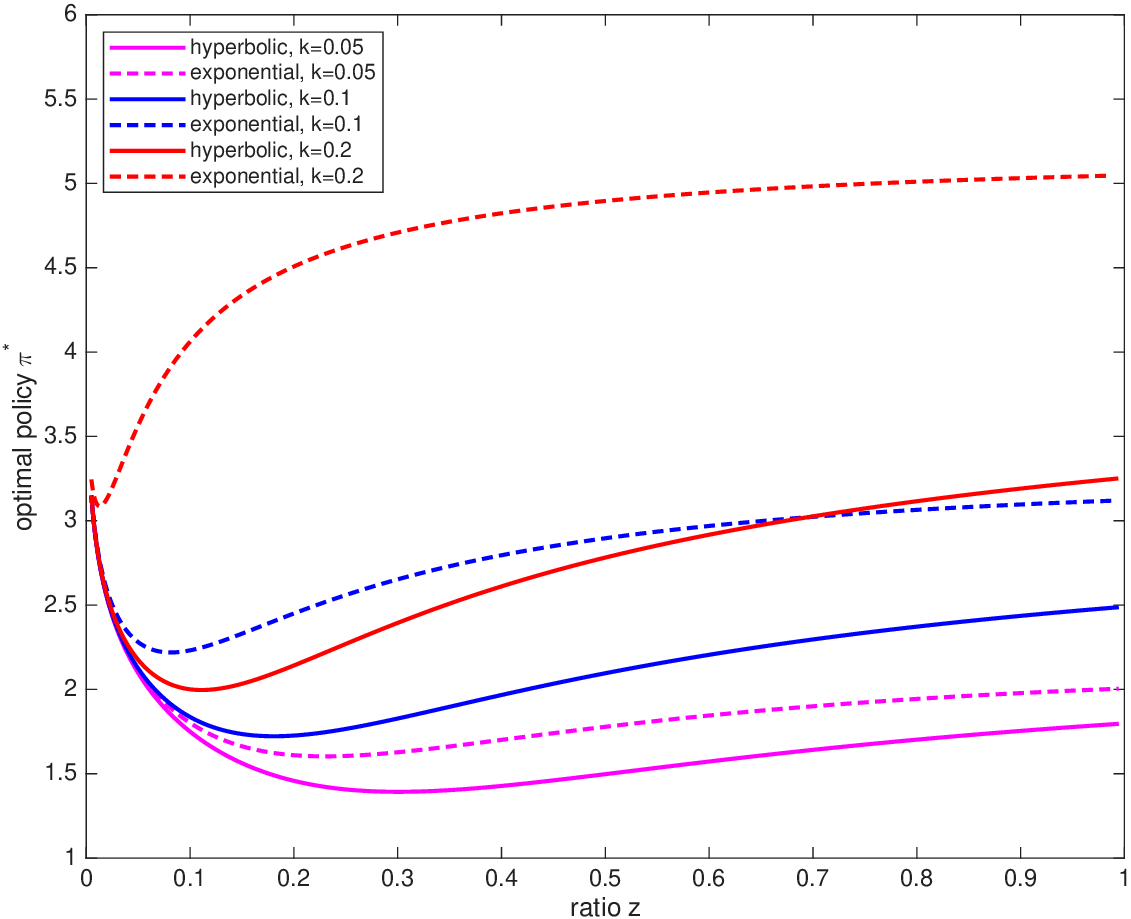}
   \includegraphics[width=7.5cm, height=6cm]{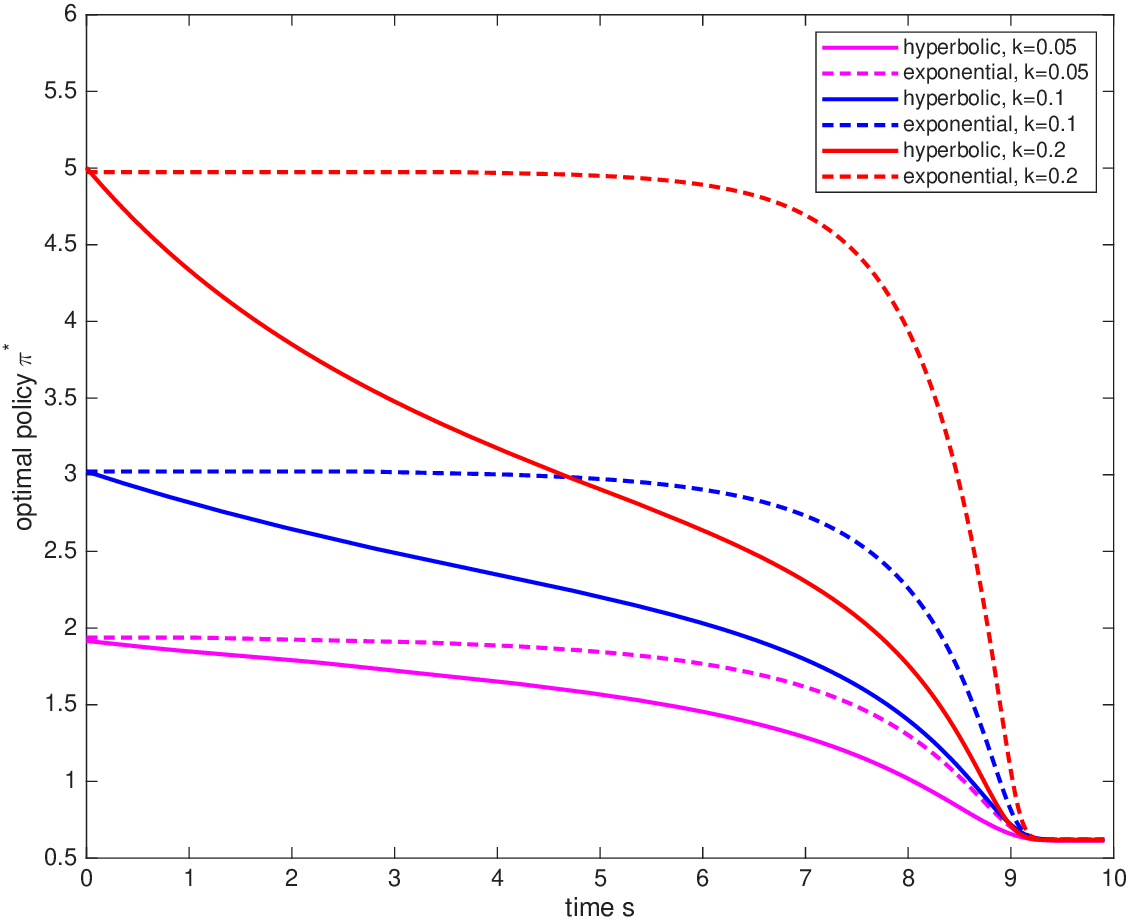}\\ 
   \includegraphics[width=7.5cm, height=6cm]{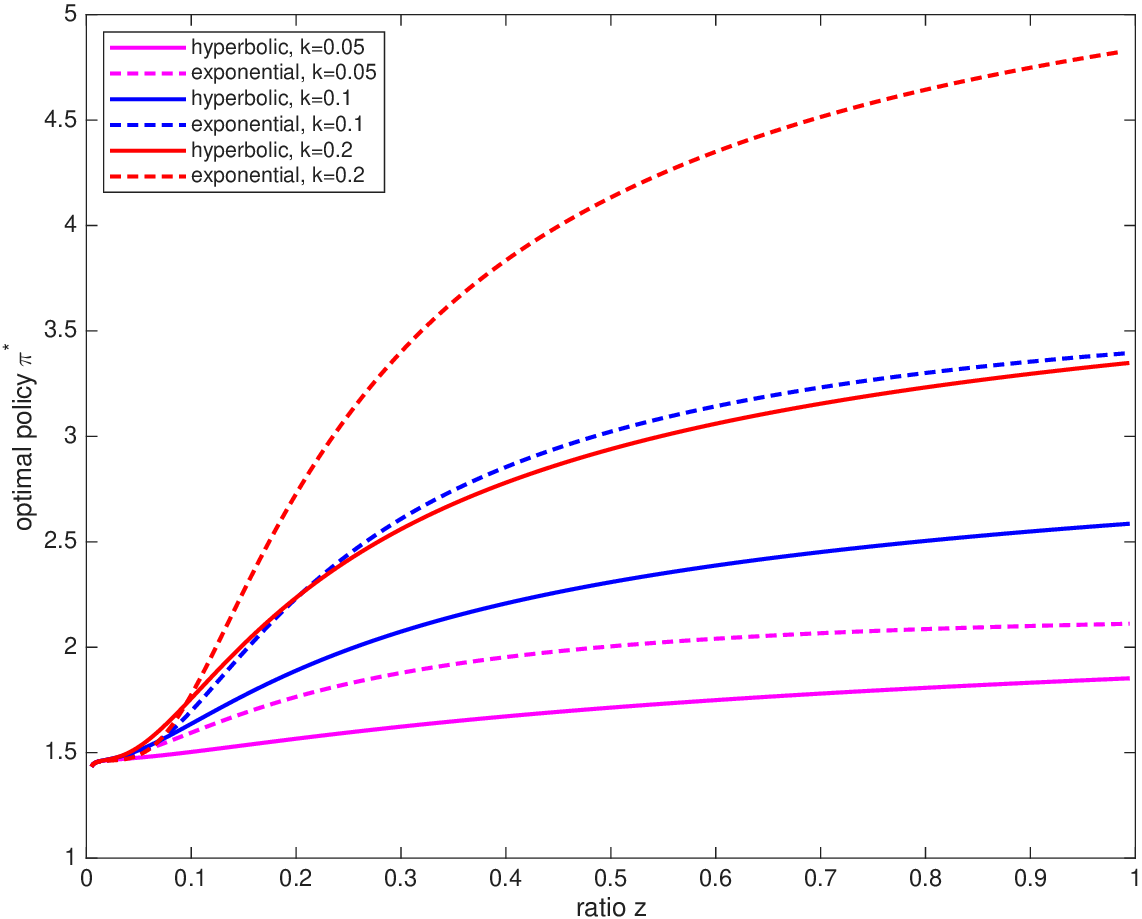}
   \includegraphics[width=7.5cm, height=6cm]{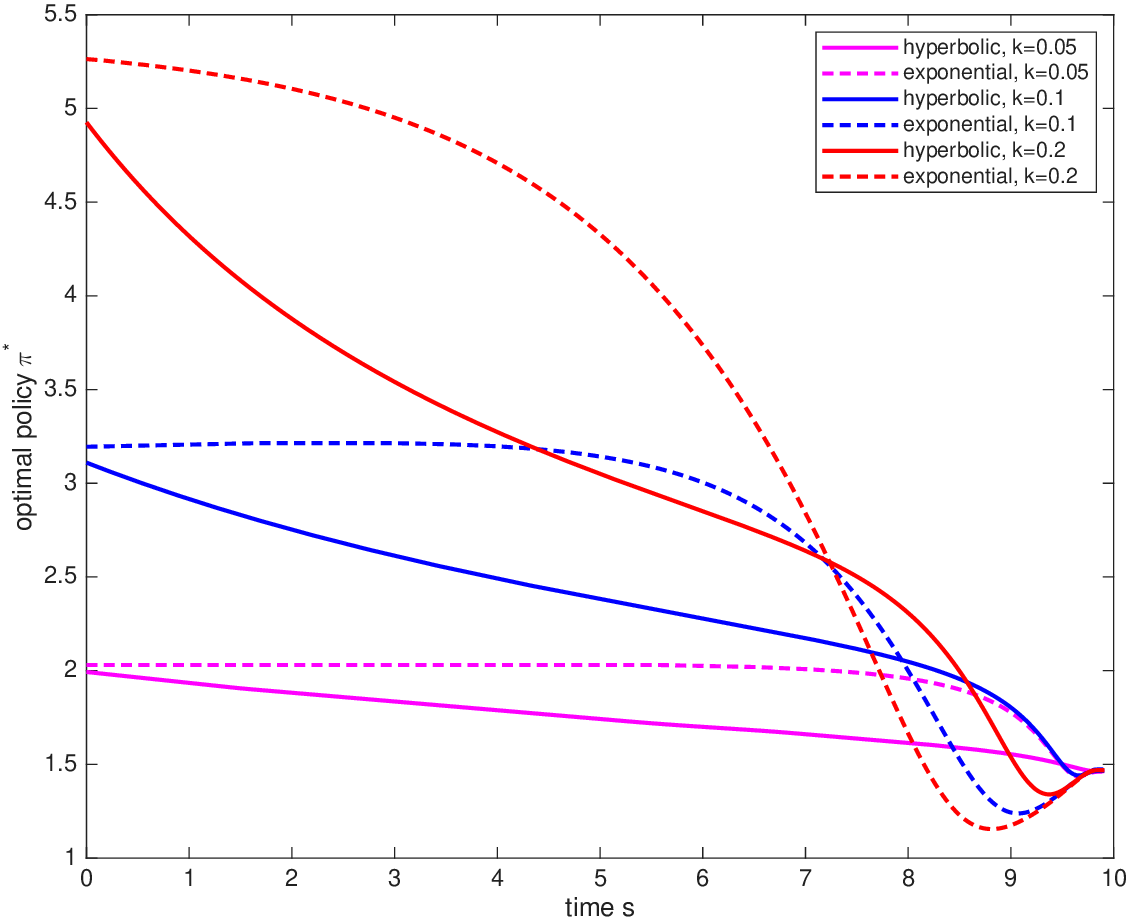}
 \caption{Optimal policy $\pi^*$ under different levels of time risk aversion
    for the CARA terminal utility (top panels) and the alternative CRRA
    terminal utility $u(z)=\sqrt{z}$ (bottom panels).
    The parameters are $\mu=0.09$, $\sigma=0.35$, $T=10$, $\gamma=2$ and
    $\alpha=\beta_2=0.01$, $\beta_1=0$.
    In the left panels, we fix $t=5$ and vary $z$; in the right panels,
    we fix $z=0.6$ and vary $t$.}
   \label{figHyperVaryk-optimalpolicy}
 \end{figure}

Figure \ref{figHyperVaryk-optimalpolicy} plots the optimal policy at $t=5$ against the funding ratio, respectively the optimal policy at $z=0.6$ over time. 
We observe that investors tend to allocate more wealth to the risky asset as the parameter $k$ increases. An investor with a higher $k$ becomes more eager to reach the goal sooner and may therefore adopt a more aggressive risky position.

The comparison between the CARA and CRRA terminal utilities helps clarify the non-monotonicity of $\pi^*$ in $z$. Under CARA terminal utility, when $z$ is small and the goal is difficult to attain, the terminal-utility component is relatively important, and the corresponding Merton fraction
$
f^{M}_{\mathrm{CARA}}(z)=\frac{\mu}{\gamma\sigma^2 z}
$
decreases with $z$. As the funding ratio increases, however, the goal becomes increasingly attainable, strengthening the incentive to use risky investment to achieve it earlier and eventually reversing the initial decline. The turning point occurs earlier for larger $k$ and under exponential rather than hyperbolic discounting, as delayed goal achievement is penalized more heavily.

Under CRRA utility
$
u(z)=\sqrt{z},
$
the corresponding Merton fraction
$
f^{M}_{\mathrm{CRRA}}=\frac{2\mu}{\sigma^2}
$
is independent of $z$, so the decreasing terminal-utility effect disappears. Consistently, the optimal policies are increasing in $z$ for all three values of $k$. Moreover, the curves are close to the common Merton benchmark when $z$ is small but separate increasingly as $z$ rises, with larger $k$ and exponential discounting inducing more aggressive investment. This comparison supports the interpretation that the non-monotonicity under CARA results from the interaction between the decreasing terminal-utility benchmark and the strengthening goal reaching incentive.


The right panels further illustrate how the role of time preference diminishes as the deadline approaches. Although the CARA and CRRA specifications exhibit different policy levels and intermediate-time dynamics, the optimal policies under different discounting specifications gradually converge near maturity. As the remaining horizon becomes shorter, risky investment has increasingly limited time to generate sufficient upside for reaching the goal, while the terminal utility becomes relatively more important in the optimization.

In particular, around $s=9$, starting from $z=0.6$, the probability of reaching the goal over the remaining horizon becomes very small. The optimization is therefore increasingly dominated by the terminal utility, and the effect of time preferences largely disappears. Accordingly, the optimal policies converge toward the corresponding Merton allocations determined by the terminal utility. Under CARA utility, this limiting allocation is
$
f^{M}_{\mathrm{CARA}}(z)=\frac{\mu}{\gamma\sigma^2z},
$
which is approximately $0.61$ at $z=0.6$, whereas under CRRA utility $u(z)=\sqrt{z}$, it is
$
f^{M}_{\mathrm{CRRA}}=\frac{2\mu}{\sigma^2}\approx1.47.
$
This explains both the convergence of the different time-preference specifications and the different limiting policy levels in the two panels. Since these Merton allocations are independent of time, the nearly constant policies observed immediately before maturity are also consistent with the corresponding terminal-utility portfolio problems.

We notice that as the preference parameter $k$ increases, the optimal policy under hyperbolic discounting diverges increasingly from its exponential counterpart, reflecting the growing difference between the two discount functions at larger time risk aversion. 
Furthermore, there is a remarkable similarity in investment behavior among investors, regardless of their levels of time risk aversion. This similarity appears when the investment goal is distant, with a high level of uncertainty regarding goal achievement, and the differences in discount functions play a relatively minor role compared to situations where the ratio $z$ is higher. Additionally, as time approaches the terminal horizon, it becomes highly likely that the goal cannot be fulfilled, and a similarity in investment behavior becomes apparent. From this, we infer that investors tend to make similar portfolio choices when the goal is set unrealistically high.

 \begin{figure}[H]
   \includegraphics[width=7.5cm, height=6cm]{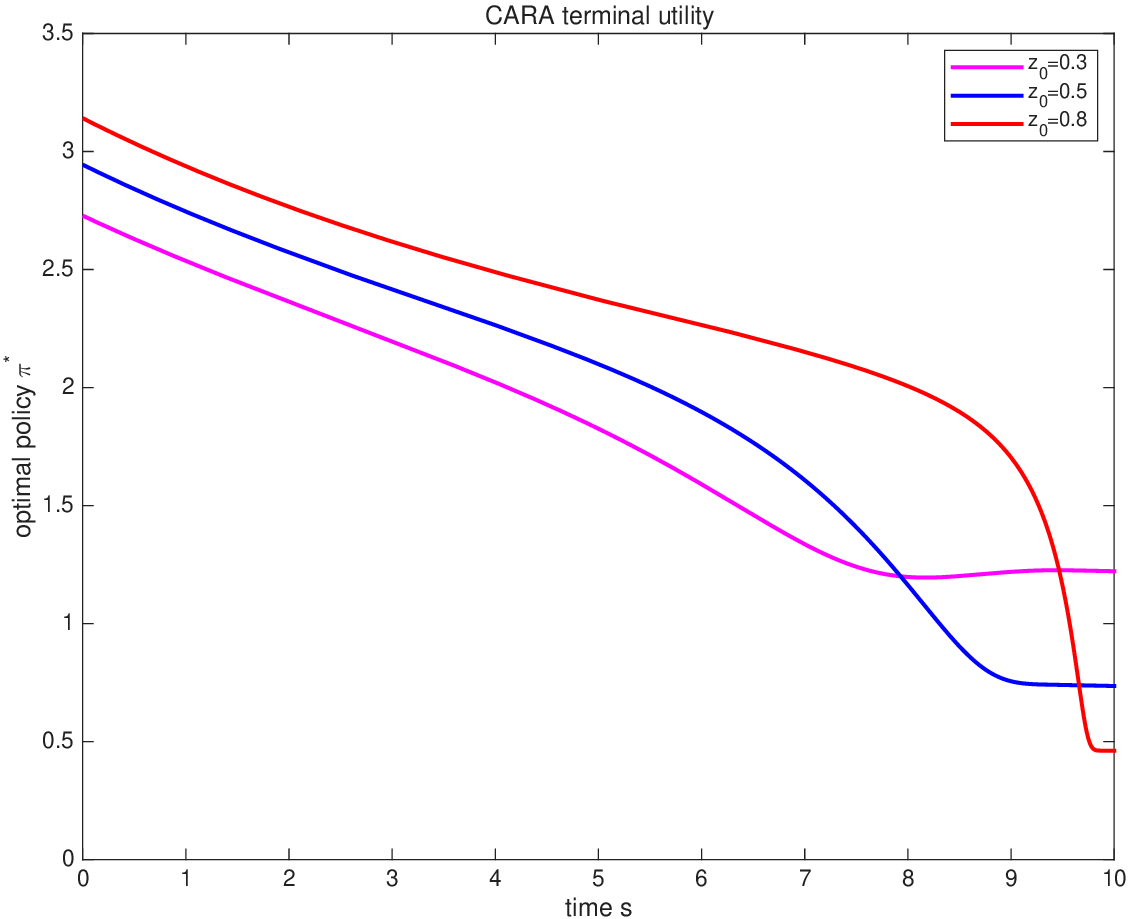}
   \includegraphics[width=7.5cm, height=6cm]{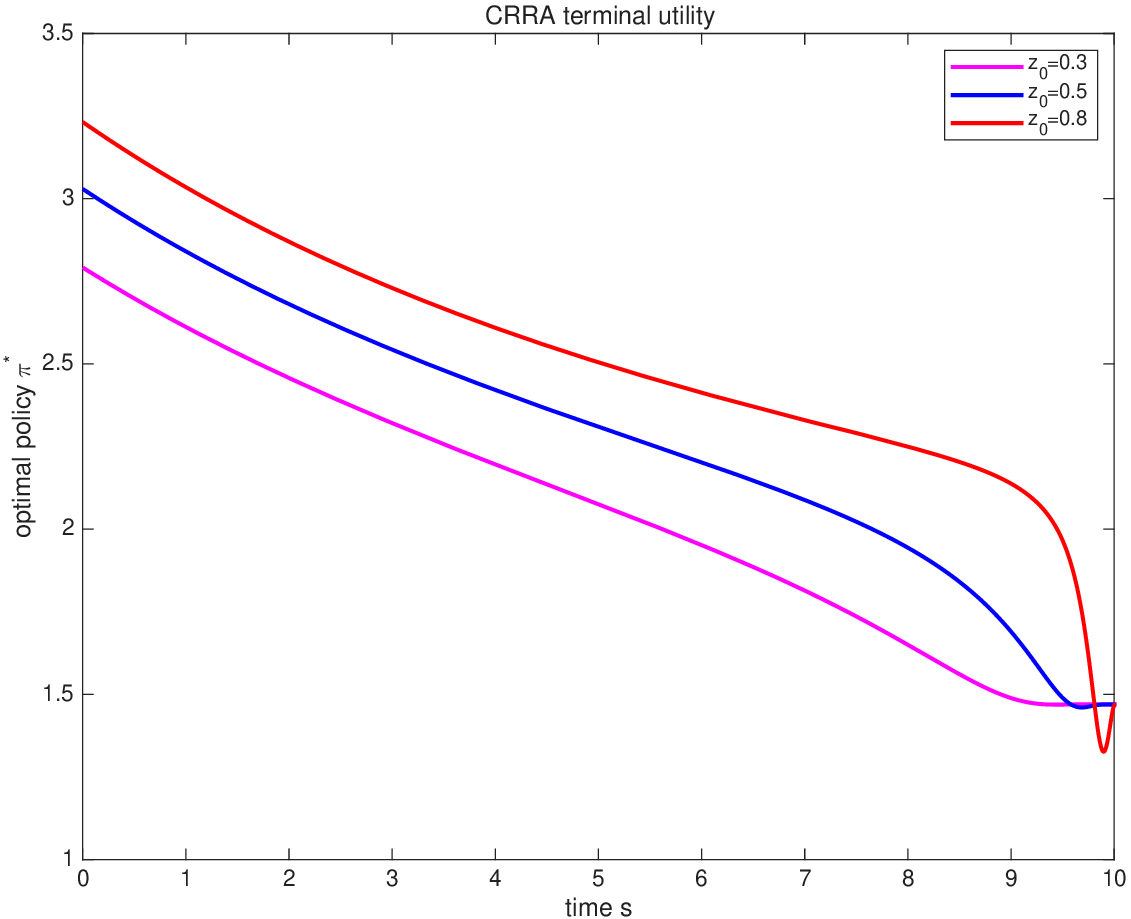}
\caption{Optimal policy $\pi^*$ over time for different initial funding ratios $z_0$ under CARA and CRRA terminal utilities. Both panels use the same hyperbolic discounting with $k=0.1$. The parameters are $\mu=0.09$, $\sigma=0.35$, $T=10$, $\gamma=2$, and $\alpha=\beta_2=0.01,\beta_1=0$.}
   \label{fig:CompareZ0}
 \end{figure}

Figure \ref{fig:CompareZ0} further illustrates how the optimal policy converges toward the policy implied by the terminal utility as maturity approaches. 
In both panels, this transition occurs progressively later as the initial funding ratio moves closer to the target. When $z_0$ is low, the remaining horizon is relatively insufficient for reaching the goal at an earlier stage, hence the goal-achievement component loses importance sooner. By contrast, for $z_0=0.8$, the investor starts closer to the target, then the goal can still be reached with a meaningful probability even relatively late in the investment horizon, allowing the goal-seeking motive to remain relevant until much closer to maturity. 

Attainability, however, is only part of the explanation. Even when the goal can still be reached close to maturity, the benefit of reaching it slightly earlier becomes small. If the goal is reached at time $\tau<T$, the payoff is $\rho(\tau)$, while reaching the target at the terminal time yields $\rho(T)$. Since $\rho(\tau)-\rho(T)\to0$ as $\tau\uparrow T$, there is little additional benefit from reaching the goal slightly before maturity.
Therefore, close to maturity, there is little additional reward from achieving the goal slightly before $T$. The incentive to adjust the portfolio specifically for earlier goal achievement consequently weakens, and the terminal utility becomes increasingly important in determining the optimal policy.

The limiting levels under CARA and CRRA provide further evidence for this interpretation. Under CRRA utility $u(z)=\sqrt{z}$, all three policies converge to approximately
$
f^M_{\mathrm{CRRA}}=\frac{2\mu}{\sigma^2}\approx1.47,
$
which is independent of the funding ratio. Under CARA utility, instead, the corresponding Merton allocation is
$
f^M_{\mathrm{CARA}}(z)=\frac{\mu}{\gamma\sigma^2z},
$
giving approximately $1.22$, $0.73$, and $0.46$ for $z_0=0.3$, $0.5$, and $0.8$, respectively, consistent with the limiting levels in the left panel. Thus, once the goal-achievement motive becomes negligible near maturity, the optimal policy converges to the corresponding classical terminal-utility allocation.

\begin{figure}[H]
   \includegraphics[width=7.5cm, height=6cm]{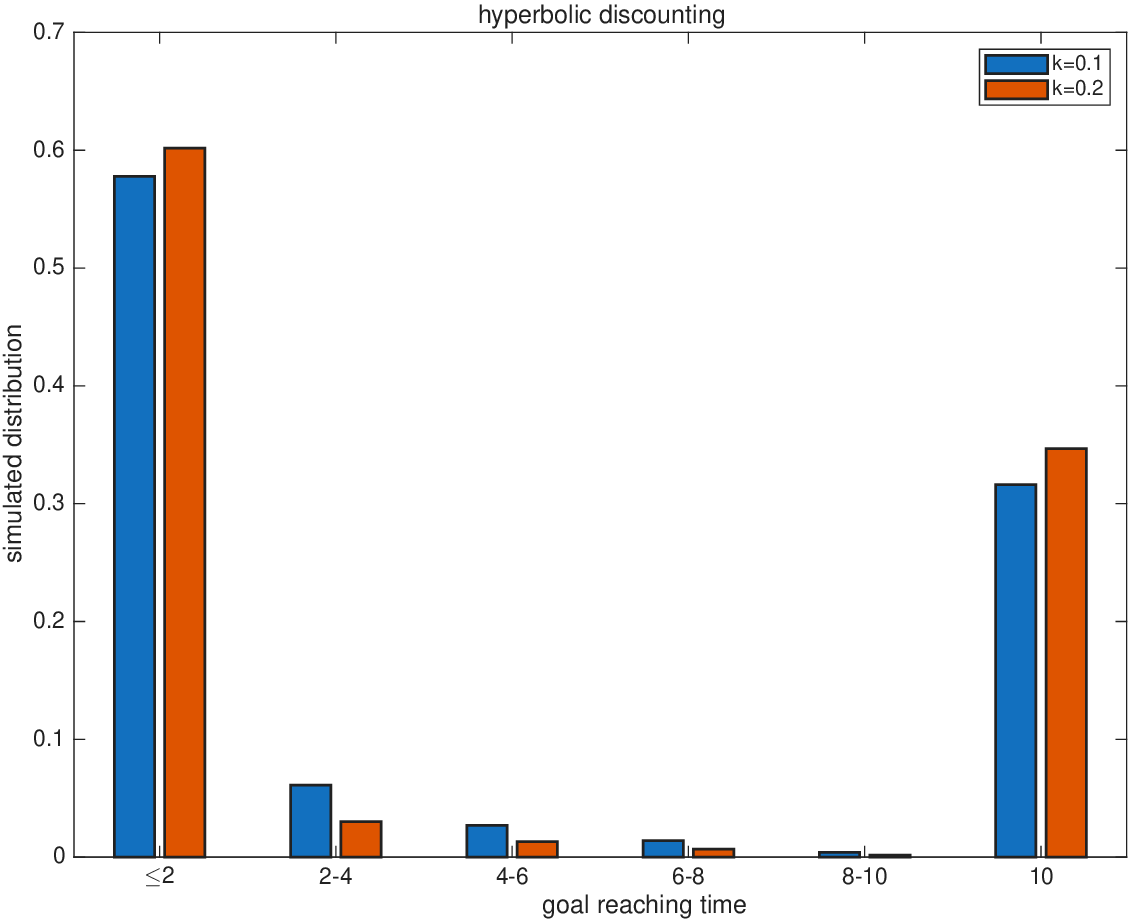} 
   \includegraphics[width=7.5cm, height=6cm]{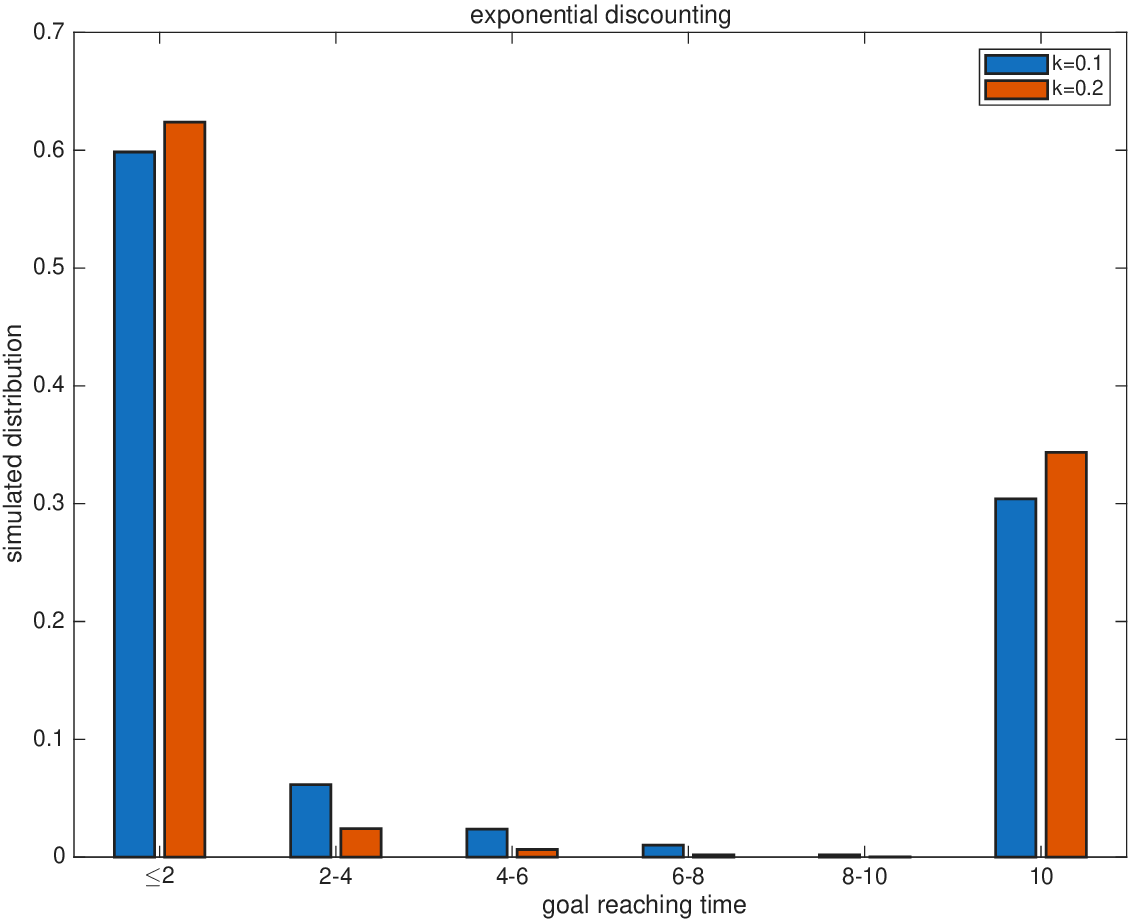}
   \caption{Simulated distribution of the goal reaching time  under different specifications. The category $10$ denotes paths on which the goal is not reached before the terminal horizon $T=10$.
   }
   \label{figEmpiricalDisVaryk}
 \end{figure}

The optimal strategies derived above naturally lead to different distributions of goal reaching times. To quantify these differences,
Figure \ref{figEmpiricalDisVaryk} presents the simulated distributions of goal reaching times under different discounting specifications and values of $k$, based on $5\times 10^5$ Monte Carlo simulations using the corresponding optimal strategies. For both discount functions, a larger $k$ increases convexity and hence strengthens delay-risk seeking, meaning that an additional zero-mean delay risk is preferred to zero additional delay. The comparison between the two discounting specifications shows that exponential discounting assigns slightly more probability mass to early goal achievement, while hyperbolic discounting leaves relatively more mass in the no-hit bin. This pattern is consistent with the marginal delay penalty: although hyperbolic discounting is more sensitive to near-term delays, it becomes less sensitive to long delays, whereas exponential discounting maintains a larger marginal penalty at longer horizons. As a result, exponential discounting provides a stronger incentive to avoid long delays, while hyperbolic discounting penalizes late achievement less severely.

Furthermore, by interpreting the simulated goal reaching time distribution under the optimal strategy as reflecting the investor's preferred timing risk, our numerical results are consistent with the notion of risk-seeking discounting in \cite{ebert2020decision}. With a convex discount function, the investor is willing to accept a higher probability of both favorable and unfavorable timing outcomes. In our setting, this appears as a more polarized goal reaching time distribution when $k$ increases: the probability of early goal achievement rises, but so does the probability of non-attainment before maturity.
Specifically, for each fixed $k$, the simulated probability of achieving the goal at interval $[0,2]$ and the deadline $T=10$ is significantly higher than intermediate values, with this difference becoming more pronounced when $k$ is higher. In instances where $k$ is higher, indicating a greater inclination towards risk-seeking behavior, investors exhibit a preference for prospects with a higher probability of achieving the goal before time $t=2$, while simultaneously having a higher probability of achieving the goal at the latest time, compared to lower values of $k$.
This implies a willingness to accept a higher potential for achieving the goal at an early stage, even if it comes at the expense of accepting increased risk of not achieving the goal ultimately.



\begin{figure}[H]
   \includegraphics[width=7.5cm, height=6cm]{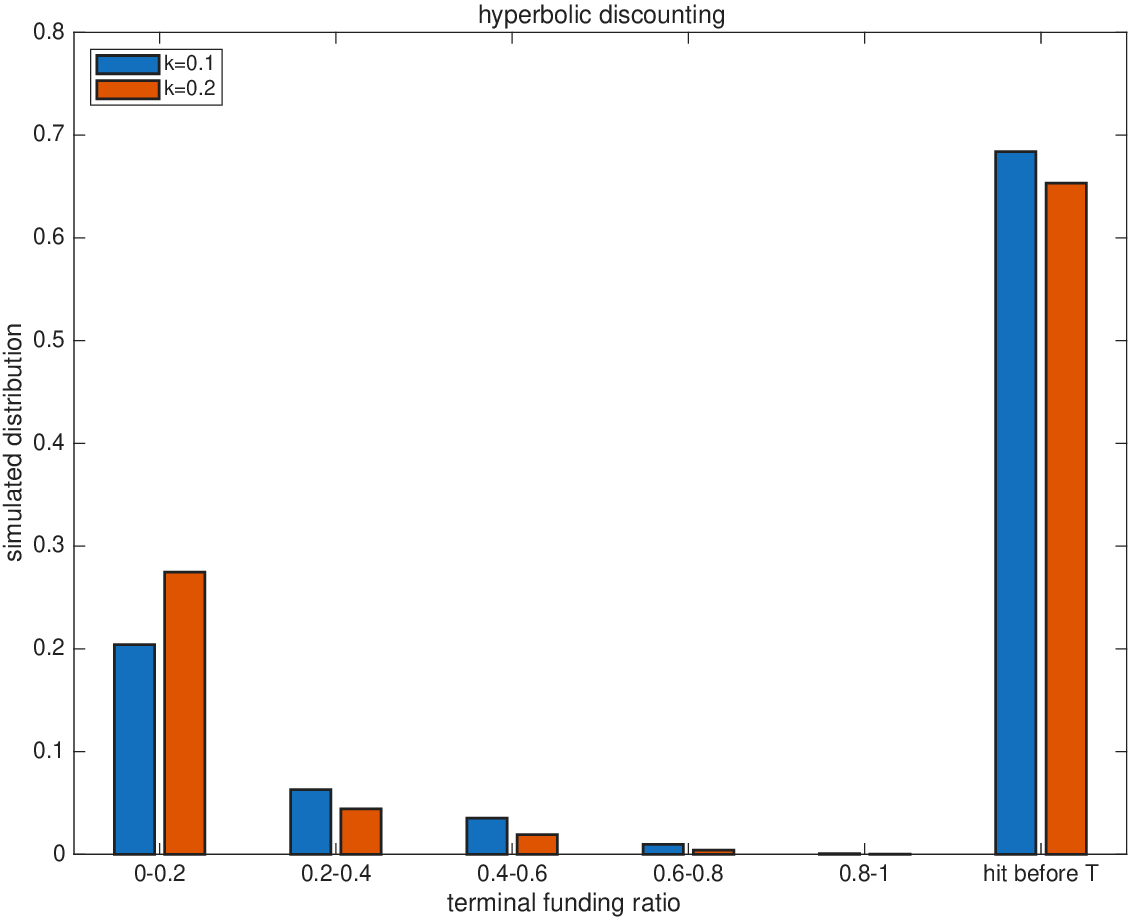} 
   \includegraphics[width=7.5cm, height=6cm]{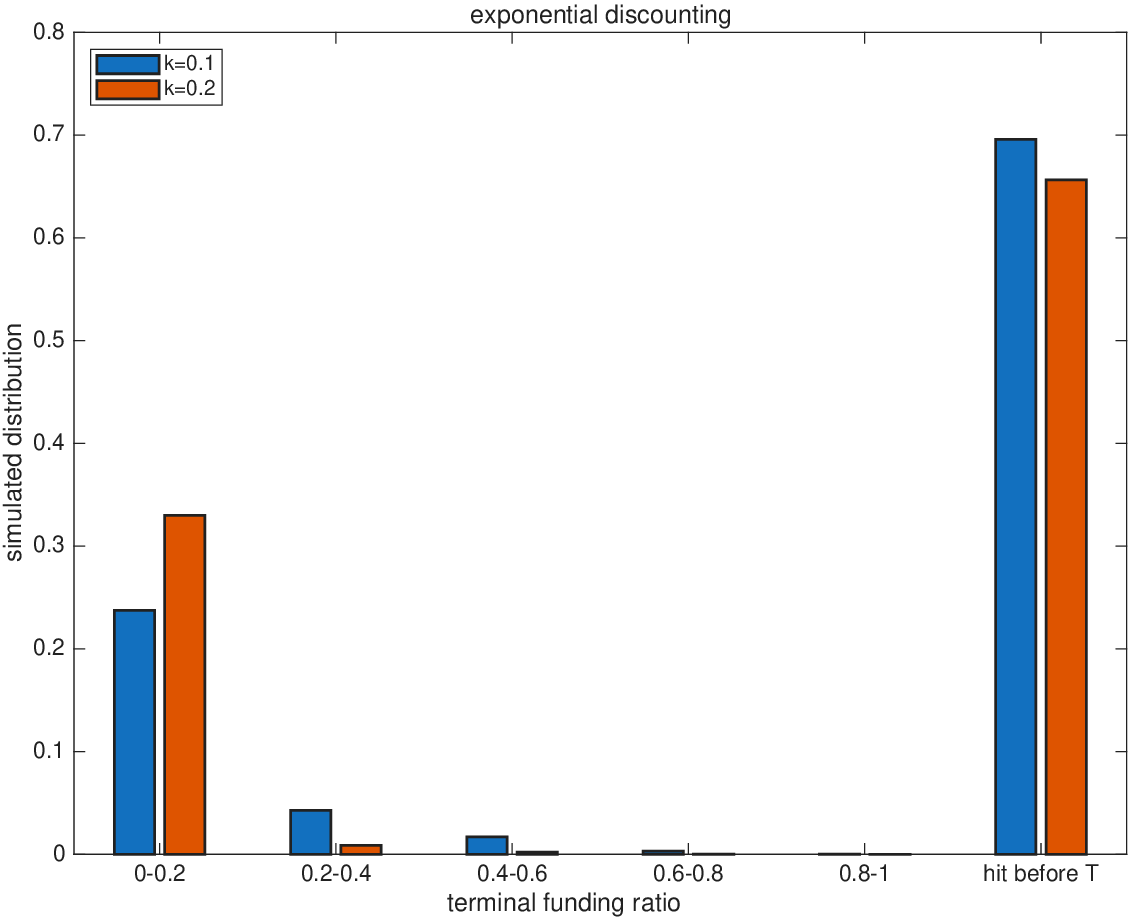}
  \caption{Simulated distributions of the terminal  ratio under different specifications
  }
   \label{figEmpiricalDisVaryk-TerminalRatio}
 \end{figure}
To provide a more complete picture of the investment performance, we show in Figure \ref{figEmpiricalDisVaryk-TerminalRatio} the simulated distribution of $Z_T$ for paths that do not reach the goal before $T$, together with the frequency of paths that hit the goal before the deadline, under both discount functions. The distribution is concentrated in two regions: a large mass on hitting the goal before maturity and a smaller mass in the low terminal-ratio region $[0,0.2)$. This reflects the goal reaching structure of the problem: favorable paths tend to cross the benchmark before $T$, while non-hit paths are more likely to correspond to unfavorable return realizations or insufficient wealth growth. Since the optimal policy accepts risk exposure to increase the probability of early goal attainment, adverse shocks along non-hit paths can lead to low terminal funding ratios.

On the one hand, consistent with Figure \ref{figEmpiricalDisVaryk}, higher $k$ strengthens time-risk seeking and increases the probability of not reaching the goal before $T$. On the other hand, a larger $k$, or a shift from hyperbolic to exponential discounting, also increases the probability mass in the lowest terminal funding region. This occurs because greater sensitivity to distant delays in goal achievement strengthens the incentive to reach the goal early, which requires taking larger risky exposure. While such exposure improves the chance of early success, it also amplifies downside risk on paths where the goal is not reached. Consequently, conditional on non-attainment before maturity, these unsuccessful paths are more likely to reflect less conservative risk-taking and to end with the lowest terminal funding ratios.



\begin{figure}[H]
   \includegraphics[width=7.5cm, height=6cm]{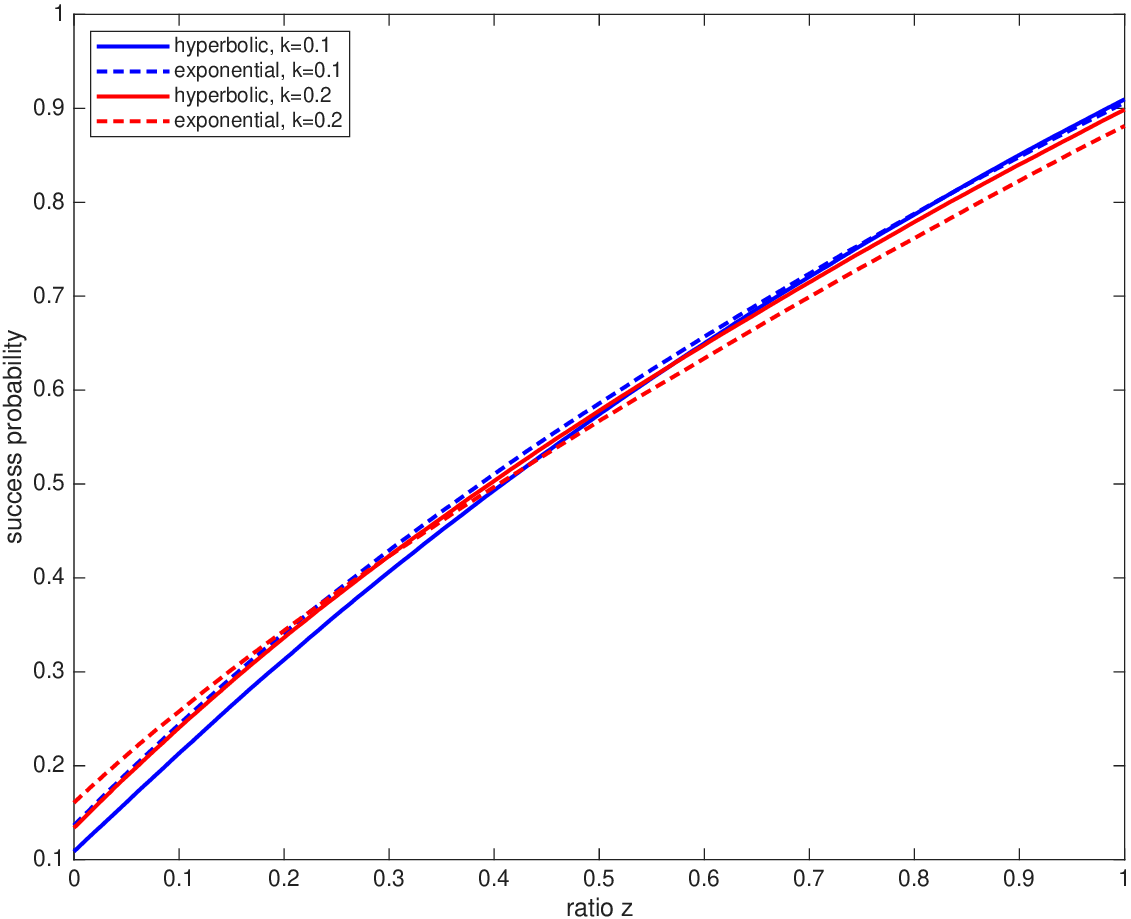}
   \includegraphics[width=7.5cm, height=6cm]{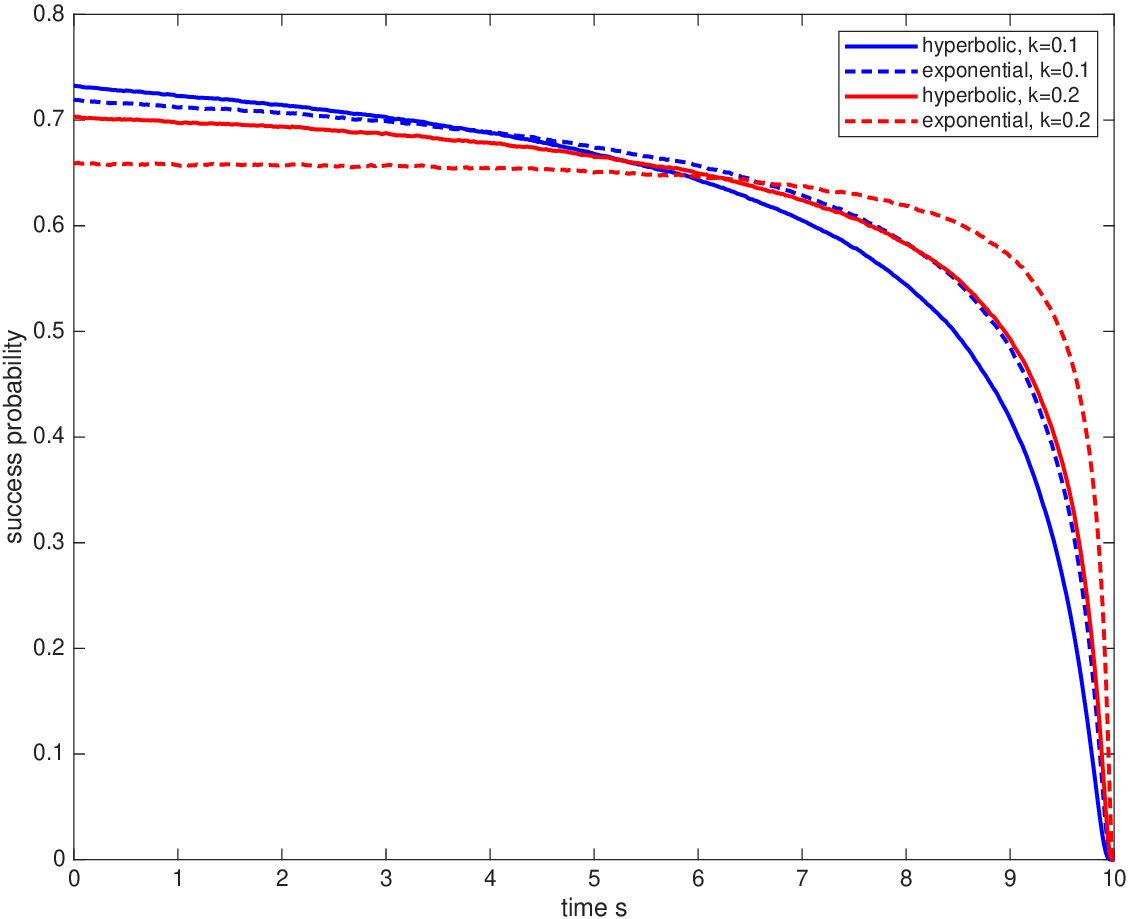}
 \caption{The probability of achieving the goal under different levels of time risk aversion, with $\mu=0.09$, $\sigma=0.35$, $T=10$, $\gamma=2$, and $\alpha=\beta_2=0.01,\beta_1=0$. In the plots varying the ratio $z$, we fix $t=5$; in the plots varying time $t$, we fix $z=0.6$. Empirical probabilities are computed from $5\times 10^5$ simulations, with grid size $5\times 10^{-4}$.}
   \label{figHyperVaryk-successProb}
 \end{figure}

 
We next investigate how the probability of reaching the goal varies with the initial state. 
Figure \ref{figHyperVaryk-successProb} reports the probability of reaching the goal before the deadline for different initial ratio levels, with the initial time fixed at $s=5$, respectively the corresponding probability for different initial times, with the initial ratio fixed at $z=0.6$.
The probability of reaching the goal before the deadline increases with the funding ratio and decreases as time approaches maturity, as expected. 
We observe from the left panel that when $z$ is low, the discounting specifications with exponential or higher $k$ generate a higher success probability. This is because the investor is relatively far from the goal, and the more aggressive risky allocations induced by larger $k$ or exponential discounting (as depicted in the left panel of Figure \ref{figHyperVaryk-optimalpolicy}) provide greater upside potential for reaching the benchmark before the deadline. As $z$ increases, however, the goal becomes more attainable, and the marginal benefit of additional risk-taking decreases. In this region, excessive risky exposure can reduce the probability of success by increasing downside risk and eroding the investor's proximity to the goal. Consequently, the probability advantage diminishes and the curves eventually reverse  for sufficiently high values of $z$. 

The right panel shows that the success probability is strongly time-dependent for a fixed ratio. The relatively flat curves up to around $s=5$ suggest that most successful paths reach the target within the earlier part of the horizon, so shortening the remaining horizon from $10$ to about $5$ has little effect. Over this region, hyperbolic discounting or a lower $k$ generally produces a higher success probability, as the less aggressive policies help preserve accumulated gains while still leaving sufficient time to reach the goal. Once the remaining horizon falls below about $5$, however, fewer paths have enough time to reach the target, and the success probability declines more rapidly. As the deadline approaches, the more impatient behavior induced by exponential discounting, particularly for $k=0.2$, leads to more aggressive risky allocations and eventually the highest success probability among the specifications considered. Nevertheless, the success probability under all specifications converges to zero at maturity.

\subsection{Sensitivity to Market Parameters}
\begin{figure}[H]
    \includegraphics[width=7.5cm, height=6cm]{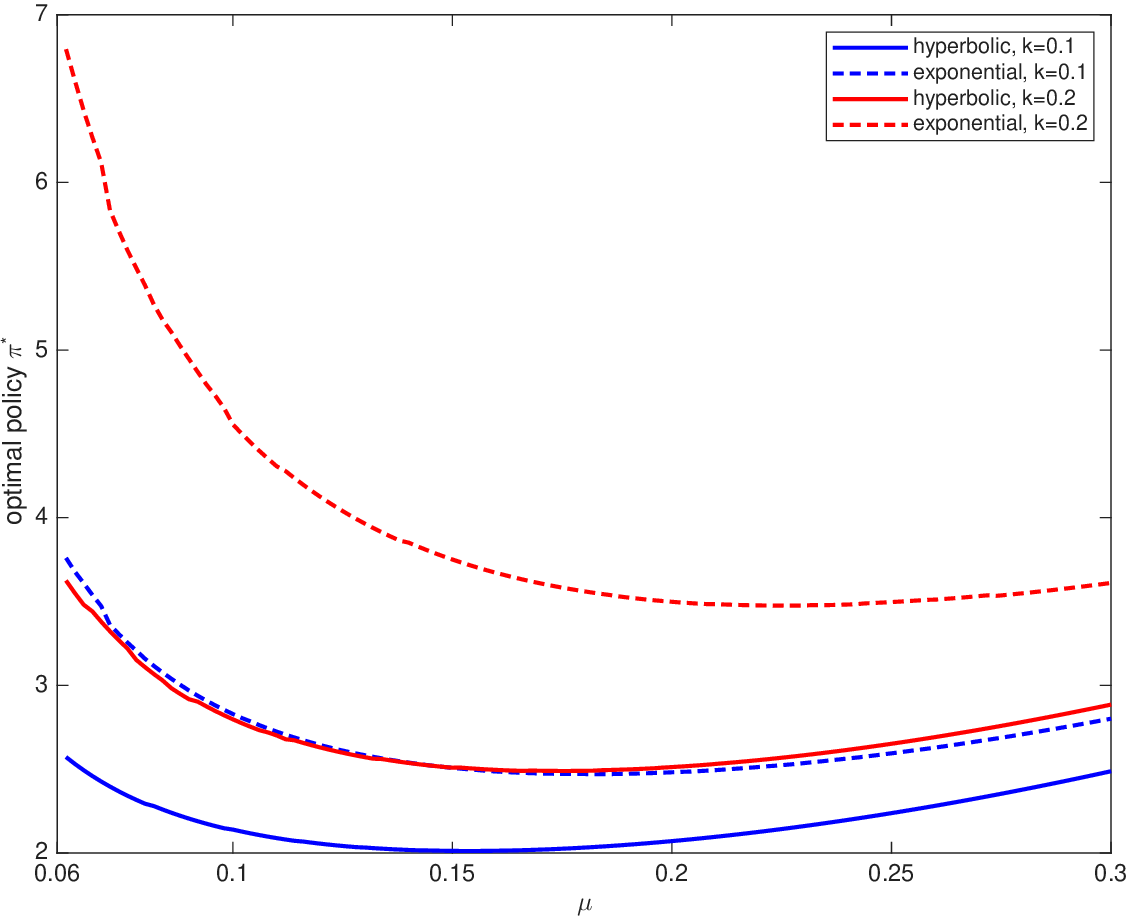}
   \includegraphics[width=7.5cm, height=6cm]{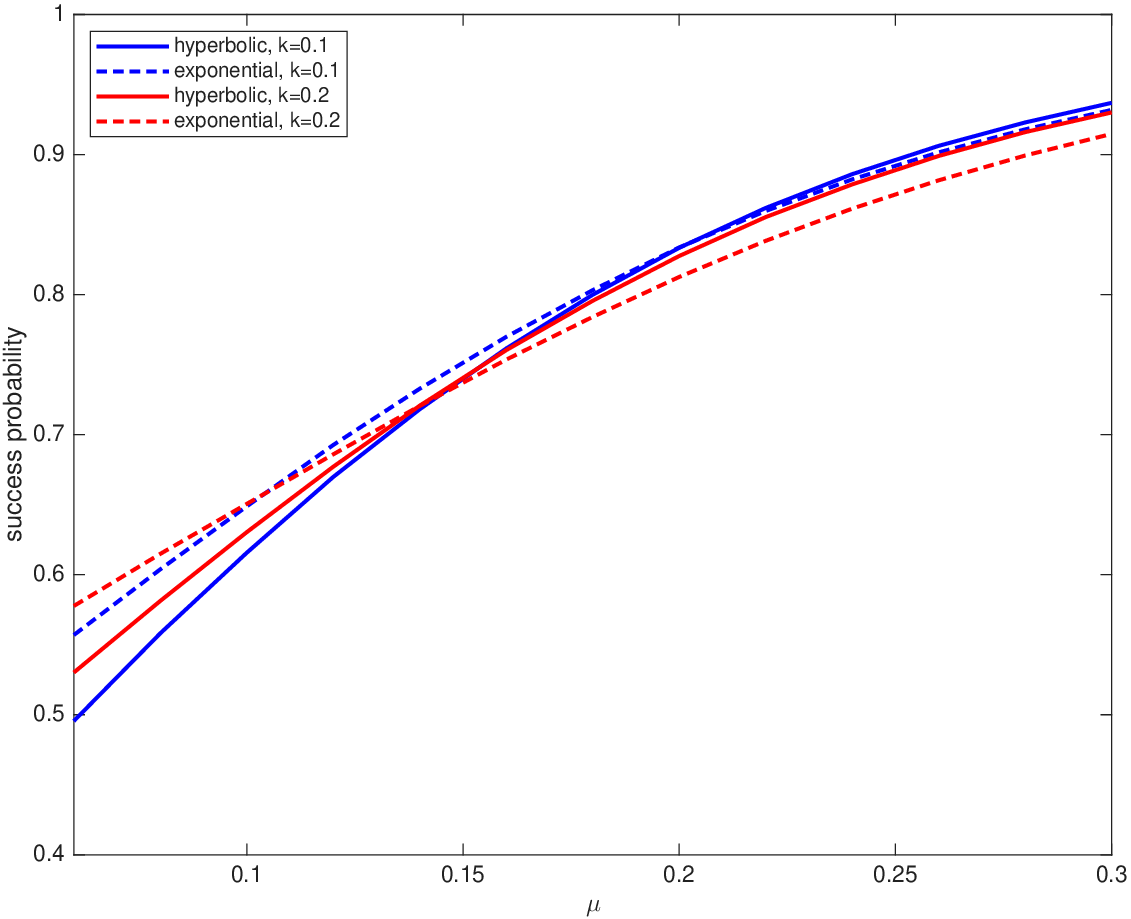} \\ 
   \includegraphics[width=7.5cm, height=6cm]{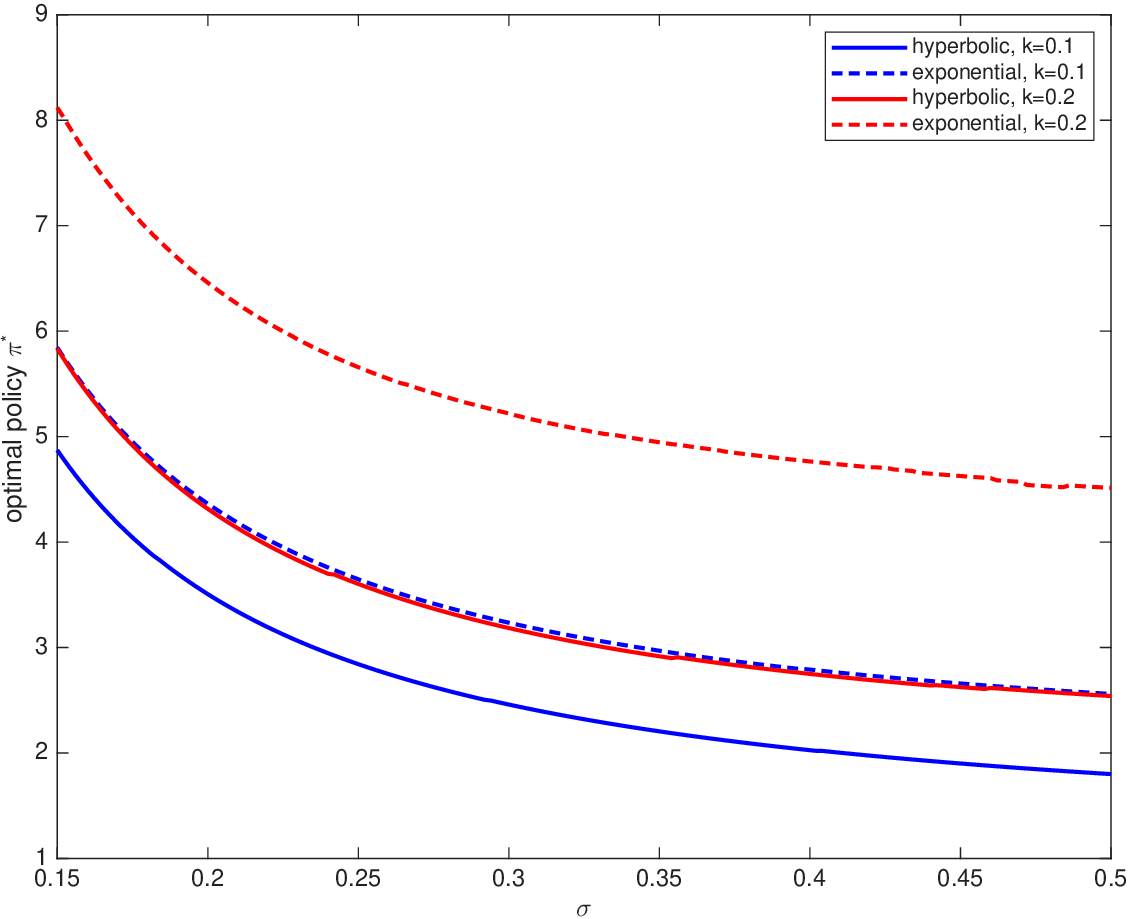}
   \includegraphics[width=7.5cm, height=6cm]{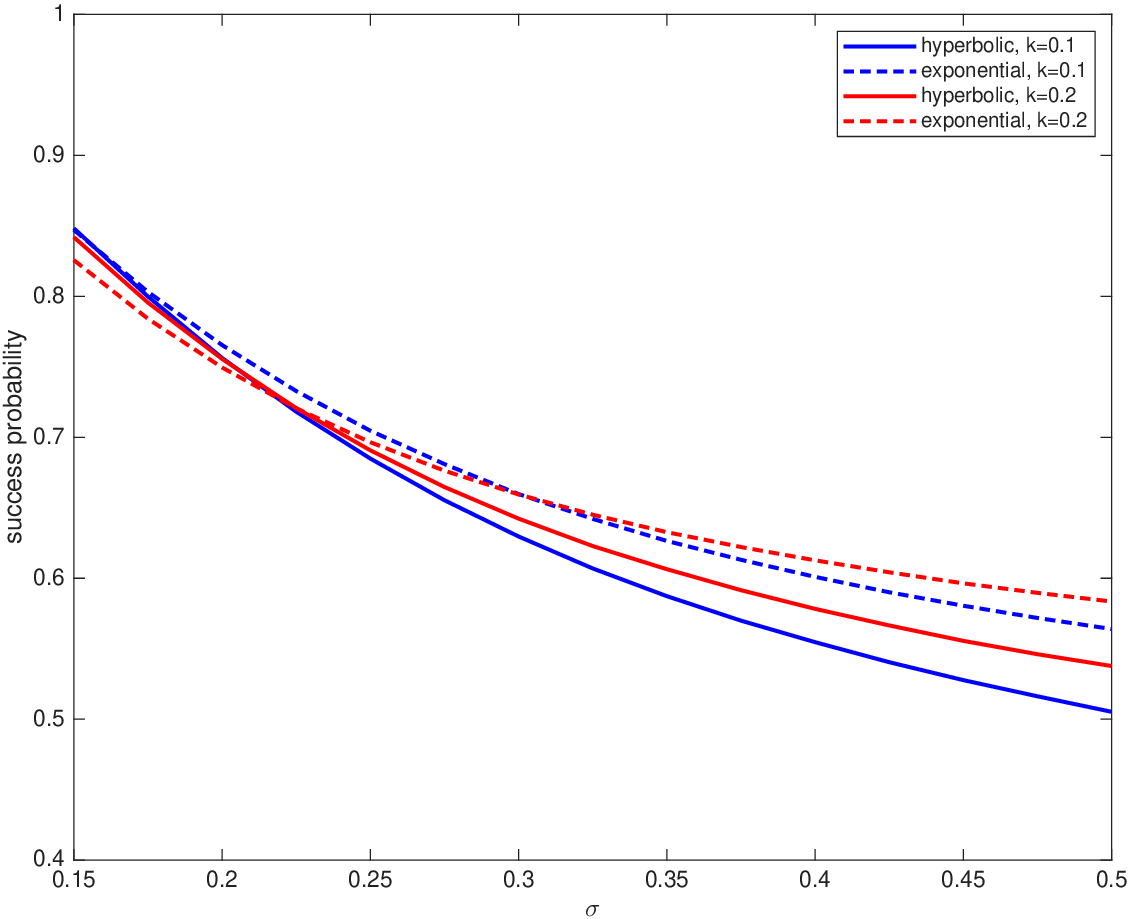}
   \caption{Optimal policy $\pi^*$ and the probability of achieving the goal under different levels of market parameters, with $(s,z)=(5,0.6), T=10$, $\gamma=2$, and $\alpha=\beta_2=0.01,\beta_1=0$. In the plots varying the drift $\mu$, we fix $\sigma=0.35$; in the plots varying volatility $\sigma$, we fix $\mu=0.09$.}
   \label{figVarymuSigma}
 \end{figure}

Next, we investigate in Figure \ref{figVarymuSigma} how market parameters influence the optimal investment under both exponential and hyperbolic preferences.
Consistent with the discussion in previous subsections, for each value of $k$, the optimal policy under exponential discounting exceeds its hyperbolic counterpart over the entire range of $\mu$ or $\sigma$. The gap is particularly pronounced when $\mu$ is small and gradually narrows as $\mu$ increases. Economically, a higher $\mu$ makes the goal easier to attain over a shorter effective horizon. Since exponential and hyperbolic discounting differ mainly in their valuation of more distant goal reaching times,  this reduces the importance of the discounting specification, causing the corresponding optimal policies to move closer together.

The two panels on the left further illustrate the impact of the discounting parameter $k$. An increase in $k$ shifts the optimal policy upward under both specifications, but the effect is substantially stronger under exponential discounting. 
This can be understood from the way $k$ changes the valuation of delayed goal achievement. A higher $k$ assigns less weight to later goal reaching times and thereby strengthens the incentive to reach the goal sooner through a more aggressive policy. Under exponential discounting, this additional penalty induced by a higher $k$ persists over a broader range of future horizons, whereas under hyperbolic discounting it is concentrated mainly on short delays and diminishes rapidly over time. Accordingly, the optimal policy is more sensitive to $k$ under exponential discounting.

We observe that the optimal allocation to the stock is first decreasing and then increasing in its drift coefficient $\mu$. Intuitively, when the drift is low, the stock return is unfavorable, and the investor's only chance of beating the stochastic goal is to gamble aggressively on potential upside. As the drift improves slightly, the stock becomes less unfavorable, so the marginal value of extreme risk-taking diminishes, and the optimal exposure decreases. However, once the drift is sufficiently high, the stock itself becomes attractive for achieving the goal, and increasing the exposure raises the chance of success. This initial decrease followed by an increase is also consistent with the pattern in Example \ref{ex2}.

In contrast, the optimal allocation decreases with volatility. A higher $\sigma$ increases the dispersion and downside risk of the wealth process without improving the expected return, thereby raising the probability of falling short of the goal. The investor therefore reduces exposure to the risky asset. This effect is most pronounced at low volatility levels and gradually weakens as $\sigma$ increases. 
When $\sigma$ is low, the goal remains attainable with a high probability. A marginal increase in volatility therefore substantially raises the likelihood of falling short of the goal, inducing the investor to reduce risky exposure considerably. When $\sigma$ is already high, however, the probability of reaching the goal has already been substantially reduced by market uncertainty. A further increase in volatility lowers the success probability by a relatively smaller amount because the investor is already operating in an environment where achieving the goal is inherently difficult. Consequently, there is less incentive to reduce risky exposure further, resulting in a flatter policy curve.

The probability of reaching the goal increases as the investment opportunity improves, whether through a higher expected return $\mu$ or a lower volatility $\sigma$. When $\mu$ is low or $\sigma$ is high, exponential discounting generally delivers a higher success probability than hyperbolic discounting for a given $k$, consistent with its more aggressive optimal policy: when expected returns are low, greater risk-taking provides additional upside potential that helps compensate for the otherwise unfavorable investment environment.
As $\mu$ increases or $\sigma$ decreases, however, the success probabilities under all discounting specifications become very close and may reverse in order. Once the investment opportunity is sufficiently favorable, additional risk-taking is less necessary and may instead increase downside risk. This also helps explain why a higher $k$ raises the success probability at low $\mu$ but lowers it at higher $\mu$. Greater impatience induces more aggressive investment, which is beneficial when the goal is otherwise difficult to attain, but may become excessive when the expected return is already high.



\subsection{Sensitivity to Goal Parameters}

\begin{figure}[H]
\includegraphics[width=7.5cm, height=6cm]{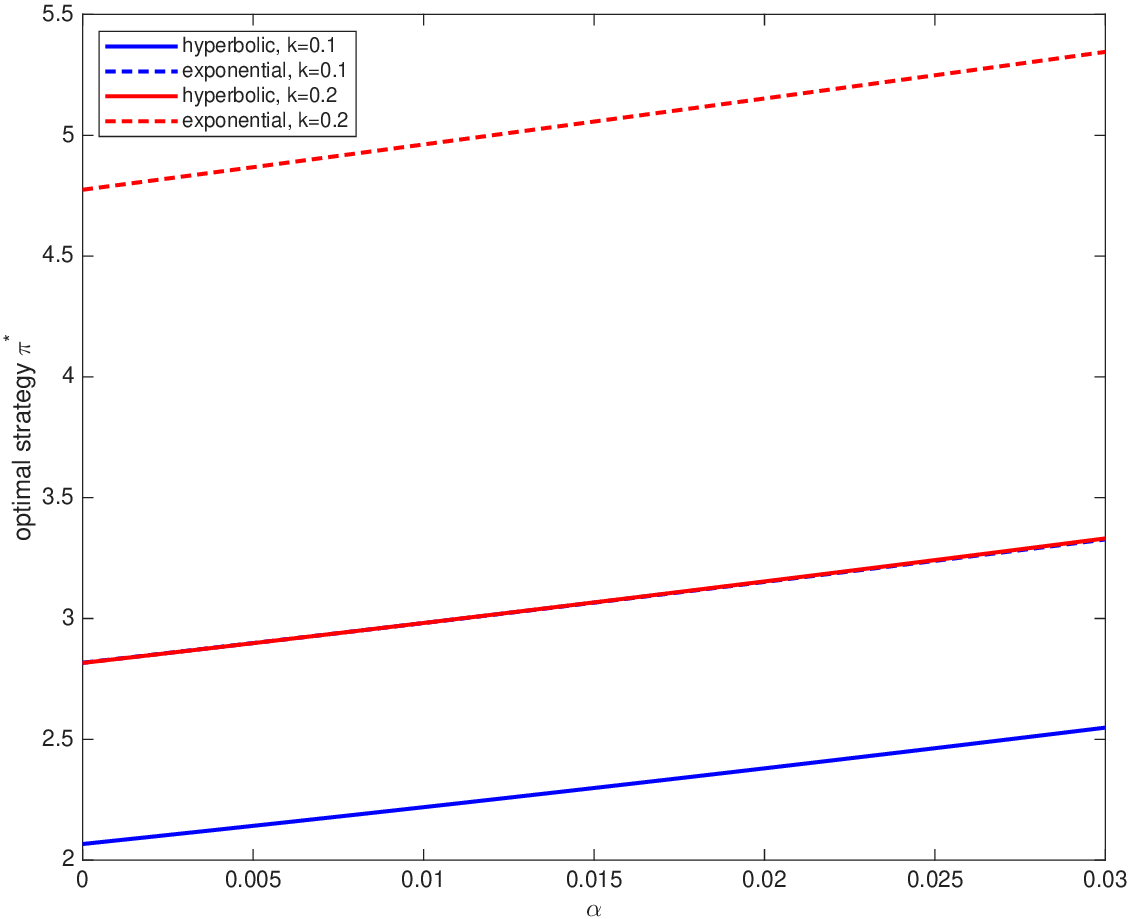}
   \includegraphics[width=7.5cm, height=6cm]{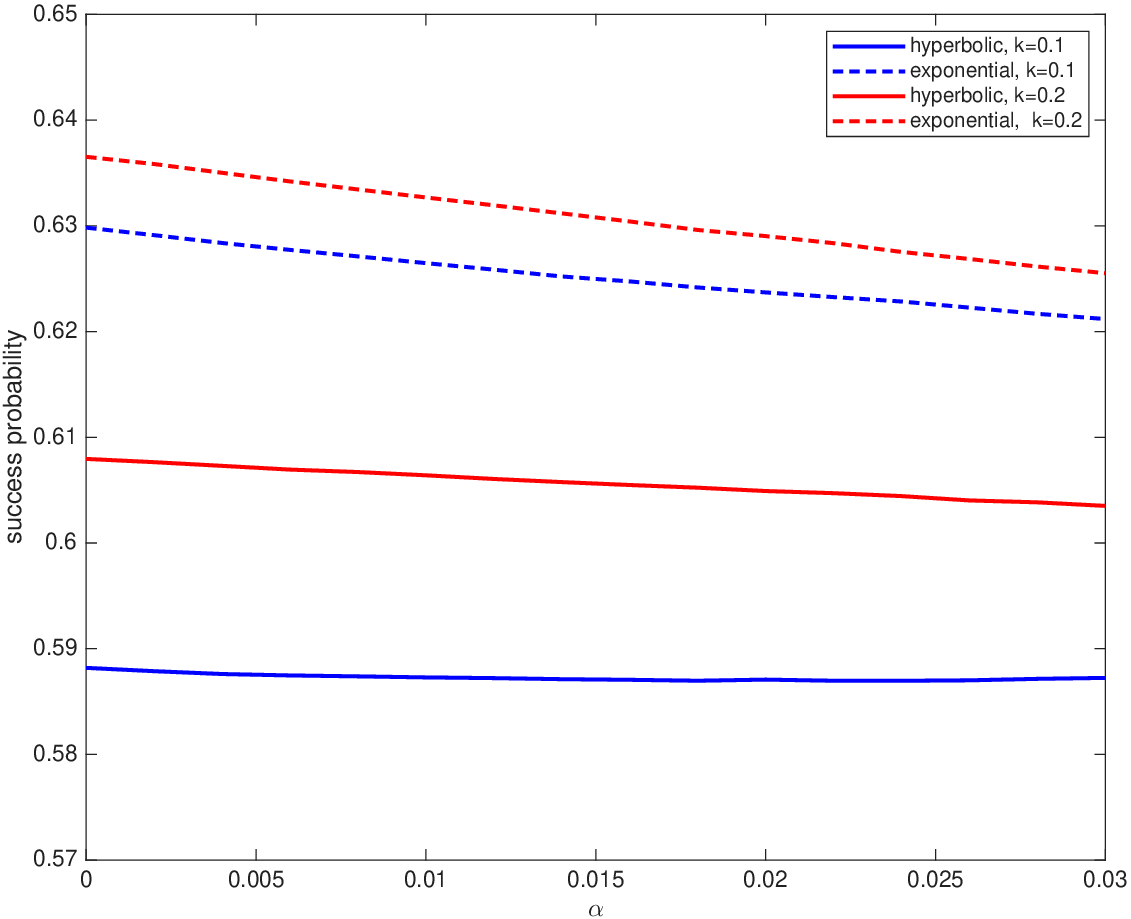} \\
    \includegraphics[width=7.5cm, height=6cm]{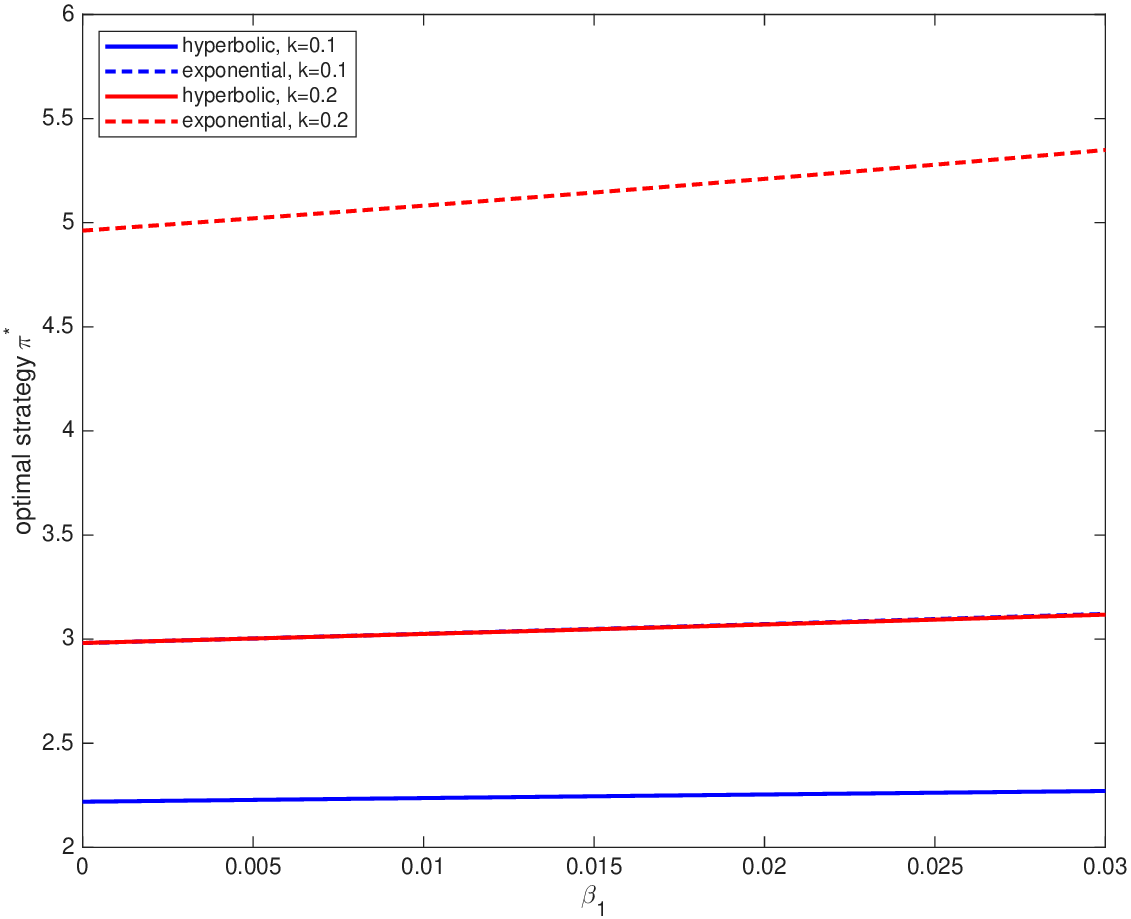}
   \includegraphics[width=7.5cm, height=6cm]{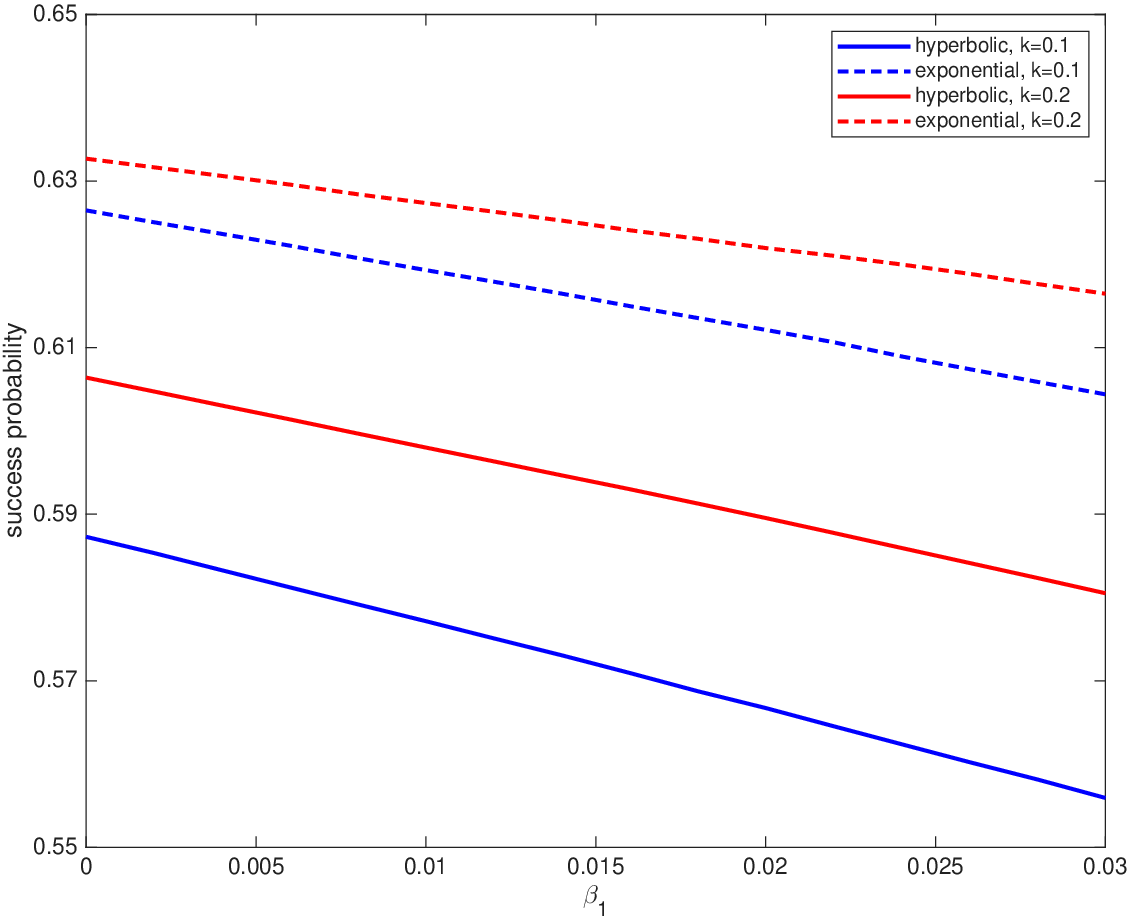} 
   \caption{Optimal policy $\pi^*$ and the probability of achieving the goal under different levels of goal parameters, with $(s,z)=(5,0.6), T=10$, $\gamma=2$, $\mu=0.09$, $\sigma=0.35$ and $\beta_2=0.01$. In the plots varying the goal drift $\alpha$, we fix $\beta_1=0$; in the plots varying goal volatility $\beta_1$, we fix $\alpha=0.01$.}
   \label{figVaryStochasticB}
 \end{figure}

To investigate the impact of a stochastic benchmark on optimal investment, we vary the associated parameters $\alpha$ and $\beta_1$ in Figure \ref{figVaryStochasticB}, and compute the optimal investment at time $s=5$ with a ratio of $z=0.6$. On the left-hand side of Figure \ref{figVaryStochasticB}, we set $\beta_1=0$ and $\beta_2=0.01$, and vary $\alpha$. On the right-hand side of Figure \ref{figVaryStochasticB}, we set $\alpha=\beta_2=0.01$, and vary only $\beta_1$, which affects both the goal volatility and its dependence with investment returns.




As shown in the upper-left panel, the optimal policy increases with $\alpha$ across all specifications. A faster-growing goal makes the target more difficult to reach, and the investor responds by increasing risky exposure in an attempt to offset this adverse effect. The ranking of the optimal policies is consistent with previous results and remains stable across $\alpha$: a larger $k$ leads to a higher risky allocation, while exponential
discounting induces a more aggressive policy than hyperbolic discounting for a given value of $k$. Note that in the left panels, the curves for exponential discounting with \(k=0.1\) and hyperbolic discounting with \(k=0.2\) lie very close to each other and are therefore difficult to distinguish visually; both curves are included.

Correspondingly, the upper-right panel shows that the success probability generally declines as $\alpha$ increases, although the change is very small under hyperbolic discounting with $k=0.1$. Thus, the increase in risky exposure is not sufficient to fully offset the adverse effect of a faster-growing goal. For a fixed value of $k$, exponential discounting yields a higher success probability than hyperbolic discounting, consistent with its more aggressive risky allocation.
However, the difference in success probabilities gradually narrows as $\alpha$ increases. Hence, exponential discounting continues to deliver a higher success probability over the range considered, but its advantage becomes smaller as the
goal becomes more difficult to reach.

The lower-left panel shows that, at state $(s,z)=(5,0.6)$, the optimal policy $\pi^*$ increases slightly with $\beta_1$. A larger $\beta_1$ increases the volatility of the goal and makes it respond more strongly to the same Brownian shock $W$ that drives the stock. Since wealth and the goal are then more exposed to the same source of risk, holding more stock allows wealth to move more closely with fluctuations in the goal and helps reduce relative fluctuations in the funding ratio. This provides a partial hedging benefit against goal uncertainty and, at the state considered here, leads to a slightly higher optimal risky allocation. As shown later, the overall effect is state-dependent and the ordering with respect to $\beta_1$ can reverse when the investor is farther from the goal or has less time remaining. Moreover, the sensitivity of the optimal policy to $\beta_1$ appears to increase with impatience, as the slope of $\pi^*$ with respect to $\beta_1$ is steeper for larger $k$ and under exponential discounting. This suggests that more impatient investors respond more strongly to increases in the common-shock component of goal risk, raising risky exposure to better track the goal and preserve the prospect of earlier goal attainment.

The lower-right panel shows that the success probability decreases with $\beta_1$ under all specifications. This negative effect reflects that a more volatile goal makes the goal harder to achieve. Although a higher $\beta_1$ strengthens the correlation between stock returns and goal movements and provides some hedging benefit, the increase in goal uncertainty dominates. The additional volatility increases the dispersion of the goal relative to the investor's wealth, reducing the probability of reaching the goal before the deadline. The decrease is smaller under exponential discounting than under hyperbolic discounting. This is consistent with the optimal policy results: exponential discounting induces a more aggressive risky allocation, which allows the investor to better offset the common-shock component of goal risk and thereby reduces the sensitivity of the success probability to $\beta_1$.

\begin{figure}[H]
   \includegraphics[width=7.5cm, height=6cm]{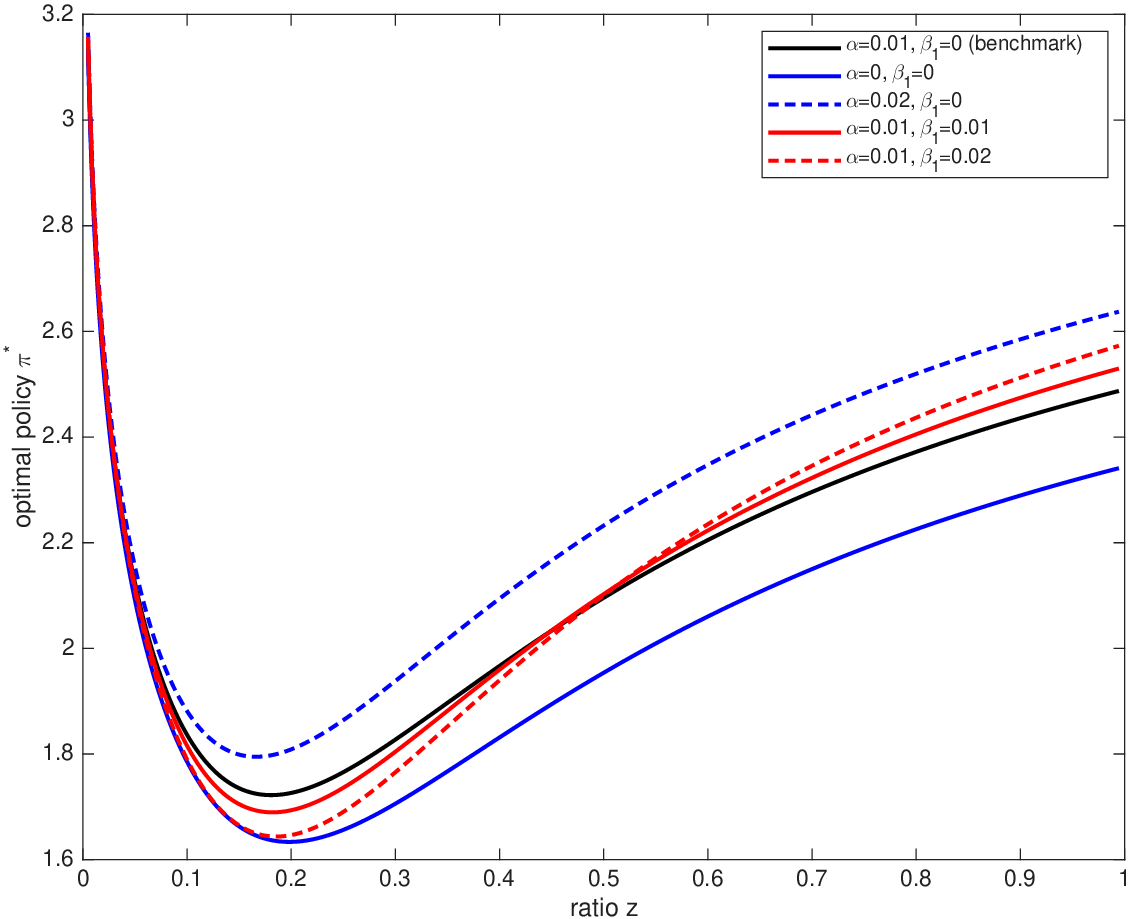}
   \includegraphics[width=7.5cm, height=6cm]{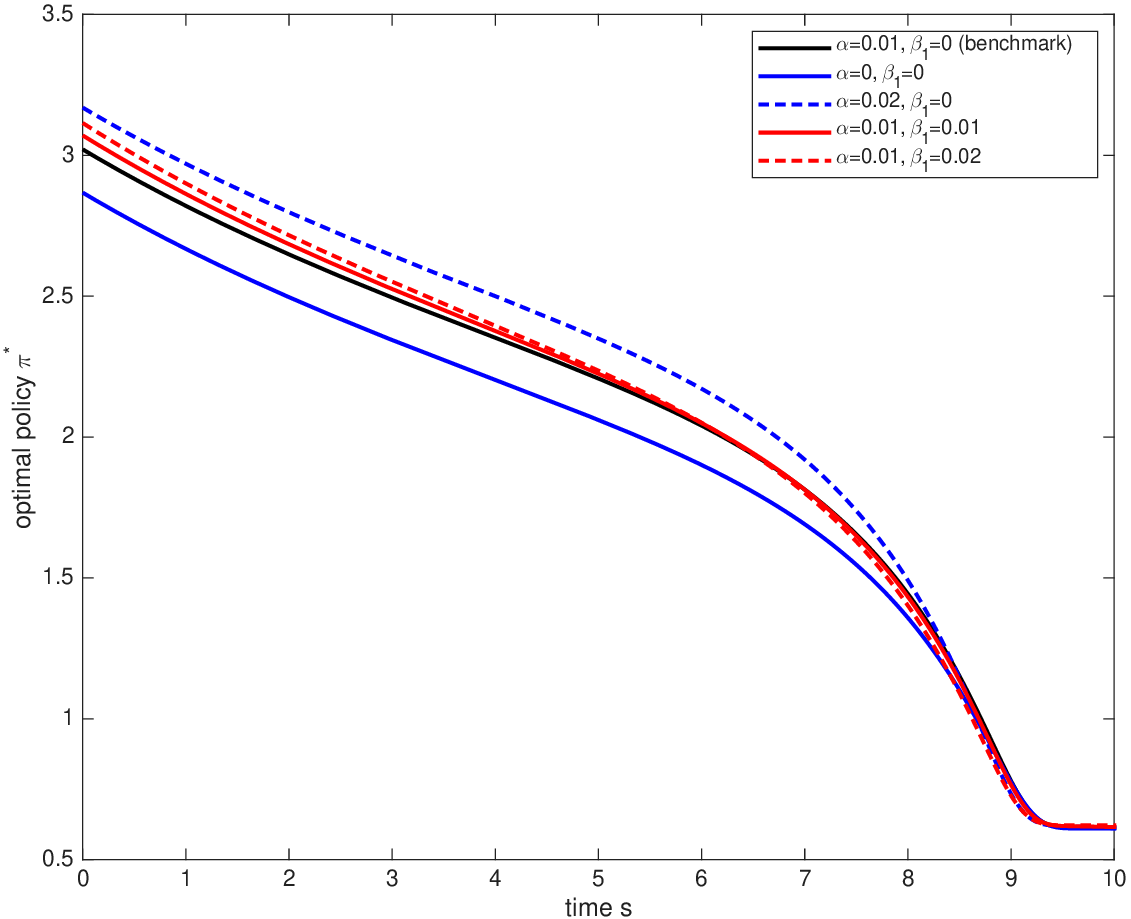}
\caption{Optimal policy $\pi^*$ as a function of the funding ratio and time under different goal parameters, with $\mu=0.09$, $\sigma=0.35$, $T=10$, $\gamma=2$, and $\beta_2=0.01$.
The benchmark case is $(\alpha,\beta_1)=(0.01,0)$, with alternative specifications
$(0,0)$, $(0.02,0)$, $(0.01,0.01)$, and $(0.01,0.02)$.
In the left panel, we fix $t=5$ and vary $z$; in the right panel, we fix $z=0.6$ and vary $t$.}
   \label{fig:CompareAlphaBeta1}
 \end{figure}

We next examine how the effects of the goal parameters vary with the funding ratio and time. Figure \ref{fig:CompareAlphaBeta1} plots the optimal policy against $z$ and $s$ under different goal parameters. A larger $\alpha$ implies a faster-growing goal and therefore makes it more difficult for the funding ratio to reach the target. Consistently, $\pi^*$ is increasing in $\alpha$ over the range considered. This effect is especially pronounced for moderate and high funding ratios in the left panel and over the earlier part of the investment horizon in the right panel, where the goal-achievement component remains relatively important compared with terminal utility.

The effect of $\beta_1$ is more state-dependent. A larger $\beta_1$ is associated with a lower optimal risky allocation when the investor is farther from the goal or has less time remaining, but with a higher allocation when the goal is easier to reach. Intuitively, a larger $\beta_1$ makes the goal respond more strongly to the same market shocks as the stock, reducing the relative movement between wealth and the goal for a given stock allocation. When the goal is relatively attainable, reaching the target becomes an immediate concern, and sufficient relative upside is valuable for pushing the funding ratio above one. A larger stock exposure can preserve sufficient upside potential when $\beta_1$ is higher, leading to a higher $\pi^*$. When the goal is difficult to attain, however, early goal achievement becomes less relevant and terminal utility plays a larger role. The investor then focuses more on the terminal funding ratio, and has less incentive to increase risk simply to generate the relative movements needed to reach the target. This is consistent with a lower $\pi^*$ for larger $\beta_1$.


\subsection{Interplay between Time and Outcome Preferences}
We next investigate the interaction between time and outcome preferences through the discounting parameter $k$ and the terminal risk-aversion parameter $\gamma$. The former governs the investor's sensitivity to delaying goal attainment, while the latter determines the investor's attitude towards terminal wealth risk.

\begin{figure}[H]
   \includegraphics[width=7.5cm, height=6cm]{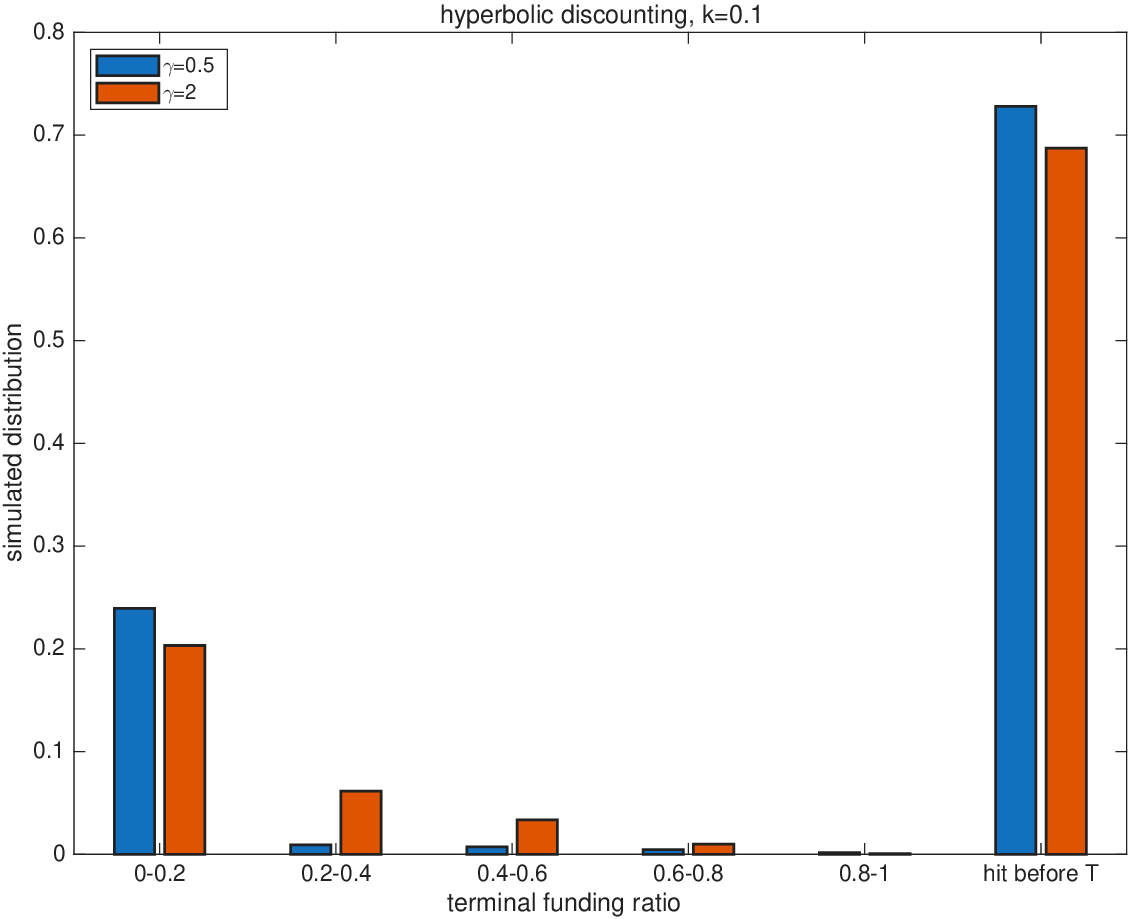} 
   \includegraphics[width=7.5cm, height=6cm]{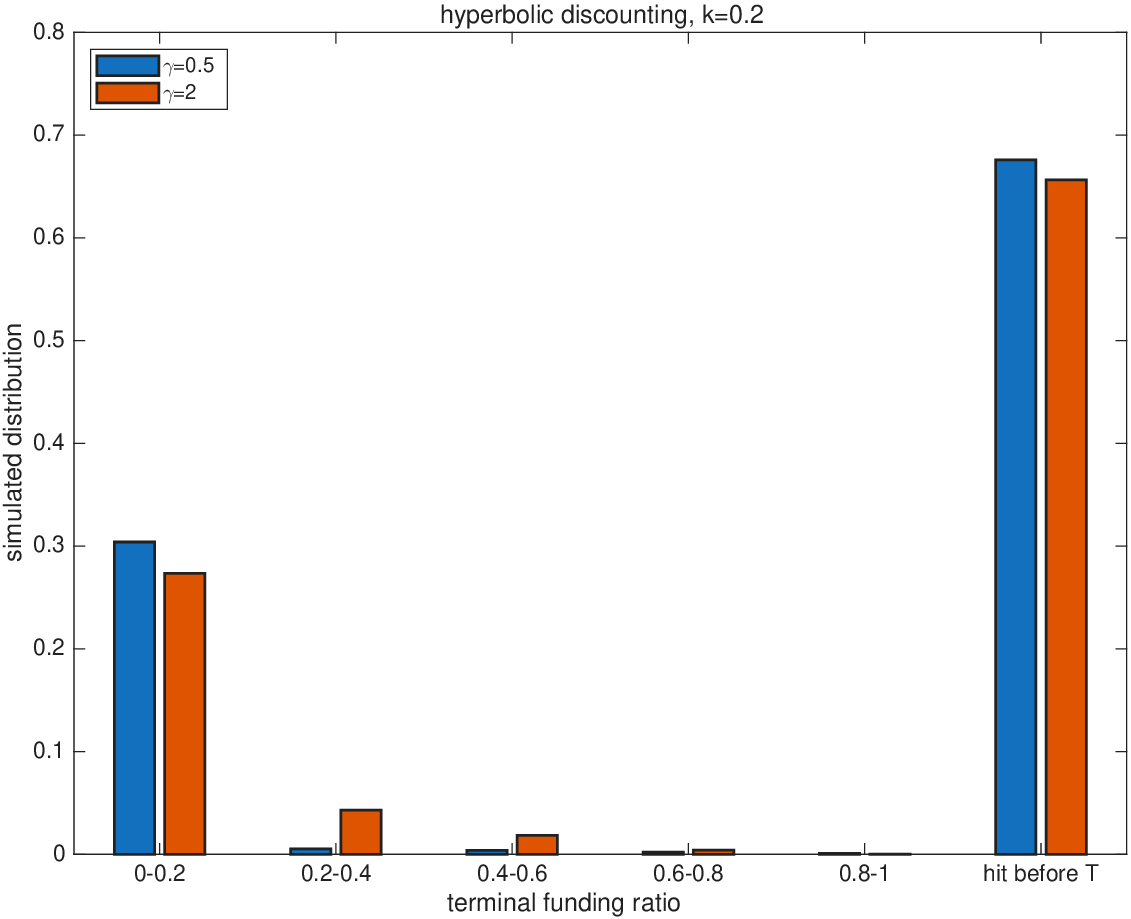}
  \caption{Interplay between impatience and terminal risk aversion in the terminal ratio distribution.  Consider $(s,z)=(0,0.6), T=10$, $\gamma=2$, $\mu=0.09$, $\sigma=0.35$ and $\alpha=\beta_2=0.01,\beta_1=0$.}
   \label{figEmpiricalDisVaryGamma-TerminalRatio}
 \end{figure}
 
Figure \ref{figEmpiricalDisVaryGamma-TerminalRatio} illustrates the simulated distributions of the terminal funding ratio under hyperbolic discounting for different values of the terminal risk-aversion parameter $\gamma$. For a fixed discounting parameter k, increasing $\gamma$ shifts probability mass away from the two extreme outcomes. Specifically, both the probability of reaching the goal before the deadline and the probability of ending with a very low terminal ratio $Z_T\in[0,0.2]$ decrease, while the probabilities of intermediate terminal ratios increase. This reflects a more risk-averse investment behavior: a larger terminal risk-aversion coefficient discourages excessively aggressive positions that may either achieve the goal or lead to substantial losses, resulting in a more conservative investment outcome with greater protection against extreme losses.

Comparing the two panels further reveals the interaction between impatience and terminal risk aversion. When $k$ increases from $0.1$ to $0.2$, the differences between the distributions corresponding to $\gamma=0.5$ and $\gamma=2$ become smaller. This suggests that the influence of impatience dominates that of terminal risk aversion. A larger discounting parameter increases the urgency of reaching the goal at earlier periods, making the incentive generated by impatience more dominant relative to terminal risk aversion. As a result, changes in $\gamma$ have a smaller impact on the optimal strategy, leading to more similar terminal wealth distributions. The two parameters govern different dimensions of risk. The parameter $\gamma$ captures aversion to uncertainty in terminal wealth, whereas $k$ determines the investor's sensitivity to delaying goal attainment. As $k$ increases, the marginal penalty for postponing the attainment of the goal becomes larger, providing a stronger incentive to reach the goal at earlier periods. Consequently, the investor is willing to accept greater uncertainty in terminal wealth in exchange for increasing the likelihood of earlier goal attainment. In this sense, a more time-risk-seeking investor behaves as if they were less outcome-risk averse, even though their intrinsic terminal risk aversion, measured by $\gamma$, remains unchanged.

\subsection{Sensitivity to the Stochastic Factor}


Consider an Ornstein--Uhlenbeck stochastic-volatility model driven by a common Brownian motion,
\[
dY_t=a(b-Y_t)\,dt+\sigma_Y\,dW_t,
\qquad
dS_t
= S_t \left( 
\left(\mu_{\mathrm{BS}}+\xi(Y_t-b)\right)dt
+\sigma_{\mathrm{BS}}\,dW_t \right),
\]
where $Y_t$ is a mean-reverting stochastic factor, and $\xi$ measures the sensitivity of the risky asset's expected return to deviations of the factor from its long-run mean. In particular, $\mu(y)=\mu_{\mathrm{BS}}+\xi(y-b)$, so that $\mu(b)=\mu_{\mathrm{BS}}$. Hence, when the stochastic factor is at its long-run mean, the drift and volatility of the risky asset coincide with those of the BS benchmark.

Note that the coefficients of the OU process are globally Lipschitz, while the corresponding market price of risk,
\(\kappa(y)
=
\frac{\mu_{\mathrm{BS}}-r+\xi(y-b)}
{\sigma_{\mathrm{BS}}},\)
is unbounded in $y$. Instead of imposing boundedness of the market price of risk, we impose the Novikov condition
\(\E\left[
\exp\left(
\frac{1}{2}\int_0^T\kappa(Y_t)^2\,dt
\right)
\right]<\infty.\)
This condition ensures that the associated stochastic exponential is a true martingale.
Writing
\(\kappa(Y_t)
=
\kappa_{\mathrm{BS}}+\lambda(Y_t-b),
\kappa_{\mathrm{BS}}
=
\frac{\mu_{\mathrm{BS}}-r}{\sigma_{\mathrm{BS}}},
\lambda=\frac{\xi}{\sigma_{\mathrm{BS}}},\)
a convenient sufficient condition is
\(T\lambda^2\frac{\sigma_Y^2}{2a}<1,\)
which is satisfied by all parameter values considered below.

We compare two values of $\xi$. The left panel considers $\xi=0.4$, under which the risky asset's expected return is less sensitive to the stochastic factor, whereas the right panel considers $\xi=0.8$, corresponding to greater sensitivity.
For numerical implementation, we truncate the state space to the bounded domain
\((s,y,z)\in[0,T]\times[y_{\min},y_{\max}]\times(0,1],\)
where \(y_{\min}=-0.04, y_{\max}=0.24.\)
These are artificial numerical boundaries rather than restrictions on the support of the OU process. Under $a=2$ and $\sigma_Y=0.03$, the stochastic factor has a small long-run dispersion around its mean. We therefore choose a sufficiently wide computational domain and verify by Monte Carlo simulation that the probability of leaving this domain over the investment horizon is negligible. 

\begin{figure}[H]
   \includegraphics[width=7.5cm, height=6cm]{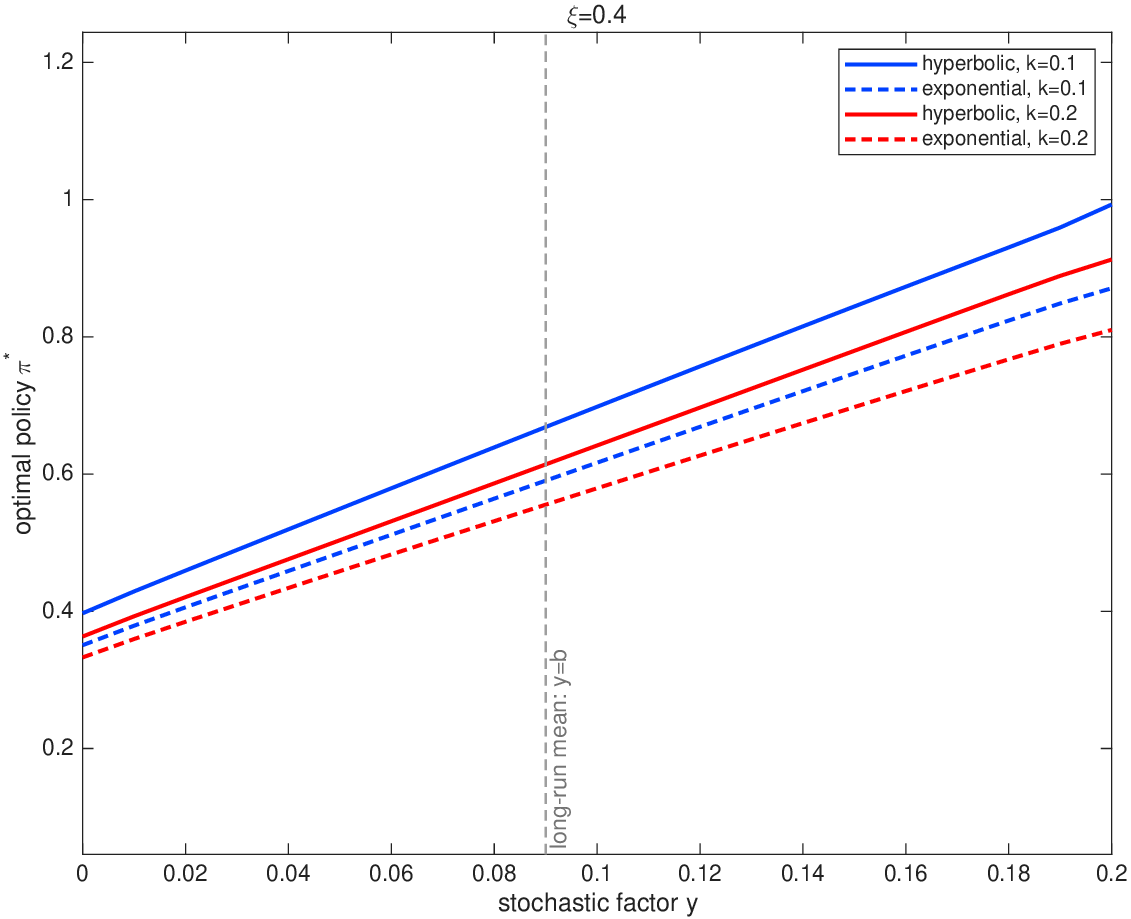} 
   \includegraphics[width=7.5cm, height=6cm]{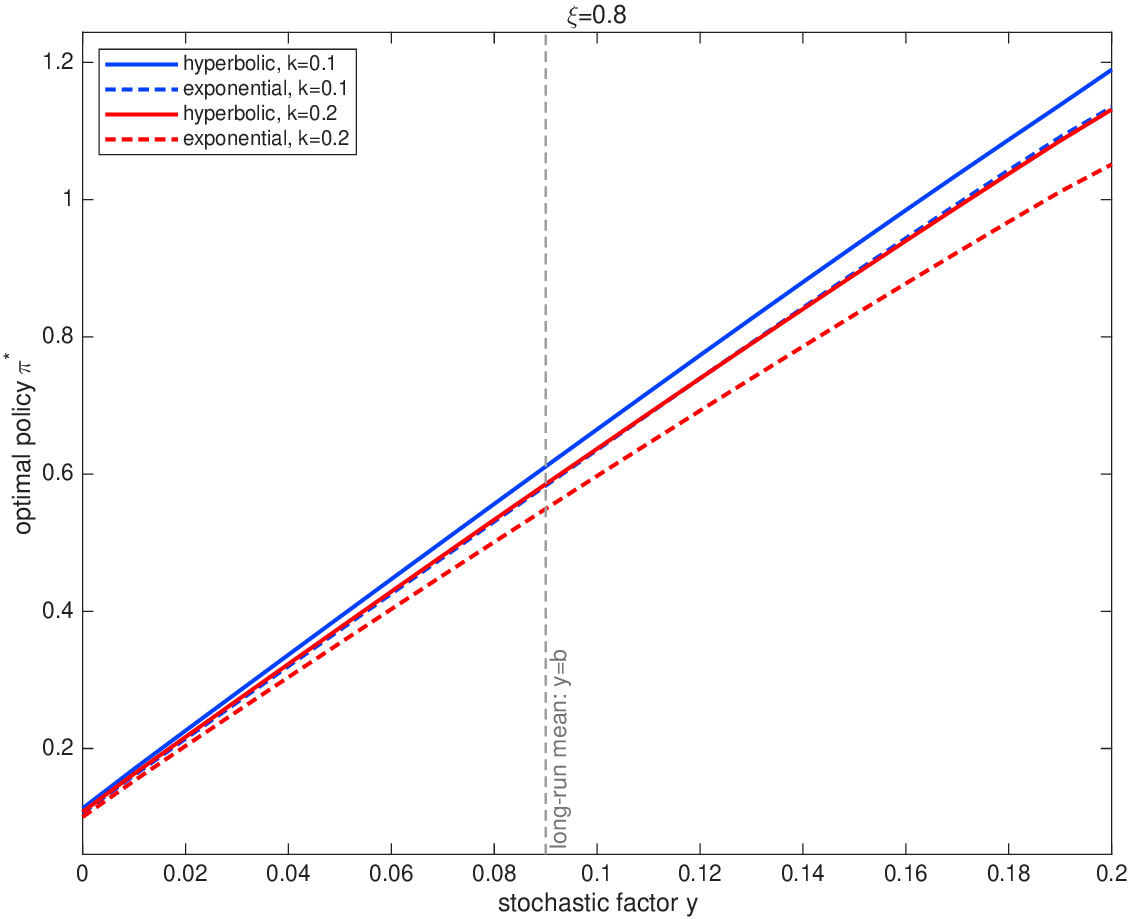}
 \caption{Optimal policy $\pi^*$ as a function of the stochastic factor $y$ under two values of $\xi$, $\xi=0.4$ and $\xi=0.8$. The policy is evaluated at $(s,z)=(5,0.6)$ with $T=10$, $\gamma=2$, $\alpha=\beta_2=0.01$, $\beta_1=0$, $a=2$, $\sigma_Y=0.03$, $\mu_{\mathrm{BS}}=b=0.09$, and $\sigma_{\mathrm{BS}}=0.35$. The vertical dashed line marks the long-run mean $y=b$, at which the risky-asset drift coincides with the BS benchmark.}
   \label{figVaryXi-OUReturnStochasticFactor}
 \end{figure}

Figure~\ref{figVaryXi-OUReturnStochasticFactor} shows that the optimal risky allocation increases with the stochastic factor $y$ for both values of $\xi$. This follows naturally from
\(\mu(y)=\mu_{\mathrm{BS}}+\xi(y-b).\)
A higher value of $y$ increases the current expected return of the risky asset and therefore leads to a larger risky allocation.
The parameter $\xi$ determines the sensitivity of the expected return to the stochastic factor. When $\xi=0.4$, changes in $y$ have a relatively moderate effect on the expected return, and the optimal policy changes more gradually with $y$. When $\xi=0.8$, the same change in $y$ produces a larger change in the expected return, resulting in a steeper policy curve: the investor reduces risky investment more strongly when $y<b$ and increases it more strongly when $y>b$.
At the long-run mean $y=b$, both specifications have the same current drift, $\mu(b)=\mu_{\mathrm{BS}}$. However, a given change in $Y$ has a larger effect on future expected returns when $\xi$ is larger, so the optimal policies need not coincide.


Across the four discounting specifications, a smaller $k$ generally leads to a larger risky allocation, and, for a given $k$, hyperbolic discounting yields a higher allocation than exponential discounting. Thus, in the present calibration, stronger impatience is associated with more conservative current investment. 
This ordering contrasts with that in the BS model, where a larger $k$ and exponential discounting generally lead to a higher risky allocation at $(s,z)=(5,0.6)$. In the BS model, a negative return shock reduces wealth but leaves future investment opportunities unchanged, so a stronger preference for early goal achievement encourages greater current risk taking. In the stochastic-factor model, however, the same negative shock reduces both wealth and $Y$, thereby moving the investor farther from the goal while also lowering subsequent expected returns. A lower risky allocation cannot prevent the deterioration in $Y$, but it limits the simultaneous loss in wealth. This protection is more valuable for a more impatient investor, since such a joint setback can substantially delay subsequent goal achievement. The resulting precautionary effect can therefore outweigh the incentive to take more risk to reach the goal sooner, reversing the ordering observed in the BS model.

Moreover, the differences across specifications widen as $y$ increases. When $y$ is low, the risky asset is less effective in advancing toward the goal, so differences in time preferences have only a limited effect on the portfolio choice. As $y$ rises, the investment opportunity improves and these differences become more pronounced. The curves are also more widely separated for $\xi=0.4$ than for $\xi=0.8$: with a larger $\xi$, the stronger direct effect of $y$ on expected returns makes the policies respond more similarly to the current factor state.

\section{Conclusion}\label{sec:Conclusion}
We study an investor who trades until she either reaches a financial goal or an exogenous deadline arrives in a continuous-time market where a stochastic factor drives the dynamics of stock prices and the financial goals.  
This framework leads to a new family of stochastic control problems. 
We, first, show regularity properties of the value function, derive Bellman's principle of optimality, and characterise the solution via an associated Hamilton-Jacobi-Bellman equation. 
We, then, provide a complete characterization of the class of discount functions leading to an analytical value function for the special case without exogenous deadline, constant goal, and no stochastic factor. 
Finally, we develop an algorithm to obtain numerical solutions. 
Optimal strategies for goal reaching problems exhibit features that are qualitatively different from the expected utility from the terminal wealth paradigm. For example, we find that optimal investment policies can be decreasing in the drift of risky assets, and that \href{}{}they need not converge to investing everything in the risk-free asset even as volatility diverges to infinity.

 \section*{Acknowledgments}
The authors are grateful to Luca De Gennaro Aquino, Giorgio Ferrari, Zhibin Liang, Zuo Quan Xu, Harry Zheng, and Xun Yu Zhou for their helpful comments and suggestions.
This work was presented at Imperial College London, the Workshop on Stochastic Control, Financial Technology, and Machine Learning at Hong Kong Polytechnic University, the First INFORMS Conference on Financial Engineering, the 7th Berlin Workshop on Mathematical Finance for Young Researchers, and the 13th World Congress of the Bachelier Finance Society.
The authors thank the participants for their comments.
This research was supported by the Gillmore Centre for Financial Technology. 
Author order is alphabetical. 
All authors are co-first authors of this paper.

\section{Appendix}
\renewcommand{\thesubsection}{\Alph{subsection}}

\subsection{Proof of Proposition \ref{prop:ContinuityofV}}
Clearly, $V$ is bounded because $\rho$ and $u$ are bounded. 
Continuity in the case $T<+\infty$ follows by Theorem 2.1, Section V, \cite{fleming2006controlled}. 
The proof of continuity for $T=+\infty$ is obtained by adapting the finite-horizon argument in \cite{fleming2006controlled}, a detailed proof is provided in \cite[Proposition 4.1]{warwick189223Yuwei}. 
\qed

\subsection{Proof of Proposition \ref{Prop:DPP}}
For simplicity, we consider the case of $T< \infty$, with the extension to $T= \infty$ being straightforward.
Consider any arbitrary $\pi\in \mathcal{A}_{s,T}$ and stopping time $s\leq \alpha\leq \widehat{\tau}^{\pi}$, since $J(\alpha,Y_{\alpha},Z_{\alpha}^{\pi})\leq V(\alpha,Y_{\alpha},Z_{\alpha}^{\pi})$, we have by the tower property of conditional expectations 
\begin{align}\label{Eqn:CanProveViscosity}
\begin{split}
J(s,y,z;\pi)&=\E \left[ \rho \left( \widehat{\tau}^{\pi} \right) + \rho (T) u \left( Z^{\pi}_{\widehat{\tau}^{\pi}} \right) \bigg\vert Y_s = y, Z_s =z \right] - \rho (T) \\&=\E\left[\E \left[ \rho \left( \widehat{\tau}^{\pi} \right) + \rho (T) u \left( Z^{\pi}_{ \widehat{\tau}^{\pi}} \right) \Bigg \vert Y_{\alpha},Z_{\alpha}^{\pi} \right] 
\Bigg \vert Y_s=y,Z_s=z\right]- \rho (T)\\&=
\E\left[J(\alpha,Y_{\alpha},Z_{\alpha}^{\pi};\pi)
 \vert Y_s=y,Z_s=z\right]
\\&\leq \E\left[V(\alpha,Y_{\alpha},Z_{\alpha}^{\pi})\vert Y_s=y,Z_s=z \right].
\end{split}
\end{align}
Since $\pi\in \mathcal{A}_{s,T}$ is arbitrary, we have 
\begin{align*}
    V(s,y,z)\leq\sup_{\pi \in \mathcal{A}_{s,T}} \E\left[V(\alpha,Y_{\alpha},Z_{\alpha}^{\pi})\big\vert Y_s=y,Z_s=z \right]. 
\end{align*}

The proof of the reverse inequality is similar to that of
\cite[Theorem 3.3]{touzi2012optimal}. We, therefore, omit the details and only provide a sketch of the argument.


For any $\epsilon>0$, there exists $(\pi^{(s,y,z),\epsilon})\in \mathcal{A}$ such that 
$J(s,y,z;\pi^{(s,y,z),\epsilon})\geq V(s,y,z)-\epsilon$, and $\pi^{(s,y,z),\epsilon}\in \mathcal{A}_{s,T}$,
for every $(s,y,z)\in \mathbb{D}$.
Since the function $(s',y',z')\rightarrow J(s',y',z';\pi^{(s,y,z),\epsilon})$ is continuous, then for any $(s,y,z)\in \mathbb{D}$, there exist positive scalars $r_{(s,y,z)}$ such that for any $(s',y',z')\in B(s,y,z;r_{(s,y,z)})$,  $J(s,y,z;\pi^{(s,y,z),\epsilon})-J(s',y',z';\pi^{(s,y,z),\epsilon})\leq \epsilon$,
where $B(s,y,z;r)=\{(s',y',z')\in \mathbb{D}: s'\in(s-r,s),\vert y-y'\vert<r, \vert z-z'\vert<r\}$.
By the Lindelöf covering theorem, we can find a countable sequence of $(s_i,y_i,z_i;r_i)_{i\geq1}$, and the associated 
sets as open covering of $\mathbb{D}$, which are denoted by $(A_{i})_{i\geq0}$ and satisfies $(\alpha,Y_{\alpha},Z_{\alpha}^{\pi})\in \cup_{i>0}A_i$, $\P-$ a.s., $A_{i}\cap A_{j}=\emptyset$ for $i\neq j$ and $J(.;\pi^{i,\epsilon})\geq V-2\epsilon$ on $A_i$, where $\pi^{i,\epsilon}:=\pi^{(s_i,y_i,z_i),\epsilon}$.

Set $A^{n}=\cup_{0\leq i\leq n}A_i, n\geq 1$, fix $\pi\in \mathcal{A}_{s,T}$, and define for $u\in[s,T]$,
$$\pi_u^{\epsilon, n}=\mathbbm{1}_{[s,\alpha]}(u)\pi_u+\mathbbm{1}_{(\alpha,T]}(u)\left(\pi_u\mathbbm{1}_{(A^n)^c}(\alpha,Y_{\alpha},Z_{\alpha}^{\pi})+\sum\limits_{i=1}^n\mathbbm{1}_{A^i}(\alpha,Y_{\alpha},Z_{\alpha}^{\pi})\pi_u^{i,\epsilon}\right).$$ Then it can be shown after some computation that $$\E \left[ \rho \left( \widehat{\tau}^{\pi^{\epsilon, n}} \right) + \rho (T) u \left( Z^{\pi}_{ \widehat{\tau}^{\epsilon, n}} \right)-\rho(T) \Bigg \vert Y_{\alpha},Z_{\alpha}^{\pi} \right]\mathbbm{1}_{A^n}(\alpha,Y_{\alpha},Z_{\alpha}^{\pi})\geq  (V(\alpha,Y_{\alpha},Z_{\alpha}^{\pi})-2\epsilon)\mathbbm{1}_{A^n}(\alpha,Y_{\alpha},Z_{\alpha}^{\pi}),$$ which, by definition of $V$, the dominated convergence theorem, and the monotone convergence theorem that further implies $V(s,y,z)\geq -2\epsilon+\E\left[V(\alpha,Y_{\alpha},Z_{\alpha}^{\pi})\bigg\vert Y_s = y, Z_s =z\right]$.
By the arbitrariness of $\pi \in \mathcal{A}_{s,T}$ and $\epsilon>0$, we have
\begin{align}\label{DPPAuxiliary}
    V(s,y,z)\geq \sup_{\pi \in \mathcal{A}_{s,T}} \E\left[V(\alpha,Y_{\alpha},Z_{\alpha}^{\pi})\big\vert Y_s=y,Z_s=z \right].
\end{align}

\qed

\subsection{Proof of Proposition \ref{Prop:MonotonicityofV}}


Applying Itô's lemma to $\log(Z_t^{\pi})$ and by means of some straightforward calculations, we obtain the following expression for $\{Z_t^{\pi}, t>s\}$ 
\begin{align*}
    Z_t^{\pi}=ze^{\int_s^t(r +  \pi_u' \widehat{\mu}(Y_u) - \alpha (Y_u) + \beta (Y_u) \beta (Y_u)'   -\frac{1}{2}\left( \pi_u' \Sigma (Y_u) -  \beta(Y_u) \right)\left( \pi_u' \Sigma (Y_u) -  \beta(Y_u) \right)'du+ \int_s^t \left( \pi_u' \Sigma (Y_u) -  \beta(Y_u) \right) dW_u},
\end{align*}
with $Z_s=z$.
Hence, for any fixed $(s,y) \in [0,T)\times \mathbb{R}^d$ and control $\pi\in \mathcal{A}_{s,T}$, $Z_t^{\pi}$ is strictly increasing in $z$, and hence $\tau_{s,z}^{\pi}$ is decreasing in $z$. 
Due to the monotonicity of $\rho$ and $u$, for any $0<z_1< z_2<1$, we have
\begin{align*}\E\left[\rho(\widehat{\tau}_{s,z_1}^{\pi})+\rho(T)u(Z_{\widehat{\tau}_{s,z_1}^{\pi}}^{\pi})\big\vert Y_s=y, Z_s=z_1\right]<\E\left[\rho(\widehat{\tau}_{s,z_2}^{\pi})+\rho(T)u(Z_{\widehat{\tau}_{s,z_2}^{\pi}}^{\pi})\big\vert Y_s=y, Z_s=z_2\right].
\end{align*}
Let $\pi_1^*$ be an optimal strategy for the problem starting from $(s,y,z_1)$. Then
\[
V(s,y,z_2)
\geq J(s,y,z_2;\pi_1^*)
>
J(s,y,z_1;\pi_1^*)
=
V(s,y,z_1).
\]
Therefore, $V(s,y,z)$ is strictly increasing in $z$, for any $s \in [0,T)$ and $y\in \mathbb{R}^{d}$.

From now on we suppose that $Y_t=y_0$ for all $t\in\mathbb{R}_{+}$. Let $0<s_1<s_2<T$ and $z\in (0,1]$ be fixed. For any $\pi \in \mathcal{A}_{s_2,T}$, the strategy of putting all one's wealth into the riskless bond over the period $[s_1,s_2)$ and subsequently following the control $\pi$ over $[s_2,T)$ is admissible. 
Given that $r\geq0$, this strategy achieves better performance than simply starting from $(s_2,z)$ with control $\pi$, provided that the benchmark $G$ is not exceeded from time $s_1$ to $s_2$, i.e., $\tau_{s_1,z}^{\pi}>s_2$. Or otherwise if $\tau_{s_1,z}^{\pi}<s_2$, then $\tau_{s_1,z}^{\pi}<\tau_{s_2,z}^{\pi}$. Overall, it follows that 
\begin{align*}
    \sup\limits_{\pi \in \mathcal{A}_{s_1, T}}\E\left[\rho(\widehat{\tau}_{s_1,z}^{\pi})+\rho(T)u(Z_{\widehat{\tau}_{s_1,z}^{\pi}}^{\pi})\big\vert Z_{s_1}=z\right]\geq \sup\limits_{\pi \in \mathcal{A}_{s_2,T}}\E\left[\rho(\widehat{\tau}_{s_1,z}^{\pi})+\rho(T)u(Z_{\widehat{\tau}_{s_2,z}^{\pi}}^{\pi})\big\vert Z_{s_2}=z\right].
\end{align*}
Hence, the value function $V(s,y_0,z)$ is decreasing in time $s$ for any $z \in (0,1)$.
\qed

\subsection{Proof of Proposition \ref{prop:VConcave}}
Fix $(s,y,z)\in\mathbb{D}$. We first show that for every $\pi\in\mathbb{R}^N,$
\begin{align}\label{GeneratorInequality}
    \mathcal{L}^{\pi}V(s,y,z)\leq0.
\end{align}

Indeed, fix $\pi\in\mathbb{R}^N$ and use the constant control $\pi$
over a sufficiently short interval starting from $s$. Let $\tau_h$ be
the minimum of $s+h$, the goal reaching time, and the first exit time
of $(Y,Z^\pi)$ from a compact neighborhood of $(y,z)$ contained in the
interior of the state space. By Bellman's principle of optimality,
\[
V(s,y,z)
\geq
\E\left[
V(\tau_h,Y_{\tau_h},Z^\pi_{\tau_h})
\mid Y_s=y,Z_s=z
\right].
\]
Applying It\^o's formula to $V$ on $[s,\tau_h]$ gives
\[
\E\left[
\int_s^{\tau_h}
\mathcal{L}^{\pi}V(t,Y_t,Z_t^\pi)\,dt
\mid Y_s=y,Z_s=z
\right]
\leq0.
\]
Dividing by $h$ and letting $h\downarrow0$, the continuity of the
coefficients and the derivatives of $V$ yields \eqref{GeneratorInequality}.

Collecting the terms in $\mathcal{L}^{\pi}V(s,y,z)$ that depend on
$\pi$, we can write
\begin{align}\label{eq:GeneratorQuadraticForm}
\mathcal{L}^{\pi}V(s,y,z)=
K(s,y,z)
+\frac12 z^2V_{zz}(s,y,z)\pi'A(y)\pi
+\pi'q(s,y,z),
\end{align}
where $K(s,y,z)$ is independent of $\pi$ and
\[
q(s,y,z)
=
z\widehat{\mu}(y)V_z(s,y,z)
-z^2V_{zz}(s,y,z)D(y)'
+zB(y)V_{yz}(s,y,z).
\]

Suppose first, by contradiction, that \(V_{zz}(s,y,z)>0.\)
Since $A(y)$ is positive definite, for every nonzero
$e\in\mathbb{R}^N$, \(e'A(y)e>0.\)
Taking $\pi=ne$ in \eqref{eq:GeneratorQuadraticForm}, we obtain
\[
\mathcal{L}^{ne}V(s,y,z)
=
K(s,y,z)
+\frac12 n^2z^2V_{zz}(s,y,z)e'A(y)e
+n e'q(s,y,z).
\]
As $n\to\infty$, the positive quadratic term dominates the linear term,
and hence \(\mathcal{L}^{ne}V(s,y,z)\longrightarrow+\infty,\)
contradicting \eqref{GeneratorInequality}. Therefore,
\(V_{zz}(s,y,z)\leq0.\)
Since $(s,y,z)$ was arbitrary, $V$ is concave in $z$ on $\mathbb{D}$.

We next show that $V_z(s,y,z)>0$. By Proposition~\ref{Prop:MonotonicityofV}, $V(s,y,\cdot)$ is strictly increasing, while the concavity established above implies that $V_z(s,y,\cdot)$ is non-increasing. Since $V$ is differentiable and increasing, $V_z(s,y,z)\geq0$. If $V_z(s,y,z_0)=0$ for some $z_0\in(0,1)$, then concavity implies $V_z(s,y,z)\leq0$ for all $z>z_0$. Hence $V_z(s,y,z)=0$ for all $z>z_0$, so $V(s,y,\cdot)$ would be constant on $(z_0,1)$, contradicting strict monotonicity. Therefore,
\(V_z(s,y,z)>0.\)

Finally, suppose that $B(y)=0$ and $\widehat{\mu}(y)\neq0$. If
$V_{zz}(s,y,z)=0$ at some $(s,y,z)\in\mathbb{D}$, then
\(q(s,y,z)
=
z\widehat{\mu}(y)V_z(s,y,z).\)
Since $V_z(s,y,z)>0$ and $\widehat{\mu}(y)\neq0$, we have
$q(s,y,z)\neq0$. Taking $\pi=nq(s,y,z)$ in
\eqref{eq:GeneratorQuadraticForm}, the quadratic term vanishes, and hence
\[
\mathcal{L}^{nq}V(s,y,z)
=
K(s,y,z)+n|q(s,y,z)|^2
\longrightarrow+\infty
\]
as $n\to\infty$. This contradicts
\eqref{GeneratorInequality}. Therefore, \(V_{zz}(s,y,z)<0.\)
Since \((s,y,z)\in\mathbb D\) was arbitrary, the result follows.
\qed

\subsection{Proof of Theorem \ref{thm:Verification}}

By the dynamics of the ratio process \eqref{RatioProcess}, Proposition \ref{Prop:DPP}, and \cite[Theorem 11.2.1]{oksendal2013stochastic}, one can write down the corresponding HJB equation for the value function $V$ in \eqref{DPE0-0}. 
This equation is to be considered in the set $\mathbb{D}$ 
with boundary data $V(s,y,1)=\rho(s)$. Additionally, in the case of a finite terminal horizon $T$, one needs to further impose boundary condition on $\{s=T\}$, namely, $V  (T,y,z)=\rho(T)u(z)$.


We may then use standard calculus for optimization with respect to $\pi$ to obtain optimal policy $\pi^{*}$ in terms of the value function, in a feedback form given by \eqref{Optimalf}.
Next, we substitute \eqref{Optimalf} back into equation (\ref{DPE0-0}), and obtain \eqref{DPE}
for $(s,y,z)\in \mathbb{D}$ with boundary condition $v(s,y,1)=\rho(s)$, and $v(T,y,z)=\rho(T)u(z)$ if given $T<+\infty$.

It remains to confirm whether the policy corresponding to each initial point, derived from solving the HJB equation \eqref{DPE0-0}, is indeed optimal. Note that the value function is bounded as argued in Proposition \ref{prop:ContinuityofV}, it is natural to seek bounded solutions to the PDE. 
Therefore, the family $\{v(\tau,Y_{\tau},Z_{\tau}^{s,z,\pi}): \tau \ {\rm{stopping}}\ {\rm{time}}, s\leq \tau \leq \widehat{\tau}_{s,z}^{\pi}\}$ is uniformly integrable for any  admissible strategy $\pi \in \mathcal{A}_{s,T}$ and one can readily apply Theorem 11.2.2 in \cite{oksendal2013stochastic} for verification.
\qed

\subsection{Proof of Corollary \ref{corollary-HJB}}
Follows immediately from simplifying expressions in Theorem \ref{thm:Verification}. 
\qed


\subsection{Proof of Theorem \ref{thm:InfHorizonCharacterization}}
We first show the vice versa statement and let $h:\mathbb{R}_+ \times \mathbb{R} \rightarrow \mathbb{R}$ be a strictly increasing solution to \eqref{PDEforFunctionh} such that  \eqref{Lowerbound} holds. Let $v$ and $\rho$ be defined through \eqref{valuefunctionAfterNormalization} and \eqref{rhoAfterNormalization} respectively. 
The condition  $v(s,1)=\rho(s)$ follows directly by integration.
By computing partial derivatives 
\begin{align*}
v_z(s,z)
&=
\frac{1}{2L}\int_0^{2\Theta(s)}
e^{-h^{(-1)}(t,z)+\frac{t}{2}}
\left(
1-
\frac{
h_{zz}\left(t,h^{(-1)}(t,z)\right)
}{
h_z\left(t,h^{(-1)}(t,z)\right)
}
\right)dt
+\frac{1}{L}e^{-h^{(-1)}(0,z)}
\\
&=
\frac{1}{2L}\int_0^{2\Theta(s)}
e^{-h^{(-1)}(t,z)+\frac{t}{2}}
\left(
1+
2\frac{
h_t\left(t,h^{(-1)}(t,z)\right)
}{
h_z\left(t,h^{(-1)}(t,z)\right)
}
\right)dt
+\frac{1}{L}e^{-h^{(-1)}(0,z)}
\\
&=
\frac{1}{L}\int_0^{2\Theta(s)}
e^{-h^{(-1)}(t,z)+\frac{t}{2}}
\left(
\frac{1}{2}
-
\frac{\partial}{\partial t}h^{(-1)}(t,z)
\right)dt
+\frac{1}{L}e^{-h^{(-1)}(0,z)}
\\
&=
\frac{1}{L}
e^{-h^{(-1)}(t,z)+\frac{t}{2}}
\Big|_{0}^{2\Theta(s)}
+\frac{1}{L}e^{-h^{(-1)}(0,z)}
\\
&=
\frac{1}{L}
e^{-h^{(-1)}(2\Theta(s),z)+\Theta(s)},
\\[0.3em]
v_{zz}(s,z)
&=
-\frac{
e^{-h^{(-1)}(2\Theta(s),z)+\Theta(s)}
}{
Lh_z\left(
2\Theta(s),
h^{(-1)}(2\Theta(s),z)
\right)
},
\\[0.3em]
v_s(s,z)
&=
-\frac{1}{2L}\kappa(s)'\kappa(s)\,
e^{-h^{(-1)}(2\Theta(s),z)+\Theta(s)}
h_z\left(
2\Theta(s),
h^{(-1)}(2\Theta(s),z)
\right).
\end{align*}
The above expressions show that $v \in C^{1,2}\left(\mathbb{R}_{+}\times(0,1]\right)$ and satisfies \eqref{DPETime}. Since $v_z>0$ and $v(s,1)=\rho(s)\leq 1$, the assumed positivity of $v$ implies that $0<v(s,z)\leq 1$ for all $(s,z)\in\mathbb{R}_+\times(0,1]$. Hence, Theorem \ref{thm:Verification} yields that $v$ is the value function.

To prove the other direction: By Proposition and \ref{Prop:MonotonicityofV} and Proposition \ref{prop:VConcave}, 
it follows that $v$, as the value function, is an increasing and strictly concave function defined for $(s,z)\in \mathbb{R}_{+}\times(0,1]$ that satisfies $(\ref{DPETime})$ and $v(0,1)=1$. By the strict concavity of $v$, we know that $v_z$ is invertible with $(v_z)^{(-1)}:\mathbb{R}_{+}\times (0,+\infty)\rightarrow (0,1]$. Then define for $(s,z)\in \mathbb{R}_{+}\times(0,1]$, the function $h$ by
\begin{align}\label{functionh}
    h(s,z)=(v_z)^{(-1)}\left(\Theta^{(-1)}(\frac{1}{2}s), e^{-z+\frac{1}{2} s}\right).
\end{align}
By $h(2\Theta(s),z)=(v_z)^{(-1)}\left(s, e^{-z+\Theta(s)}\right)$,
(\ref{functionh}) is equivalent to $v_z(s,z)=e^{-h^{(-1)}(2\Theta(s),z)+\Theta(s)}$.

Next, we obtain by differentiating that
\begin{align*}
    h_z(s,z)=-\frac{e^{-z+\frac{1}{2} s}}{v_{zz}(\Theta^{(-1)}(\frac{s}{2}), h(s,z))}>0,
\end{align*}
thus function $h$ is increasing in the space variable. Also, one can show that $h$ satisfies PDE (\ref{PDEforFunctionh}) by \cite[Lemma 27]{musiela2010portfolio} 
and computing the associated derivatives. 

Furthermore, equations $(\ref{DPETime})$ and $(\ref{functionh})$ yield
\begin{align*}
    v_s(s,z)=-\frac{1}{2} \kappa(s)' \kappa(s) e^{-h^{(-1)}(2\Theta(s),z)+\Theta(s)}h_z(2\Theta(s),h^{(-1)}(2\Theta(s),z)).
\end{align*}
Direct integration yields that 
\begin{align*}
v(s,z)=\int_0^{s}v_s(t,z)dt-\int_z^{1}v_z(0,x)dx+v(0,1),
\end{align*}
after plugging in the form of $v_s, v_z$ and $v(0,1)=1$, we have
\begin{equation*}
\begin{aligned}
    v(s,z)=1-\frac{1}{2} \int_0^{s}\kappa(t)' \kappa(t) e^{-h^{(-1)}(2\Theta(t),z)+\Theta(t)}h_z\left(2\Theta(t),h^{(-1)}(2 \Theta(t),z)\right)dt+\int_1^ze^{-h^{(-1)}(0,x)}dx.
\end{aligned}
\end{equation*}
It can be shown, after some simple transformation (change of variable), that 
\begin{align*}
    v(s,z)=1-\frac{1}{2}\int_0^{2\Theta(s)}e^{-h^{(-1)}(t,z)+\frac{t}{2}}h_z\left(t,h^{(-1)}(t,z)\right)dt+\int_1^ze^{-h^{(-1)}(0,x)}dx,
\end{align*}
and moreover,
\begin{align*}
    v(s,1)=1-\frac{1}{2}\int_0^{2\Theta(s)}e^{-h^{(-1)}(t,1)+\frac{t}{2}}h_z\left(t,h^{(-1)}(t,1)\right)dt.
\end{align*}

Consider $L$ as defined by \eqref{Lowerbound}. Noting that $1-L=\lim\limits_{s\rightarrow +\infty}v(s,1)=\rho(+\infty)=0,$ hence $L=1$. 
It follows that $v(s,z)$ admits the representation given in \eqref{valuefunctionAfterNormalization}.
Since $\rho(s) = v(s,1)$, the corresponding expression for $\rho$ follows immediately.

The expression for the optimal strategy follows from \eqref{Optimalf} and simplifying
\begin{align*}
\pi^*(s,z)=-\Sigma (s)^{-1}\widehat{\mu} (s) \left(\frac{v_z(s,z)}{zv_{zz}(s,z)}\right)=\Sigma(s)^{-1} \widehat{\mu} (s) \frac{h_z\left(2\Theta(s), h^{(-1)}(2\Theta(s),z)\right)}{z}.
\end{align*}

\qed

\subsection{Proof of Proposition \ref{SufficientCondforLfinite}} 
We prove only the case of $\lim\limits_{s\to+\infty}\Theta(s)=\infty$ herein, as in the case of $\lim\limits_{s\to+\infty}\Theta(s)$ being finite, $L<\infty$ follows immediately.
Let $z(t) := h^{(-1)}(t,1),$
so that $h(t,z(t)) = 1$ for all $t \ge 0$.  
Since 
$1 = h(t,z(t)) \ \ge \ \frac{C}{\gamma} e^{\gamma z(t) - \frac{1}{2} \gamma^2 t},$
taking logarithms yields
\[
\gamma z(t) - \frac{1}{2} \gamma^2 t + \log\!\Big(\frac{C}{\gamma}\Big) \le 0,
\]
hence
\begin{equation}\label{eq:zt-upper}
z(t) \ \le \ \frac{\gamma}{2} t + \frac{1}{\gamma} \log\!\Big( \frac{\gamma}{C} \Big) 
=: \frac{\gamma}{2} t + K_0.
\end{equation}

Now, the integrand can be written as
\[
e^{-z(t) + \frac{t}{2}} \, h_z\big(t, z(t)\big)
= e^{-z(t) + \frac{t}{2}} \left[
C e^{\gamma z(t) - \frac{1}{2} \gamma^2 t} + \sum_{i=1}^N \partial_z h^{(i)}\big(t, z(t)\big)
\right].
\]
Using \eqref{eq:zt-upper},
$
e^{-z(t) + \frac{t}{2}} C e^{\gamma z(t) - \frac{1}{2} \gamma^2 t}
= C \, e^{(\gamma-1) z(t) - \frac{1}{2} (\gamma^2 - 1) t}
\le C \, e^{(\gamma-1) K_0} \, e^{ - \frac{1}{2} (\gamma - 1) t}.
$
Since $\gamma > 1$, this bound is exponentially decaying and integrable on $[0,\infty)$.

By assumption, there exist $\gamma_i>\gamma-1$, $K_i$ and $T_{K_i}$
such that for all $t \ge T_{K_i}$,
\[
\partial_z h^{(i)}\big(t, z(t)\big) 
\ \le \ K_i \, e^{\gamma_i z(t) - \frac{1}{2} \gamma_i^2 t}.
\]
Thus, using the same calculation as for the first term,
\[
e^{-z(t) + \frac{t}{2}} \partial_z h^{(i)}\big(t, z(t)\big)
\ \le\ K_i \, e^{(\gamma_i-1) K_0} e^{ - \frac{1}{2} (\gamma_i - 1)(\gamma_i + 1-\gamma) t},
\quad t \ge T_{K_i}.
\]
Therefore, the remainder term is also exponentially decaying in $t$.

Since there exist constant $T_K>0$ such that each term is integrable on $[T_K,\infty)$, and on $[0,T_K]$ the integrand is bounded because $z(t)$ and $h_z(t,z(t))$ are continuous.  
Hence
\[
\int_0^\infty e^{-h^{(-1)}(t,1) + \frac{t}{2}} \, h_z\!\left(t, h^{(-1)}(t,1)\right) dt < \infty.
\]
By the monotone convergence theorem and $2\Theta(s) \to \infty$ as $s \to \infty$, we conclude that $L<\infty$.

\qed

\subsection{Proof of Proposition \ref{Prop:Convergence}}

Consider any $T>0$ and $(s,y,z)\in [0,T]\times\mathbb{R}^{d}\times(0,1]$. 
The reward functional $J^T(s,y,z;\pi)$ of the finite-horizon problem is given by (\ref{eq:Objective}), and the reward functional $J^{\infty}(s,y,z;\pi)$ of the infinite-horizon problem reduces to $J^{\infty}(s,y,z;\pi)=\E\left[\rho(\tau^{\pi})\vert Y_s=y, Z_s=z\right]$.
Therefore, for each fixed investment strategy $\pi \in\mathcal{A}_{s, \infty}$, the difference between $J^T(s,y,z;\pi)$ and $J^{\infty}(s,y,z;\pi)$ is computed by
\begin{equation}\label{DiffJ^TJ^infty}
    \begin{aligned}
    \big\vert J^T(s,y,z;\pi)-J^{\infty}(s,y,z;\pi)\big\vert=&
    \bigg\vert\E\left[\mathbbm{1}_{\{\tau^{\pi}\leq T\}}\rho(\tau^{\pi})+\mathbbm{1}_{\{\tau^{\pi}> T\}}\rho(T)u(Z_{T}^{\pi})-\rho(\tau^{\pi})\vert Y_s=y, Z_s=z\right]\bigg\vert\\=&\bigg\vert\E\left[\mathbbm{1}_{\{\tau^{\pi}> T\}}(\rho(T)u(Z_{T}^{\pi})-\rho(\tau^{\pi}))\vert Y_s=y, Z_s=z\right]\bigg\vert
    \\ \leq &\E\left[\mathbbm{1}_{\{\tau^{\pi}> T\}}\big\vert\rho(T)u(Z_{T}^{\pi})-\rho(\tau^{\pi})\vert\big\vert Y_s=y, Z_s=z\right]
    \\=&\rho(T)\E\left[\mathbbm{1}_{\{\tau^{\pi}> T\}}\Bigg\vert u(Z_{T}^{\pi})-\frac{\rho(\tau^{\pi})}{\rho(T)}\bigg\vert\Bigg\vert Y_s=y, Z_s=z\right]
    \\\leq& \rho(T).
\end{aligned}
\end{equation}

The last inequality holds because the terminal function \( u \) is valued between $(0,1]$, and the function \( \rho \) is strictly decreasing, thus \( \frac{\rho(\tau^{\pi})}{\rho(T)} \in (0,1) \) if \( \tau^{\pi} > T \). Letting $T\rightarrow \infty$ we obtain, since $\rho(T)\rightarrow 0$, that for any $\pi \in\mathcal{A}_{s, \infty}$, $$\big\vert J^T(s,y,z;\pi)-J^{\infty}(s,y,z;\pi)\big\vert \rightarrow 0.$$

Consider a sequence of admissible policies $\{\pi_n^T\}_{n=1}^{\infty}$ in $\mathcal{A}_{s, T}$, and respectively a sequence of admissible policies $\{\pi_n^{\infty}\}_{n=1}^{\infty}$ in $\mathcal{A}_{s, \infty}$, such that $\lim\limits_{n\rightarrow \infty}J^T(s,y,z;\pi_n^T)=V^{T}(s,y,z)$, and\break$\lim\limits_{n\rightarrow \infty}J^{\infty}(s,y,z;\pi_n^{\infty})=V^{\infty}(s,y,z)$.

On the one hand, $V^T(s,y,z)=\lim\limits_{n\rightarrow \infty}J^T(s,y,z;\pi_n^{T})\geq \lim\limits_{n\rightarrow \infty}J^T(s,y,z;\pi_n^{\infty})$ for any $T>0$. Thus,  
$$\lim\limits_{T\rightarrow \infty} V^T(s,y,z)\geq 
\lim\limits_{T\rightarrow \infty}\lim\limits_{n\rightarrow \infty}J^T(s,y,z;\pi_n^{\infty})=\lim\limits_{n\rightarrow \infty}J^{\infty}(s,y,z;\pi_n^{\infty})=V^{\infty}(s,y,z).$$ 
On the other hand, since $V^{\infty}(s,y,z)=\lim\limits_{n\rightarrow \infty}J^{\infty}(s,y,z;\pi_n^{\infty})\geq \lim\limits_{n\rightarrow \infty}J^{\infty}(s,y,z;\pi_n^{T,e})$, for any $T>0$ and any $\pi_n^{T,e}\in\mathcal{A}_{s, \infty}$ satisfying $\pi_n^{T,e}(\widehat{s},\widehat{y},\widehat{z})=\pi_n^{T}(\widehat{s},\widehat{y},\widehat{z})$ on $(\widehat{s},\widehat{y},\widehat{z})\in[s,T)\times\mathbb{R}^{d}\times (0,1]$. Then we have
$$V^{\infty}(s,y,z)\geq \lim\limits_{n\rightarrow \infty}
\lim\limits_{T\rightarrow \infty}J^T(s,y,z;\pi_n^{T,e})=\lim\limits_{T\rightarrow \infty}V^T(s,y,z).$$
Combining the above two inequalities, we obtain $\lim\limits_{T\rightarrow \infty} V^T(s,y,z)=V^{\infty}(s,y,z)$ and concludes the proof.
\qed

\subsection{Proof of Proposition \ref{prop:ConvergenceStrat}}
Consider any $(s,y,z)\in [0,\infty)\times\mathbb{R}^{d}\times(0,1]$. 
For every $\varepsilon>0$, choose $T_0>s$ sufficiently large such that $\rho(T_0)\le \frac{1}{3}\eta$.
Then, for any $T\ge T_0$, we have from \eqref{DiffJ^TJ^infty} that $J^T(s,y,z;\pi)\le J^\infty(s,y,z;\pi)+\rho(T)$ for any admissible $\pi\in\mathcal{A}$.
Taking the supremum yields
\begin{align*}
\sup_{{d(\pi,\pi^{*,\infty})\ge\varepsilon}}J^T(s,y,z;\pi)\;\le\;\sup_{{d(\pi,\pi^{*,\infty})\ge\varepsilon}}J^\infty(s,y,z;\pi)+\rho(T).
\end{align*}

Equation \eqref{DiffJ^TJ^infty} also yields $J^T(s,y,z;\pi^{*,\infty})\ge J^\infty(s,y,z;\pi^{*,\infty})-\rho(T)$.
Hence, 
\begin{align*}
    &J^T(s,y,z;\pi^{*,\infty})-\sup_{d(\pi,\pi^{*,\infty})\ge\varepsilon}J^T(s,y,z;\pi)\\\ge &J^{\infty}(s,y,z;\pi^{*,\infty})-\sup_{d(\pi,\pi^{*,\infty})\ge\varepsilon}J^\infty(s,y,z;\pi)-2\rho(T)\\\ge&\frac{1}{3}\eta>0,
\end{align*}
and the optimizer $\pi^{*,T}$ must therefore satisfy $d(\pi^{*,T},\pi^{*,\infty})<\varepsilon$ for all $T\ge T_0$. 
By the arbitrariness of $\varepsilon>0$, we have $d(\pi^{*,T},\pi^{*,\infty})\to0$ as $T\to\infty$.
\qed

\bibliography{BibFile}

\end{document}